\documentclass[twocolumn,10pt]{IEEEtran}
\usepackage{acronym}
\usepackage{cite}
\usepackage{graphicx}
\usepackage{amsmath}
\usepackage{booktabs}
\usepackage{threeparttable}
\usepackage{algorithm}
\usepackage{algpseudocode}
\usepackage{xcolor}
\usepackage{multirow}
\usepackage{tabularx} 
\usepackage{dblfloatfix} 
\usepackage{verbatim}
\usepackage{bm}
\usepackage{makecell}
\usepackage{empheq} 
\usepackage{underoverlap} 
\usepackage{newtxmath}
\usepackage{subcaption}
\usepackage{hyperref}

\usepackage{enumitem} 
\usepackage{calc}

\usepackage{textcomp}
\def\BibTeX{{\rm B\kern-.05em{\sc i\kern-.025em b}\kern-.08em
		T\kern-.1667em\lower.7ex\hbox{E}\kern-.125emX}}

\newcommand{\yf}{{\mathbf{y}}}

\newcommand{\Yf}{\mathbf{Y}}

\newcommand{\st}{\mathbf{s}}

\newcommand{\xt}{\mathbf{x}}
\newcommand{\Xt}{\mathbf{X}}
\newcommand{\xf}{{\mathbf{x}}}

\newcommand{\Zt}{\mathbf{Z}}
\newcommand{\St}{\mathbf{S}}

\newcommand{\Ht}{\mathbf{H}}

\newcommand{\av}{\mathbf{a}}

\newcommand{\Pc}{\mathbf{P}}

\DeclareMathOperator*{\argmax}{argmax} 

\newcommand{\bb}{} 

\begin{document}

\title{Enabling Joint Communication and Radar Sensing in Mobile Networks - A Survey}

\author{{J. Andrew Zhang,~\IEEEmembership{Senior~Member,~IEEE}, Md Lushanur Rahman, Kai Wu, Xiaojing Huang,~\IEEEmembership{Senior~Member,~IEEE}, \\
		Y. Jay Guo,~\IEEEmembership{Fellow,~IEEE}, Shanzhi Chen~\IEEEmembership{Fellow,~IEEE}, and Jinhong Yuan,~\IEEEmembership{Fellow,~IEEE}} 
	\thanks{The paper was accepted for publications in IEEE Communications Surveys and Tutorials in Oct. 2021.}
\thanks{This work is partially supported by the Australian Research Council under grant No. DP210101411.}
\thanks{J. Andrew Zhang, Md Lushanur Rahman, Kai Wu, Xiaojing Huang and Y. Jay Guo are with the Global Big Data Technologies Centre (GBDTC) and the School of Electrical and Data Engineering, The University of Technology Sydney (UTS), Sydney, NSW 2007, Australia. Email: \{Andrew.Zhang; Kai.Wu; Xiaojing.Huang; Jay.Guo\}@uts.edu.au; lushanur.rahman@gmail.com.}
\thanks{Shanzhi Chen is with China Academy of Telecommunications Technology (CATT), Beijing, 100191, China. Email: chensz@cict.com.}
\thanks{Jinhong Yuan is with the University of New South Wales, Sydney, NSW 1466, Australia. Email: j.yuan@unsw.edu.au.}
}

\maketitle

\begin{abstract}
 Mobile network is evolving from a communication-only network towards one with joint communication and radar/radio sensing (JCAS) capabilities, that we call perceptive mobile network (PMN). Radio sensing here refers to information retrieval from received mobile signals for objects of interest in the environment surrounding the radio transceivers, and it may go beyond the functions of localization, tracking, and object recognition of traditional radar. In PMNs, JCAS integrates sensing into communications, sharing a majority of system modules and the same transmitted signals. \bb{The PMN is expected to provide a ubiquitous radio sensing platform and enable a vast number of novel smart applications, whilst providing non-compromised communications. In this paper, we present a broad picture of the motivation, methodologies, challenges, and research opportunities of realizing PMN, by providing a comprehensive survey for systems and technologies developed mainly in the last ten years.} Beginning by reviewing the work on coexisting communication and radar systems, we highlight their limits on addressing the interference problem, and then introduce the JCAS technology. We then set up JCAS in the mobile network context and envisage its potential applications. We continue to provide a brief review of three types of JCAS systems, with particular attention to their differences in design philosophy. We then introduce a framework of PMN, including the system platform and infrastructure, three types of sensing operations, and signals usable for sensing. Subsequently, we discuss required system modifications to enable sensing on current communication-only infrastructure. Within the context of PMN, we review stimulating research problems and potential solutions, organized under nine topics: performance bounds, waveform optimization, antenna array design, clutter suppression, sensing parameter estimation, resolution of sensing ambiguity, pattern analysis, networked sensing under cellular topology, and sensing-assisted communications. \bb{We conclude the paper by listing key open research problems for the aforementioned topics and sharing some lessons that we have learned.} 
\end{abstract}
\section{Introduction}\label{sec:intro}

\subsection{Abbreviations}

A list of the major abbreviations used in this paper is provided in Table \ref{table:abbv}.

\begin{table}[t]
	
\footnotesize
	\centering
	\caption{List of abbreviations.}
	\begin{tabular} {ll}
		
		\hline
		\textbf{Abbreviations} & \textbf{Meanings} \\
		\hline
		AoA
		&
		Angle of arrival 
		\\
		\hline
		AoD
		&
		Angle of departure
		\\
		\hline
		ANM
		&
		Atomic norm minimization
		\\
		\hline
		BBU
		&
		Baseband unit
				\\
		\hline
		BS
		&
		Base-station
		\\
		\hline
		CACC
		&
		Cross-antenna cross-correlation
		\\
		\hline
			CPI &
		Coherent processing interval
		\\
		\hline
		CRAN
		&
		Cloud radio access network
		\\
		\hline
		CS
		&
		Compressive sensing
		\\
		\hline
		CSI
		&
		Channel state information
		\\
		\hline
		C\&S
		&
		Wireless communications and \\
		& radar/radio sensing
		\\
		\hline
		DFRC
		& Dual-function(al) radar communications
		
		\\
		\hline
			DFT
			&
		Discrete Fourier Transform
		\\
		\hline
		DMRS
		&
		Demodulation reference signals
		\\
		\hline		
		FDD
		&
		Frequency division duplexing
		\\
		\hline
		FMCW
		&
		Frequency-modulated continuous-wave
		\\
		\hline
		GMM
		&
		Gaussian mixture model
		\\
		\hline
		IFFT
		&
		Inverse fast Fourier transform
		\\
		\hline
		IoT
		&
		Internet of things
		\\
		\hline
		JCAS
		&
		Joint communications and radio/radar sensing
		\\
		\hline
		LFM
		&
		Linear frequency modulation 
		\\
		\hline
		LFM-CPM
		&
		LFM-continuous phase modulation
		\\
		\hline
		LOS
		& Line of sight
		\\
		\hline
		MAC
		&
		Medium access
		\\
		\hline
		MI
		&
		Mutual information
		\\
		\hline
		MIMO
		&
		Multiple-input and multiple-output
		\\
		\hline
		MISO
		&
		Multiple-input and single-output
		\\
		\hline
		MMSE
		&
		Minimum mean-square error
		\\
		\hline
		MMV
		&
		Multi measurement vector
		\\
		\hline
		mmWave
		&
		Millimeter wave		\\
		\hline
		NLOS
		&
		None line of sight
		\\
		\hline
		NR
		&
		New radio
		\\
		\hline
		OFDM
		&
		Orthogonal frequency-division multiplexing
		\\
		\hline
		OFDMA
		&
		Orthogonal frequency-division multiple access
		\\
		\hline
		PHY
		&
		Physical
		\\
		\hline
		PAPR
		&
		Peak-to-average power ratio
		\\
		\hline
		PMN
		&
		Perceptive mobile network
		\\
		\hline
		PDSCH
		&
		Physical downlink shared channel
		\\
		\hline
		PUSCH
		&
		Physical uplink shared channel
		\\
		\hline
		PRB
		&
		Physical resource-block		\\
		\hline
		RFID
		&
		Radio-frequency identification
		\\
		\hline
		RIP
		&
		Restricted isometry property
		\\
		\hline
		RIS
		&Reconfigurable intelligent surface
			\\
		\hline

		RMA
		&
		Recursive moving averaging
		\\
		\hline
		RMSE
		&
		Root mean square error
		\\
		\hline
			RRU
		&
		Remote radio unit
		\\
		\hline
		Rx&
		Receiver
			\\
		\hline
		SC
		&
		Single carrier
		\\
		\hline
		SDMA
		&
		Spatial division multiple access
		\\
		\hline
		SISO
		&
		Single input single output
		\\
		\hline
		SRS
		&
		Sounding reference signals
		\\
		\hline
		SSB
		&
		Synchronization signal and broadcast blocks
		\\
		\hline
		TDD
		&
		Time-division duplexing
		\\
		\hline
		Tx & Transmitter
			\\
		\hline
		UE
		&
		User equipment
		\\
		\hline
		ULA
		&
		Uniform linear array
		\\
		\hline
		V2V
		&
		Vehicle to vehicle
		\\
		
		\hline
	\end{tabular}
	\label{table:abbv}
\end{table}

\subsection{Background}

Wireless communication and radar sensing (C\&S) have been advancing in parallel for decades, yet with limited intersections.  They share many commonalities in terms of signal processing algorithms, devices and, to a certain extent, system architecture. This has recently motivated significant research interests in the coexistence, cooperation, and joint design of the two systems \cite{RN32, 8828016, RN15,7465731,liu2019joint,Hassanien_2019, 8828030, 8827589}. 

The coexistence of communication and radar systems has been extensively studied in the past decade, with a focus on developing efficient interference management techniques so that the two individually deployed systems can operate smoothly without interfering with each other \cite{li2016optimum,6331681,8267374,7536903,7898445}. Although radar and communication systems may be co-located and even physically integrated, they transmit two different signals overlapped in time and/or frequency domains. They operate simultaneously by sharing the same resources cooperatively, with a goal of minimizing interference to each other. Great efforts have been devoted to mutual interference cancellation in this case, using, for example, beamforming design in \cite{7898445}, cooperative spectrum sharing in \cite{8835700},  opportunistic primary-secondary spectrum sharing in \cite{6331681}, and dynamic coexistence in  \cite{7536903}. However, effective interference cancellation typically has stringent requirements on the mobility of nodes and information exchange between them. The spectral efficiency improvement is hence limited in practice.

Since the interference in coexisting systems is caused by transmitting two separate signals, it is natural to ask whether one can use a single transmitted signal for both communications and radar sensing. Radar systems typically use specifically designed waveforms such as short pulses and chirps, which enable high power radiation and simple receiver processing  \cite{4383615}. However, these waveforms are not necessary for radar sensing. \textit{Passive radar} or \textit{passive sensing} is a good example of exploring diverse radio signals for sensing \cite{Hack14,Gogineni14, Abdullah2016, 8666692}. In principle, the objects to be sensed can be illuminated by any radio signal of sufficient power, such as  TV signals \cite{5393298}, WiFi signals \cite{6020778}, and mobile (cellular) signals \cite{RN48, RN43, doi:10.1002/wat2.1289}.  This is because the propagation of radio signals is always affected by environmental dynamics such as transceiver movement, surrounding objects' movement and profile variation, and even weather changes. Hence the environmental information is encoded to the received radio signals and can be extracted by using passive radar techniques. However, there are two major limitations with passive sensing. Firstly, the clock phases between transmitter and receiver are not synchronized in passive sensing and there is always an unknown and possibly time-varying timing offset between the transmitted and received signals. This leads to timing and therefore ranging ambiguity in the sensing results, as well as causing difficulties in aggregating multiple measurements for joint processing. Secondly, the sensing receiver may not know the signal structure. As a result, passive sensing lacks the capability of interference suppression, and it cannot separate multiuser signals from different transmitters. Of course, the radio signals are not optimized for sensing in any way.

Traditional radar is evolving towards more general \textit{radio sensing}. We prefer the term radio sensing to radar due to its generality and comprehensiveness. Radio sensing here can widely refer to retrieving information from received radio signals, other than the communication data modulated to the signal at the transmitter. It can be achieved through the measurement of both \textit{ sensing parameters} related to location and moving speed, such as time delay, angle-of-arrival (AoA), angle-of-departure (AoD), Doppler frequency and magnitude of a multipath signal, and \textit{physical feature parameters} (such as inherent pattern signals of devices/objects/activities), using radio signals. The two corresponding processing activities are called \textit{sensing parameter estimation} and \textit{pattern recognition} in this paper. In this sense, radio sensing refers more to general sensing techniques and applications using radio signals, like video sensing using video signals. Radio sensing involves more diverse applications such as object, activity and event recognition in Internet of Things (IoT), WiFi and 5G networks \cite{article4}. In  \cite{7465731}, the authors described the ubiquitous use of wireless technologies such as WiFi, Bluetooth, FM radio and mobile cellular networks, as signals of opportunity in the implementation of IoT. These radio signals are transmitted by an existing infrastructure and are not specifically designed for the sensing purpose. In \cite{8794643}, the authors surveyed works on WiFi sensing where WiFi signals can be used for people and behavior recognition in an indoor environment. In \cite{article23}, it is shown that other radio signals, such as RFID and ZigBee, can also be used for activity recognition. These publications demonstrate the strong potential of using low-bandwidth communication signals for radio sensing applications. 

\begin{table*}[th]
	\caption{Comparison of C\&S with separated waveforms, coexisting C\&S, passive sensing, cognitive radio, and JCAS.} 
	\centering
	\bgroup
	\def\arraystretch{1.5}
	\begin{tabular}{|m{1.5cm}|m{4.5cm}|m{4cm}|m{6cm}|} 
		\hline 
		\textbf{Systems} & \centering{\textbf{Signal Formats and Key Features}} & \centering{\textbf{Advantages}} & \hspace{1.5cm}\textbf{Disadvantages}    \\ 
		\hline \hline
		\textbf{C\&S with Separated Waveforms (e.g., \cite{RN14})} & - C\&S signals are separated in time, frequency, code and/or polarization; 
		
		- C\&S hardware and software are partially shared.  & - Small mutual interference;
		
		- Almost independent design of  C\&S waveforms.
		& - Low spectrum efficiency; 
		
		- Low order of integration; 
		
		- Complex transmitter hardware.	   \\ 
		
		\hline 
		\textbf{Coexisting C\&S (e.g.,\cite{li2016optimum,7898445})} & C\&S use separated signals but share the same resource.  & Higher spectrum efficiency & - Interference is a major issue;
		
		- Nodes cooperation and complicated signal processing are typically required. \\ 
		
		\hline 
		\textbf{Passive sensing (e.g., \cite{Hack14,Gogineni14, Abdullah2016, 8666692})} & - Received radio signals are used for sensing at a specifically designed sensing receiver, external to the communication system; 
		
		- No joint signal design at transmitter. & - Without requiring any change to existing infrastructure;
		
		- Higher spectrum efficiency.  & - Require dedicated sensing receiver;
		
		- Timing ambiguity; 
		
		- No waveform optimization.
		
		- Non-coherent sensing and limited sensing capability when signal structure is complicated and unknown, e.g., incapable of separating multi-user signals from different transmitters; 
		\\ 
		\hline
		\textbf{Cognitive Radio (e.g.,\cite{6331681})} & Secondary systems coexist with primary ones by sensing spectrum holes or via interference mitigation. & - Improved spectrum efficiency; 
		
		- Negligible impact on the operation of primary systems.  & Performance of secondary systems cannot be guaranteed. They also have higher complexities due to requirement for spectrum sensing and potential interference suppression.
		\\
		
		\hline 
		\textbf{JCAS (e.g., \cite{8828030, 8550811,liu2019joint,Ma20,luong2021radio})} & A common transmitted signal is jointly designed and used for C\&S.  & - Highest spectral efficiency; 
		
		- Fully shared transmitter and largely shared receiver; 
		
		- Joint design and optimization on waveform, system and network; 
		
		- ``Coherent sensing''.
		&  - Requirement for full-duplex or equivalent capability of a receiver co-locating with the transmitter; 
		
		- Sensing ambiguity when transmitter and receiver are separated without clock synchronization.
		\\ 
		\hline 	
	\end{tabular}
	\egroup
	
	\label{tab-compareschemes}
\end{table*}

\textit{Joint communication and radar/radio sensing} (JCAS) \cite{RN9,RN24,8550811} is emerging as an attractive solution for integrating communication and sensing into one system. It has also been known under different terms, such as Radar-communications (RadCom) \cite{RN32}, joint radar (and) communications (JRC) \cite{8828030,liu2019joint,Feng20}, joint communications (and) radar (JCR) \cite{Kumari17, Kumari20}, dual-function(al) radar communications (DFRC) \cite{Hassanien16, Liu18, Ma20}, and more recently, integrated sensing and communications (ISAC). Here, we use JCAS in more of the sense of DFRC, which jointly designs and uses a single transmitted signal for both communication and sensing. This means that the majority of transmitter modules can be shared by C\&S. Most of the receiver hardware can also be shared, but receiver processing, particularly the baseband signal processing, is typically different for C\&S.  Via joint design, JCAS can also potentially overcome the two aforementioned limitations in passive sensing. These properties make JCAS significantly different from existing spectrum sharing concepts such as cognitive radio, the aforementioned coexisting communication-radar systems, and ``integrated'' systems using separated waveforms \cite{RN14} where communication and sensing signals are separated in resources such as time, frequency and code, despite the two functions may physically be combined in one system. In Table \ref{tab-compareschemes}, we briefly compare the signal formats and key features, advantages, and disadvantage of five types of systems: C\&S with separated waveforms, coexisting C\&S, passive sensing, cognitive radio, and JCAS.

The initial concept of integrated C\&S systems may be traced back to the 1960s \cite{liu2019joint}, and had been primarily investigated for developing multi-mode or multi-function military radars. In the early days, most of such systems belong to the type C\&S with separated waveforms, as detailed in Table \ref{tab-compareschemes}. There has been limited research on JCAS for domestic systems before 2010. In the past ten years, JCAS has been studied based on both simple point-to-point communications such as vehicular networks \cite{Kumari17, 8057284,8550811,8722599} and complicated mobile/cellular networks \cite{8827589, Liu18, Liu18mumimo, 8741844}. The former can find great applications in autonomous driving, while the latter may revolutionize the current communication-only mobile networks. 

JCAS has the potential of integrating radio sensing into large-scale mobile networks, creating what we call \textit{Perceptive Mobile Networks} (PMNs) \cite{RN9, RN30,8062279,8905229,8827589}. \bb{By ``perceptive'', we mean the added capability of perceiving the environment via radio vision and inference to existing mobile networks. Such perception can go far beyond localization and tracking, enabling the mobile network to ``see'' and understand the environment.} Evolving from the current mobile network, the PMN is expected to serve as a ubiquitous radio-sensing network, whilst providing uncompromising mobile communication services. It can be built on top of existing mobile network infrastructure, without requiring significant changes on network structure and equipment.  It will unleash the maximum capabilities of mobile networks, and avoid the prohibitively high infrastructure costs of building separate wide-area radio sensing networks. With a large coverage, the integrated communication and sensing  capabilities are expected to enable many new applications for which current sensing solutions are either impractical or too costly.

\subsection{Potential Sensing Applications of PMNs}

Large-scale sensing is becoming increasingly important for the growth of our industry and society \cite{RN9, RN30,8062279}. It is a critical enabler for disruptive IoT applications and a diverse range of smart initiatives such as smart cities and smart transportation \cite{article4}. Unfortunately, its adoption is severely constrained by the high infrastructure costs due to the limited coverage areas of existing sensors. For example, seamless camera surveillance over expansive areas will be prohibitively expensive due to the sheer number of cameras and communication links required to connect them. In addition, there are significant privacy concerns. 

PMN is able to provide simultaneous communication and radio sensing services, and it can potentially become a ubiquitous solution for radio sensing because of its larger broadband coverage and powerful infrastructure. Its joint and harmonized communication and sensing capabilities will increase the productivity of our society, and facilitate the creation and adoption of a vast number of new applications that no existing sensors can efficiently enable. Some earlier work on passive sensing using mobile signals has demonstrated its potentials. For example, \cite{RN48}, \cite{RN43} and \cite{doi:10.1002/wat2.1289} used GSM-based radio signals for traffic monitoring, weather prediction and remote sensing of rainfall, respectively. The perceptive network can be widely deployed for both communication and sensing applications in transport, communications, energy, precision agriculture, and security, where existing solutions are either infeasible or inefficient. It can also provide complementary sensing capabilities to existing sensor networks, with its unique features of day-and-night operation and see-through of fog, foliage, and even solid objects. 

There have been numerous WiFi sensing demonstrators developed and reported in literature, for applications concerning safety, security, health and entertainment \cite{8794643}. The PMN has more advanced infrastructure than WiFi sensing, including larger antenna array, larger signal bandwidth, more powerful signal processing, and distributed and cooperative base-stations. In particular, with massive multiple-input and multiple-output (MIMO), the PMN equivalently possesses a massive number of “pixels” for sensing. This enables radio devices to resolve numerous objects at a time and achieve sensing results with much better resolution. 

\begin{table}[tb]
	\caption{\bb{Potential sensing applications of PMN.}} 
	\centering
	\def\arraystretch{1.2}
	\begin{tabular}{|m{1.3cm}|m{6.7cm}|} 
	\hline
	\textbf{Application Areas} & \hspace{1.5cm}\textbf{Cases and Examples} \\
	\hline
Smart Transportation &  - Real-time city-wide vehicle classification and tracking;

- Vehicle speed measurement;

- On-road parking space detection;  

- Sensing assistant to autonomous driving;

- Drone monitoring and management.

 \\
\hline

Smart City & 
- Extensive on-street and open space surveillance for security and safety;

- Low-cost automatic street lighting systems;

- Crowd management for major events and emergency evacuation;

- Integrated personal navigation and safety services provided by PMN and smart mobile devices.

 \\
	\hline
Smart Home	& - (Through-the-wall) localization and tracking;

- Human behavior recognition and fall detection;

- Monitoring of biomedical signals such as respiration patterns;

- Human presence detection and radio fence.

    \\
	\hline
Industrial IoT & - Localization and tracking of vehicles, equipment, and workers;

- Surveillance and proximity detection;

- Object recognition and authentication;

- Gesture recognition for equipment operation.
 
  \\
	\hline
Environ-mental Sensing & - Factory emissions and pollution monitoring; 

- Rainfall monitoring and flooding prediction;

- Animal migration monitoring;  

- Monitoring of migratory birds and insects.
 
  \\
	\hline
Sensing-assisted Comms	& - Radio signal propagation mapping and site survey;

- Beam tracking and predictive beamforming;

- Sensing-seeded encrypted communications;

- Sensing assisted resource optimization for communications. 

  \\
	\hline
\end{tabular}

\label{tab-application}
\end{table}

\begin{figure}[t]
	\centering
	\includegraphics[width=\linewidth]{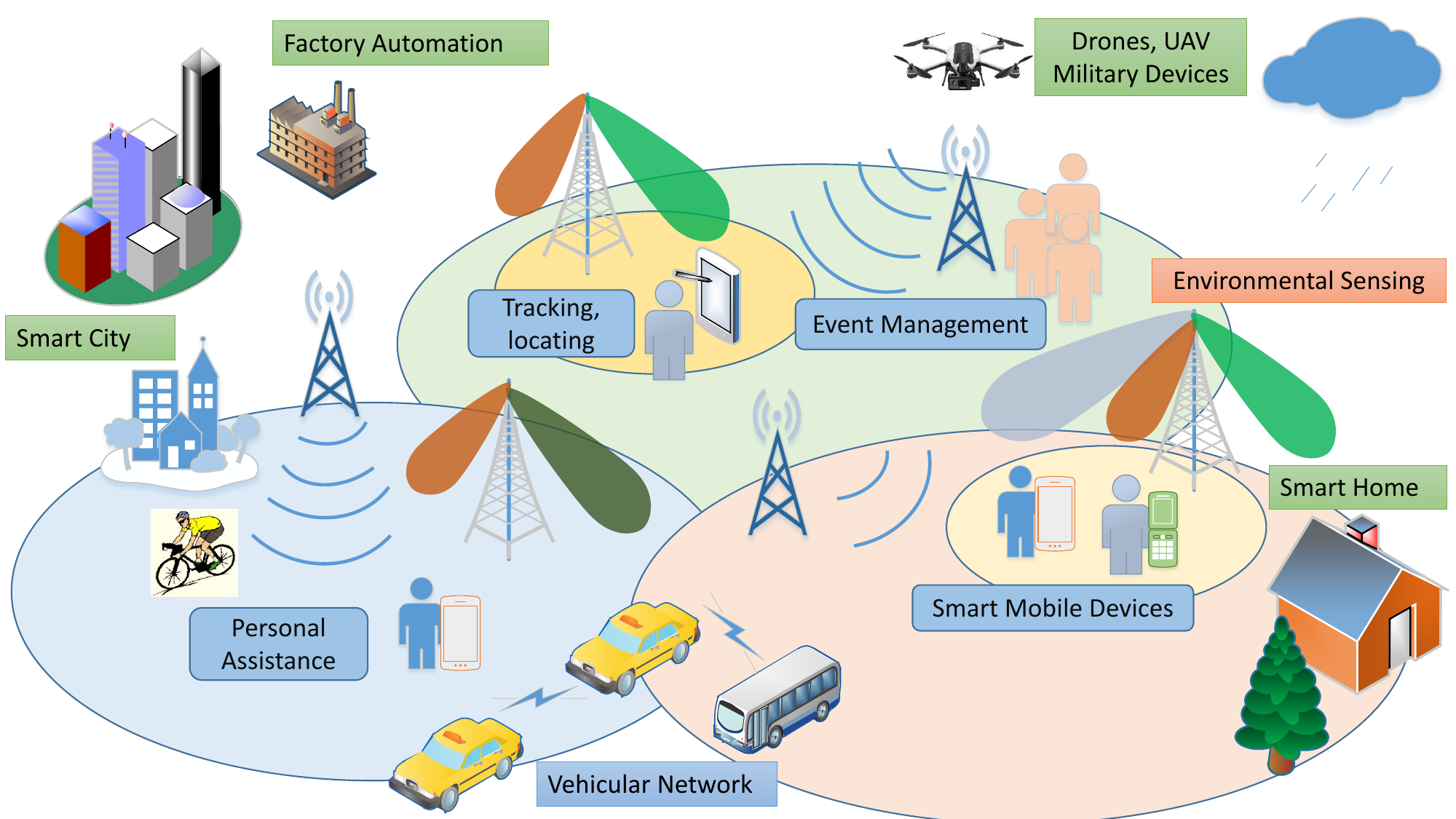}
	\caption{Applications and use cases of PMN, with integrated communication and sensing capabilities.}
	\label{fig-system1}
\end{figure}

\bb{Some of the sensing applications that can be enabled by PMN are illustrated in Fig. \ref{fig-system1}. They may be classified as several major areas, such as smart transportation, smart city, smart home, industrial IoT, environmental sensing, and sensing-assisted communications. More specific examples of these applications are listed in Table \ref{tab-application}. Detailed discussions on some of these applications are available from \cite{liu2019joint}.}

\subsection{Contributions and Structure of this Paper}
This paper provides a comprehensive survey on the state-of-the-art research on PMN that realizes JCAS technology in mobile networks. \bb{different from some existing overview articles~\cite{RN32, RN14, 7782415, RN15, article4,liu2019joint,Hassanien_2019, 8828016,8828030,Feng20,Ma20,luong2021radio}, as elaborated on in Table \ref{tab-compdiff}, we focus on JCAS techniques that are tailored to cellular/mobile networks, and provide in-depth review, observations, and insights from the signal-processing perspective. This paper is distinctive in terms of its main topics of coverage, as can be clearly seen from Table \ref{tab-compdiff}.} Based on our own extensive research experience and outputs, this article intends to provide a clear picture on what the PMN will look like and how it may evolve from the current communication-only network from the viewpoints of both infrastructure and technology. It also provides many more technical details on JCAS technologies, particularly on the receiver side processing. More specifically, in this survey, we consider mobile-network-specific JCAS challenges and solutions, associated with heterogeneous network architecture and components, sophisticated mobile signal format, and complicated signal propagation environment. We refer to complicated mobile signals as those with modulations of orthogonal frequency-division multiple access (OFDMA) and multiuser-MIMO (or spatial division multiple access, SDMA). We discuss major challenges and required changes to system infrastructure for the paradigm shift from communication-only mobile network to PMN with integrated communication and sensing, and provide a comprehensive review of existing technologies and open research problems, in order to address these challenges within the framework of PMN. 

\begin{table*}[t]
	\centering
	\def\arraystretch{1.3}
	\caption{\bb{A comparison of the scope and topics of coverage between existing related overview articles and this survey. The letters A to J represent the following topics: A - PMN Framework and evolution; B - Performance Bounds; C - Waveform design and optimization; D - Antenna array design; E - Clutter Suppression; F - Sensing parameter estimation; G - Resolution of Sensing ambiguity; H - Pattern analysis; I - Networked Sensing; J - Sensing-assisted Communications. We use the numbers 0 to 2 to represent the extent of coverage, with 0 for no coverage, 1 for limited review, and 2 for broad and extensive coverage. \textbf{It is important to note that every listed paper has its own focus, merits, and contributions, as summarized in the Scope column. The listed topics of coverage are limited to those considered in our paper, and other topics discussed in the papers being compared are not appraised here. So a low score does not disparage the overall contribution of the paper in any sense.}}} 
	\label{tab-compdiff}
\begin{tabular}{|m{0.8cm}|m{8.5cm}|m{0.3cm}|m{0.3cm}|m{0.3cm}|m{0.3cm}|m{0.3cm}|m{0.3cm}|m{0.3cm}|m{0.3cm}|m{0.3cm}|m{0.3cm}|}
	\hline
\textbf{Papers}	& \centering\textbf{Scope} & A &B&C&D&E&F&G&H&I&J  \\
	\hline
	\cite{RN32} & Probably the first overview article on JCAS that provides some basic signal models and receiver signal processing techniques for sensing using both single carrier and multicarrier communication signals. &0&0&1&0&0&0&0&0&0&0\\
	\hline
	\cite{RN15} & Introduces the estimation-rate metric for radar, and offers a general review on radar-communications convergence, including coexisting, cooperation and co-design, by referring to this metric and the communication-rate metric.  Limited coverage on co-design (JCAS).&0&2&0&0&0&0&0&0&0&0  \\
	\hline	
\cite{liu2019joint} & Provides a detailed introduction to the applications and evolution of JCAS technologies, and a brief overview of coexisting C\&S and DFRC (JCAS) systems. A detailed mmWave DFRC system is also proposed. &0&1&2&1&1&0&0&0&0&0  \\
	\hline
	
	\cite{8828016} & Reviews recent work on coexistence between C\&R systems, including signal models, waveform design, and signal processing techniques. A very good introduction to signal models. &0&0&1&0&0&1&0&0&0&0  \\
	\hline
	
\cite{Hassanien_2019}	& Provides a very good introduction to radar-centric JCAS systems, with a focus on signal embedding techniques.&0&0&1&0&0&0&0&0&0&0 \\
	\hline
\cite{8828030}	& Focuses on mmWave JCAS systems. Provides a detailed introduction to signal modeling and waveform design, and an excellent comparison between pulse-modulated continuous wave and OFDMA JCAS systems.&0&0&1&0&0&0&0&0&0&0 \\
	\hline
	\cite{Ma20}	& Reviews JCAS technologies in the context of autonomous vehicles, with a focus on signal modeling and waveform design particularly index modulation.&0&0&2&1&0&0&0&0&0&0 \\
	\hline
\cite{Feng20}	& Reviews technologies in coexistence, cooperation,
co-design and collaboration of C\&R systems. &0&1&2&1&0&0&0&0&0&1\\
	\hline
\cite{luong2021radio}	&  Offers a comprehensive review of radio resource management in JCAS systems, such as spectrum sharing, power allocation, and interference management. Countermeasures
to the security issues in JCAS are also discussed. &0&1&2&1&0&0&0&0&0&1\\
\hline
This  survey	&  Reviews mobile-network-specific JCAS challenges and solutions; Provides detailed review of network structure, evolution, and key technologies from the signal processing perspective for PMN. &2&2&2&2&2&2&2&1&1&2 \\
	\hline
\end{tabular}
\end{table*}

 \begin{figure*}[h!]
	\centering
	\includegraphics[width=1.02\linewidth]{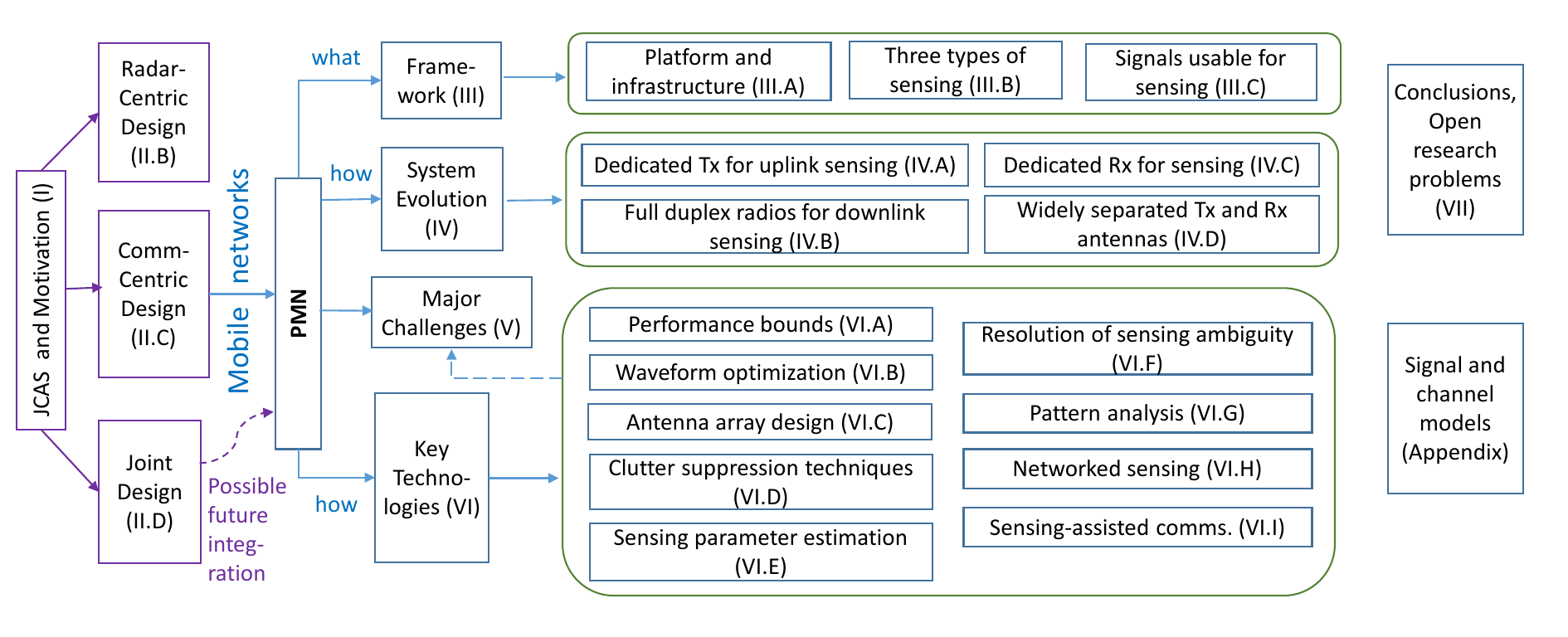}
	\caption{\bb{Main structure of the article.} }
	\label{fig-articlestr}
\end{figure*}

The overall structure of the paper is illustrated in Fig. \ref{fig-articlestr}. The rest of this paper is organized as follows.
\begin{itemize}
	\item In Section \ref{sec:relatedwrk}, we first discuss the difference between communication and radar waveforms. \bb{We then briefly review the research on three types of JCAS systems: (1) radar-centric design which realizes communication function in a primary radar system, (2) communication-centric design which realizes radio sensing function in a primary communication system, and (3) joint design without being constrained to an underlying system. Particular attention is paid to how the three types of JCAS systems overcome the waveform difference to meet the different requirements for C\&S. Note that the PMN is an example of communication-centric design that realizes radio sensing in primary mobile communication networks.}
	\item  \bb{In Section \ref{sec:intro2}, we provide detailed discussions on what PMN is. This is about the framework of a PMN, specifying how sensing may be integrated at the basic network and signal levels. This framework is built on the current structure and signals of existing mobile networks, by identifying and endowing their roles in sensing. We review works on such a framework of a PMN, including system architecture, three types of unified sensing options, and signals usable for sensing. Signal and channel models on PMN are also provided in the Appendix.}
	\item \bb{In Section \ref{sec:structure}, we further discuss the required modifications to system infrastructure, in order to evolve the current communication-only network to PMN. We review both long-term full-duplex option and three near-term options that enable JCAS in PMNs without requiring significant network modifications, particularly for time-division duplexing (TDD) systems.}
	\item Section \ref{sec-challenge} discusses three major research challenges, as well as research opportunities, in PMNs, including sensing parameter extraction, joint design and optimization, and networked sensing.
	\item In Section \ref{sec:wlk}, we provide a comprehensive review of technologies that have been developed to address the above challenges and beyond, and discuss open research problems and potential solutions. This part of the review is organized under nine topics: performance bounds, joint waveform optimization to balance the performance of communication and sensing, antenna array design, clutter suppression, sensing parameter estimation, resolution of sensing ambiguity, pattern analysis, networked sensing under cellular topology, and sensing-assisted communications. 
	\item Finally, conclusions are drawn in Section \ref{sec:conclu}. We also provide a table that summarizes the matureness and research difficulties for each of the major technical areas, and highlight their key open research problems.
\end{itemize}

\section{Three Types of JCAS Systems}\label{sec:relatedwrk}

Based on the design priority and the underlying signal formats, the current JCAS systems may be classified into the following three categories, namely:
\begin{itemize}
\item \textit{Radar-centric design}: Realizing communication function in a primary radar system (or integrating communication into radar);
\item \textit{Communication-centric design}: Realizing radio/radar sensing function in a primary communication system (or integrating radar into communication); and
\item \textit{Joint design and optimization}: technologies without constrain to underlying systems and signals.
\end{itemize}
In the first two categories, the design and research focus are typically on how to realize the other function based on the signal formats of the primary system, with the principle of not significantly affecting the primary system. Slight modifications and optimizations may be applied to the system and signals. The last category considers the design and optimization of the signal waveform, system and network architecture, without bias to either communication or sensing, aiming at fulfilling the desired applications only. PMNs belong to the second class, aiming to evolve a communication-only network towards one with integrated communication and sensing. 

\begin{figure*}[t]
	\centering
	\includegraphics[width=0.8\linewidth]{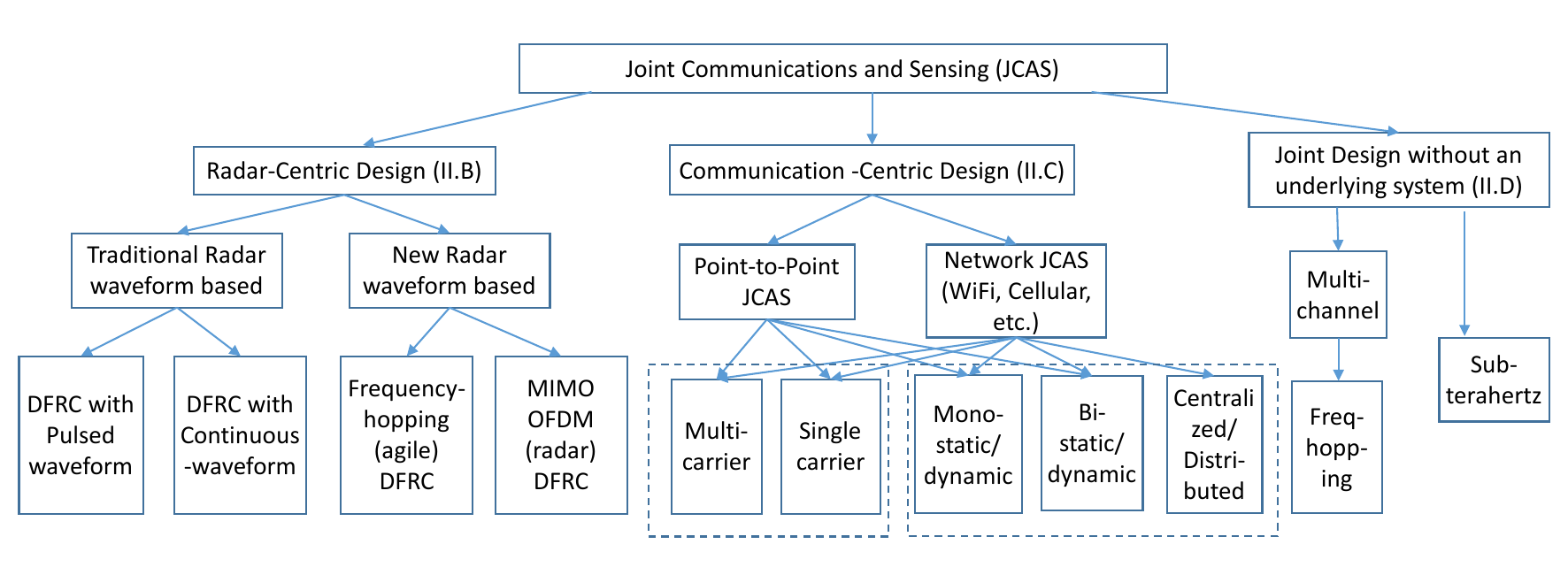}
	\caption{A classification of the three types of JCAS systems, in terms of their technical scope.}
	\label{fig-3jcas}
\end{figure*}

Next, we first briefly discuss the major differences between traditional communication and radar signals, which are important for understanding the design philosophy of the three categories of JCAS systems. We then provide a brief review on the recent research progress in each of the categories, referring to the classification of the three categories of JCAS systems in terms of their technical scope, as shown in Fig. \ref{fig-3jcas}. {A brief summary of selected recent research works on each of the JCAS categories, as well as co-existing C\&S, is provided in Table \ref{table:litersurv}.}

\begin{table*}[t!]
	\begin{threeparttable}
		\def\arraystretch{1.3}
		\centering
		\caption{Exemplified research on Radar-Communication coexistence and JCAS solutions in the literature.}
		\begin{tabular} {|m{1.7cm}|m{1.5cm}|m{6.2cm}| m{6.5cm}|}
			
			\hline
			\textbf{Category} & \centering\textbf{Reference} & \centering\textbf{Model} & \hspace{1.5cm}\textbf{Main Contributions} \\
			\hline
			\hline
				\multirow{4}{1.7cm}{Realizing Communication in Primary Radar Systems}
			&
			\cite{RN16}
			&
			Multi-carrier quasi-orthogonal
			LFM-CPM waveform of MIMO radar utilized for multiuser communication in JCAS 
			&
			Shared waveform design, signal processing and performance evaluation for the radar and communication subsystems  
			\\
			\cline{2-4}
			&
			\cite{8378784}
			&
			Integration of communication into MIMO radar 
			&
			Transmit and receive beamforming design for full-duplex communication in MIMO radar 
			\\
			\cline{2-4}
			
			&
			
			\cite{8365175} 
			
			&
			
			JCAS realized in primary MIMO radar 
			
			&
			
			Communication signals embedded into the radar transmit waveform for JCAS
			\\
			\cline{2-4}
			
			&
			\cite{7455269}
			&
			Carrier frequency
			of the radar signal used for modulating communication
			information.
			&
			Random step frequency signal used
			in designing a JCAS system
			\\
			\hline
			
		\multirow{5}{1.7cm}{Realizing Sensing in Primary Communication Systems}

			&
			\cite{Kumari17}
			&
			IEEE 802.11ad mmWave V2V communication system utilized as automotive radars.
			&
			Radar detection as well as range and velocity estimation by leveraging standard WLAN receiver algorithms and classical pulse-Doppler radar algorithms
			\\
			\cline{2-4}
			
			&
			
			\cite{8057284}
			&
			
			IEEE 802.11 OFDM communications waveform used for sensing in vehicular network
			
			&
			
			Brute-force optimization performed based on received mean-normalized channel energy for radar range estimation
			\\
			\cline{2-4}
			
			&
			
			\cite{Braun14}
			&
			
			OFDM communication signals used for radar sensing
			
			&
			
			Signal processing aspects of OFDM radar by periodogram and ESPRIT algorithms
			\\
			\cline{2-4}
			
			&
			
			\cite{8683655}
			&
			
			Generalized multicarrier radar and OFDM communication system are converged where subsets of subcarriers assigned to the radar or the communications tasks.
			
			&
			
			Multicarrier waveform proposed for JCAS
			\\
			\cline{2-4}
			
			&
			\cite{5760698}, \cite{7952790}
			
			&
			IEEE 802.11p OFDM communication waveform in vehicular networks used for radar sensing
			&
			Processing of delay and Doppler information by applying the ESPRIT method
			\\
			
			\hline

		\multirow{5}{1.7cm}{Joint Design Without an Underlying System}
			&
			
			\cite{RN32}
			
			&
			
			Implementation of digital beamforming radar and MIMO
			communications integration at 24 GHz ISM band for intelligent transportation applications
			
			&
			
			Continuous single-carrier and multicarrier waveforms design, as well as techniques for range, Doppler and DoA estimation by classical Fourier transform-based approach and MUSIC algorithms
			\\
			\cline{2-4}
			&
			
			\cite{8741844}
			
			&
			
			Multiple input, single output
			(MISO) joint multiple access channel topology considered for JCAS operation
			
			&
			
			Successive interference cancellation (SIC) for communication, and Pulse-Doppler processing techniques are employed to extract Doppler and range estimates in radar operation.
			\\

			\cline{2-4}
			
			&
			
			\cite{7485311, 7944212, 7421368}
			&
			
			Joint radar-communications integrated receiver simultaneously perform radar target parameter estimation and decode a communications signal
			
			&
			
			Matched filtering clutter suppression techniques related to modern mobile networks based JCAS
			\\
			\cline{2-4}
			
			&
			
			\cite{8747215, 8683591, 8828030, kumari2019adaptive, 8642926}
			&
			
			mmWave JCAS systems. 
			
			&
			
			Beamforming design, waveform design, adaptive mmWave waveform design
			for automotive applications  
			\\
			\cline{2-4}

			&
			\begin{center}
				\cite{Liu18,  Liu18mumimo, 8445869}
			\end{center}
			&
			
			Waveform design for MIMO radar and multi-user MIMO communication
			
			&
			
			Several optimization-based waveform designs for given radar beam patterns and under constant modulus constraints and similarity constraints (SC)
			\\
			
			\hline

	\multirow{2}{1.7cm}{Co-existence of C\&S System}

			&
			\begin{center}
				\cite{8835700,6331681}, \cite{7898445}
			\end{center}
			
			&
			
			Spectrum sharing between downlink MU-MIMO communication
			and co-located MIMO radar on the same frequency band.
			
			&
			
			Transmit beamforming optimization solution given for both shared and separated antenna deployment and for both perfect and imperfect CSI.
			\\
			\cline{2-4}

			&
			
			\cite{7536903}
			
			&
			
			Coexistence between a co-located MIMO radar
			system and a LTE communications system.
			
			&
			
			Spectrum sharing techniques introduced for JCAS operation and a genetic algorithm used to solve the optimization problem
			\\
			\hline
		\end{tabular}
		\label{table:litersurv}
	\end{threeparttable}
\end{table*}

\subsection{Major Differences between C\&S Signals}\label{sec-diffcs}

\begin{figure}[t]
	\centering
	\includegraphics[width=1\linewidth]{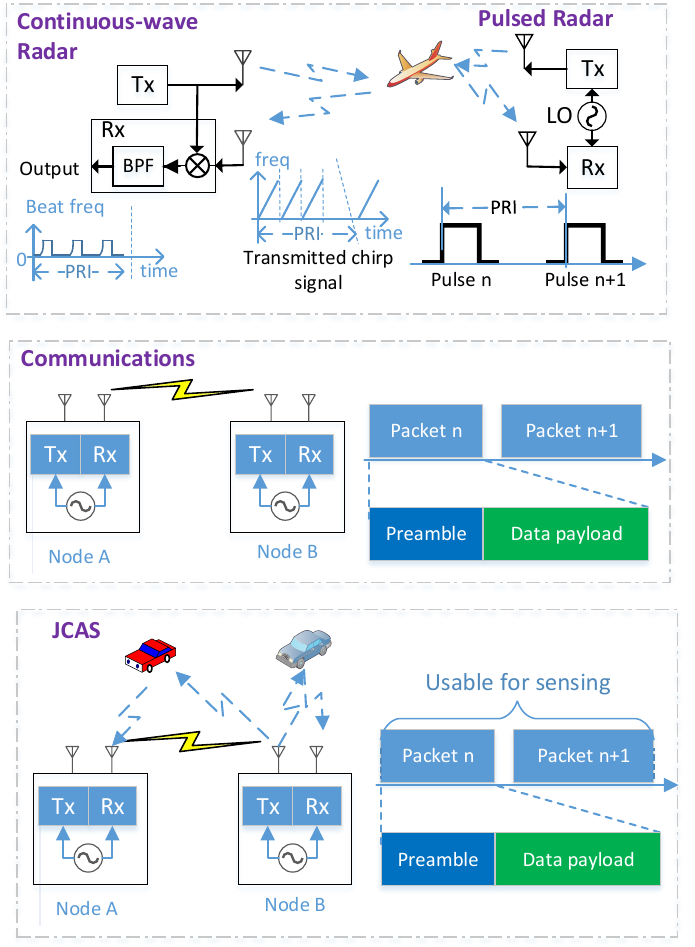}
	\caption{Illustration of basic pulse and continuous-wave radar, communication systems, and JCAS systems. Tx stands for transmitter; Rx for Receiver; PRI for pulse repetition interval; and BPF for bandpass filter.}
	\label{fig-radcom}
\end{figure}

Fig. \ref{fig-radcom} presents the simplified transceivers and signal structures of C\&R to illustrate their major differences.

Conventional radar systems include pulsed and continuous-wave radars \cite{RN14, RN15, mimoradarbook}, as shown in Fig. \ref{fig-radcom}. In pulsed radar systems, short pulses of large bandwidth are transmitted either individually or in a group, followed by a silent period for receiving the echoes of the pulses. Continuous-wave radars transmit waveforms, such as chirp, continuously, typically scanning over a large range of frequencies. In either system, the waveforms are typically non-modulated. These waveforms are used in both SISO and MIMO radar systems, with orthogonal waveforms used in MIMO radars \cite{4383615,mimoradarbook}. 

In most of radar systems, low peak-to-average power ratio (PAPR) is a desired feature for the transmitting signal, which enables high efficiency power amplifier and long-range operation. The transmitting waveform is also desired to have an ambiguity function with steep and narrow mainlobes, which is the correlation function of the received echo signals and the local template signal \cite{mimoradarbook, Hassanien_2019}. These waveforms are designed to enable low-complexity hardware and signal processing in radar receivers, for estimating key sensing parameters such as delay, Doppler frequency and angle of arrival. However, they are not indispensable for estimating these parameters. A pulsed radar receiver typically samples the signal at a high sampling rate twice of the transmitted pulse bandwidths, or at a relatively lower sampling rate at the desired resolution of the delay (ranging); while the receiver of a continuous-wave radar, e.g., frequency modulated continuous wave (FMCW) radar, typically samples signals of ``beat'' frequency at a rate much smaller than the scanning bandwidth, proportional to the desired detection capability of the maximal delay. \bb{Here, the beat frequency equals the difference between the frequencies of the echo signal and the transmitted signal that is used as the input to the local oscillator of the receiver, and contains the range information.} Due to their special signal form and hardware, radar systems generally cannot support very high-rate communications, without significant modifications to the waveforms and/or receiver structure \cite{liu2019joint,Hassanien_2019}. 

Comparatively, communication signals are designed to maximize the information-carrying capabilities. They are typically modulated, and modulated signals are typically appended with non-modulated intermittent training signals in a packet, as can be seen from Fig. \ref{fig-radcom}. To support diverse devices and communication requirements, communication signals can be very complicated. For example, they can be discontinuous and fragmented over time and frequency domains, have high PAPR, have complicated signal structures due to advanced modulations applied across time, frequency, and spatial domains. 

Although being designed without considering the demand for sensing, communication signals can potentially be used for estimating all the key sensing parameters. However, different from conventional channel estimation, which is already implemented in communication receivers, sensing parameter estimation requires extraction of the channel composition rather than channel coefficients only. Such detailed channel composition estimation is largely limited by the hardware capability. The complicated communication signals are very different to conventional radar and demand new sensing algorithms. There are also practical limits in communication systems, such as full-duplex operation and asynchronization between transmitting node and receiving node, which requires new sensing solution to be developed. We note that the detailed information on the signal structure, such as resource allocation for time, frequency and space, and the transmitted data symbols, can be critical for sensing. For example, the knowledge on signal structure is important for coherent detection. In comparison, most passive radar sensing can only perform non-coherent detection with the unknown signal structure, and hence only limited sensing parameters can be extracted from the received signals with degraded performance \cite{Hack14,Gogineni14}. 

The differences and benefits of JCAS in comparison with individual radar or communication system are summarized in Table \ref{table:com}. 

\begin{table*}[t]
	
	\def\arraystretch{1.2}%
	\centering
	\caption{Comparison between Radar, Communications and JCAS.}
	\label{table:com}
	\begin{tabular}{|m{1.7cm}|m{4.6cm}|m{4.6cm}|m{5.2cm}|}		
		\hline		
		\textbf{Specifications} & \centering{\textbf{Radar}} & \centering{\textbf{Communications}} & \hspace{1.5cm} {\textbf{JCAS System}} 
		\\
		\hline
		\textbf{Signal Waveform}
		&
		Typically simple unmodulated single-carrier signals occupying large bandwidth; Pulse or continuous-wave frequency modulated; Orthogonal if multiple spatial streams and orthogonality can be realized in one or more domains of time, frequency, space and code; typically low peak-to-average power ratio (PAPR). Radars with advanced waveforms such as OFDM and frequency hopping is also emerging.
		
		&
		Mix of unmodulated (pilots/training sequences) and modulated data symbols; Complicated signal structure and resource usage; advanced modulations, e.g., OFDMA and multiuser-MIMO; High PAPR.
		
		&
		JCAS can use both traditional radar and communication signals, with appropriate modifications to support both C\&S and optimize their performance jointly.  
		
		\\
		\hline
		\textbf{Transmission Power}
		&
		Typically high in large-scale and long-range radar; low in short-range radar such as FMCW radar used in vehicular networks.
		
		&
		Typically low, supporting linkage distance up to a few kilometers.
		
		&        
		Communications integrated into Radar can achieve very long link distance. Sensing integrated into a single communication device can only support short range, but overall JCAS can cover very large areas due to the wide coverage of communication networks.
		\\
		\hline
		\textbf{Bandwidth}
		&
		Large signal bandwidth. Resolution proportional to bandwidth. But the bandwidth of the output signal in FMCW radar may be narrow, depending on the signal propagation distance.
		&
		Typically much smaller than radar.
			
		&
       mmWave signals are very promising for JCAS, due to large signal bandwidth and limited propagation. In addition, sensing applications do not have to rely on large bandwidth, such as known WiFi sensing examples.
		
		\\
		\hline
		\textbf{Signal Band}
		&
		X, S, C and Ku
		
		&
		sub-6 GHz and mmWave bands
		
		&
		Have an impact on operation distances and resolution capabilities of JCAS.
		\\
		\hline
		\textbf{Transmission Capability (Duplex)}
		&
		Full-duplex (continuous-wave radar) or half-duplex (pulse radar) 
		
		&
		Co-located transmitter and receiver typically cannot operate on the same time and frequency block. Communications are in either TDD or frequency-division duplex (FDD).
		
		&
		Full-duplex is a favourite condition, but not essential.
		
			\\
		\hline
		\textbf{Clock Synchronization}
	&
	Transmitter and receiver are clock-locked.
	& 
	Co-located transmitter and receiver share the same clock, but non- co-located nodes typically do not. 
	& 
	Clock-level synchronization removes ambiguity in sensing parameter estimation, but is not essential for some sensing applications.
		\\
		\hline
		
	\end{tabular}
	
\end{table*}

\subsection{Radar-Centric Design: Realizing Communication in Primary Radar Systems}

Radar systems, particularly military radar, have the extraordinary capability of long-range operation, up to hundreds of kilometers. Therefore, a major advantage of implementing communication in radar systems is the possibility of achieving long-range communications, with much lower latency compared to satellite communications. However, the achievable data rates for such systems are typically limited, due to the inherent limitation in the radar waveform. In \cite{RN13}, authors implemented a combined radar and communication system based on a software-defined radar platform, in which the radar pulses are used for communication. Research work in \cite{RN15} and \cite{RN66} shows that communication network establishment can be possible for both static and moving radars used in the military and aviation domains. Adaptive transmit signals from airborne radar mounted unmanned vehicles can also be used to simultaneously sense a scene and communicate sensed data to a receiver at the ground base station. The objective of such systems is to establish low latency, secure and long-range communications on top of existing radar systems. Such JCAS systems have been mainly called as dual-function radar-communications (DFRC).

Realization of communication in radar systems has traditionally been based on either pulsed or continuous-wave radar signals. Hence information embedding is one of the major challenges. For example, in \cite{7455269}, a random step frequency signal is used in designing a JCAS system where the carrier frequency of
the radar signal is used for modulating communication
information. In \cite{RN16}, the authors show that the quasi-orthogonal multicarrier linear frequency modulation-continuous phase modulation (LFM-CPM) waveform radiated by a MIMO radar can be applied for communications with multiple users. For more information on embedding communication information to radar signals, the readers can refer to \cite{Hassanien_2019} which provides an excellent review on this topic.

Integrating communications into radar systems with new radar waveforms has also been investigated, such as the MIMO-OFDM radar \cite{Lin15} and frequency-agile (frequency hopping) radar \cite{Huang20IM,Wu20waveform}. Their signal formats are closer to modern communication systems, and hence can be potentially better integrated for information transmission. Such systems typically apply \textit{index modulation} to embed communication information into the radar waveform. \bb{Here, index modulation embeds information to various combinations and/or permutations of signal parameters over space, time, frequency and code domains \cite{Hassanien_2019,Ma20, 9288847}. One example is to use the indexes of subcarriers and transmitting antennas to carry the information. The main advantage of applying it in radar-centric JCAS is that index modulation does not change the basic radar waveform and signal structure, and has negligible influence on radar operation.} 

What is missing here in the literature is the communication protocol design and receiver signal processing. Communication protocols, particularly medium access (MAC) layer protocol and physical layer frame structure, are well designed in communication systems. However, the design of communications protocols which can be fitted into radar signals is not straightforward. The main challenges lie on the requirement that communication protocol design shall be seamlessly integrated into radar operation.  Some early work is reported in \cite{Wuicc20}, where a frame structure is proposed for JCAS with frequency-hopping continuous-wave radar signals. Based on the frame structure, channel estimation techniques are then developed without knowing the frequency hopping sequence at the communication receiver. Nevertheless, a complete receiver signal processing for extracting the information embedded in radar waveform is not well studied yet.

\subsection{Communication-Centric Design: Realizing Sensing in Primary Communication Systems}
This is the category of JCAS systems that the PMN belongs to, and we will provide a comprehensive survey on it in the rest of this paper. Here, we briefly review the research in this category. Considering the topology of communication networks, systems in this category can be classified into two sub-categories, namely, those realizing sensing in point-to-point communication systems particularly for applications in vehicular networks, and those realizing sensing in large networks such as mobile networks. Depending on how the transmitter and sensing receiver are spatially distributed, in terms of sensing, these systems are analogue to traditional mono-static, bi-static and multi-static radars. 

\bb{Two fundamental problems in integrating sensing into communications are: (1) how to realize full-duplex operation in a mono-static setup where the sensing receiver and transmitter are co-located, and (2) how to remove the clock asynchronization impact in a bi-static or multi-static setup due to typically unlocked clocks between spatially separated transmitters and (sensing) receivers. Full-duplex here means that the receiver and transmitter work at the same time over the same frequency band. For a mono-static radar, full-duplex operation is avoided in pulsed radar via temporally separating the transmitting and receiving timeslots, leading to blind spots in near-field sensing; for FMCW radar, it is realized via using the transmitted signal as the input to the local oscillator to suppress the leakage signal from the transmitter, which leads to the output of the beat-frequency signal with little information on the transmitted signal. Modern communication systems primarily transmit continuous waveform and have un-modulated sinusoidal signals as the input to the oscillator. Hence both radar methods are not practical in communication systems, unless a dedicated sensing receiver hardware similar to FMCW radar is integrated. In the long term, full-duplex technologies, as have been widely investigated for communications, would be a desired solution for mono-static sensing. Initial studies have also been conducted for JCAS systems, such as in \cite{8805161}. However, the technology is still immature for practical applications. There are near-term sub-optimal solutions, which will be detailed in Section \ref{sec:structure}. For bi-static and multi-static radars, clock synchronization is typically realized via wired connections or locking to the GPS signals \cite{1370671}. These methods are feasible for some communication setup, as will be discussed in Section \ref{sec:structure}. In the presence of clock asynchronization, it is also possible to apply signal processing techniques to overcome it, which will be elaborated in Section \ref{sec-resamb}.
}

There have been quite a few works on sensing in vehicular networks using IEEE 802.11 signals.
In \cite{5760698}, the authors implemented active radar sensing functions into a communication system with OFDM signals for vehicular applications. The presented radar sensing functions involve Fourier transform algorithms that estimate the velocity of multiple reflecting objects in IEEE 802.11.p based JCAS system. In \cite{Kumari17}, automotive radar sensing functions are performed using the single carrier (SC) physical (PHY) frame of IEEE 802.11ad in an IEEE 802.11ad millimeter wave (mmWave) vehicle to vehicle (V2V) communication system. In \cite{8057284}, OFDM communication signals, conforming to IEEE 802.11a/g/p, are used to perform radar functions in vehicular networks. More specifically, a brute-force optimization algorithm is developed based on received mean-normalized channel energy for radar ranging estimation. The processing of delay and Doppler information with IEEE 802.11p OFDM waveform in vehicular networks is shown in \cite{7952790} by applying the ESPRIT method. 

There has been rapidly increasing JCAS work reported for modern mobile networks. In \cite{Braun14}, some early work on using OFDM signals for sensing was reported. In \cite{Wang18}, sparse array optimization is studied for MIMO JCAS systems. Sparse transmit array design and transmit beampattern synthesis for JCAS are investigated in \cite{WANG2018223}, where antennas are assigned to different functions. In \cite{yliu17}, mutual information for an OFDM JCAS system is studied, and power allocation for subcarriers is investigated based on maximizing the weighted sum of the mutual information for C\&S. In \cite{Liu18mumimo}, waveform optimization is studied for minimizing the difference between the generated signal and the desired sensing waveform. In \cite{Rong17}, the multiple access performance bound is derived for a multiple antenna JCAS system. In \cite{8683655}, a multicarrier waveform is proposed for dual-use radar-communications, for which interleaved subcarriers or subsets of subcarriers are assigned to the radar or the communications tasks. These studies involve some key signal formats in modern mobile networks, such as MIMO, multiuser MIMO, and OFDM. In \cite{RN9, RN30,8062279,8905229,8827589}, the authors systematically studied how JCAS can be realized in mobile networks by considering their specific signal, system and network structures, and how radar sensing can be done based on modern mobile communication signals. Based on reported results in the literature and our own  experience and vision on this technology, we provide a comprehensive review of existing techniques and open research problems under the framework of PMNs in the following sections.

\subsection{Joint Design Without an Underlying System}

Although there is no clear boundary between the third category of technologies and systems and the previous two categories, there is more freedom for the former in terms of signal and system design. That is, JCAS technologies can be developed without being limited to existing communication or radar systems. In this sense, they can be designed and optimized by considering the essential requirements for both communication and sensing, potentially providing a better trade-off between the two functions.
 
The mmWave and (sub-)Terahertz-wave JCAS systems are great examples of facilitating such joint design. On one hand, with their large bandwidth and short wavelength, mmWave and Terahertz signals provide great potentials for high date-rate communications and high-accuracy sensing. On the other hand, mmWave and Terahertz systems are emerging. They are yet to be widely deployed, and the standards for Terahertz systems are yet to be developed. Millimeter and Terahertz JCAS can facilitate many new exciting applications, both indoor and outdoor. Existing research on mmWave JCAS has demonstrated its feasibility and potentials in indoor and vehicle networks \cite{8747215, 8683591,8828030, kumari2019adaptive, 8642926,RN24,8550811,Luo19jcas}. The authors in \cite{8828030} provide an in-depth signal processing aspect of mmWave-based JCAS with an emphasis on waveform design for JCAS systems. Future mmWave JCAS for indoor sensing is envisioned in \cite{8747215}. Hybrid beamforming design for mmWave JCAS systems is investigated in \cite{8683591}. An adaptive mmWave waveform structure is designed in \cite{kumari2019adaptive}. Design and selection of JCAS waveforms for automotive applications are investigated in \cite{8642926}, where comparisons between phase-modulated continuous-wave JCAS and OFDMA-based JCAS waveforms are provided, by analyzing the system model and enumerating the impact of design parameters. In \cite{8550811,Luo19jcas}, multibeam technologies are developed to allow C\&S at different directions, using a common transmitted signal. Beamforming vectors are designed and optimized to enable fast beam update and achieve balanced performance between C\&S. In \cite{elbir2021terahertzband}, the beamforming design for Terahertz massive MIMO JCAS systems is investigated.

Another example is multi-channel JCAS systems where one or more channels are used at a time, and multiple channels are occupied over a period of signal transmission. One specific example is the frequency hopping system, such as the existing Bluetooth system where the operating frequency channel is changed over different packets. Multi-channel systems can offer an overall large signal bandwidth for sensing, while without increasing the instantaneous communication bandwidth. This can largely reduce the hardware cost, and also match with the spectrum usage of communication systems well. Works on combining multi-channel signals for sensing have been reported for passive radar in, e g., \cite{9104263, 8423070}. The key challenge is how to remove or reduce the imperfections and distortions from the received signals for each channel and then concatenate them together for sensing. For JCAS, an additional important problem is how to design the signals to make such concatenation easier, while balancing the performance of communication and sensing. 

\subsection{Advantages of JCAS Systems}
With harmonized and integrated communication and sensing functions, JCAS systems are expected to have the following advantages:
\begin{itemize}
\item \textbf{Spectral Efficiency:} Spectral efficiency can ideally be doubled by completely sharing the spectrum available for wireless communication and radar \cite{RN14}, \cite{RN13}, \cite{8267374}, \cite{8365175 };
 \item \textbf{Beamforming Efficiency:} Beamforming performance can be improved through exploiting channel structures obtained from sensing, for example, quick beam adaption to channel dynamics and beam direction optimization \cite{RN19, RN7, 8539689, WinNT,RN26};
\item \textbf{Reduced Cost/Size:} Compared to two separated systems, the joint system can significantly reduce the cost and size of transceivers \cite{RN14}, \cite{7782415}, \cite{Wang18};
\item \textbf{Mutual Benefits to C\&S:} C\&S can benefit from each other with the integration. Communication links can provide better coordination between multiple nodes for sensing; and sensing provides environment-awareness to communications, with potentials for improved security and performance.
\end{itemize}

\subsection{\bb{Summary of Key Research Problems}}

\bb{We conclude this section by briefly summarizing some key research problems and the associated challenges for the three types of JCAS systems. The summary is presented in Table \ref{tab-threejcas}. For communication-centric design, we will provide more details, particularly those related to mobile networks, in later sections. }

\begin{table*}[th]
	\def\arraystretch{1.4}
		\centering
		\caption{\bb{Key research problems in three types of JCAS systems and the associated challenges.}}
		\label{tab-threejcas}
		\begin{tabular}{|m{1.5cm}|m{7cm}|m{8cm}|}
			
			\hline
				\begin{center} \textbf{Category}	\end{center} & \begin{center} \textbf{Key Research Problems} \end{center} & \begin{center} \textbf{Challenges} \end{center} \\
			\hline
			
			\multirow{3}{1.5cm}{\textbf{Radar-Centric (II.B)}}
		
			&
			Embed information to radar signals \cite{7455269,RN16, Hassanien_2019}, particularly those methods that can lead to higher data rates \cite{Huang20IM,Wu20waveform}.
			& 
			- Without significantly affecting radar operation, such as the radar ambiguity function and PAPR;
			
			- Modulation constellation and codebook design to maximize the Euclidean distance between constellation points.			
			\\
			\cline{2-3}
		
			&
			Design frame structure and communication protocol based on the selected information embedding method  \cite{Wuicc20}.
			&
			- Conventional preamble for communications has a special pattern, which may be a major problem when being realized with radar waveform;
			
			- Communication protocols may introduce latency and low duty cycle to radar sensing.
			\\
			\cline{2-3}

			&
			
			Develop signal reception and processing technologies, including channel estimation, equalization, and demodulation schemes, particularly for JCAS based on new generations of radar systems \cite{Wu20waveform,wu2020integrating,Huang20IM}. 
			
			&
			- Extraction of communication signals from the beat frequency signal of conventional FMCW radar is challenging, as the beat signal is range-dependent;
			
			- Receiver processing needs to be tailored to the information embedding method and frame structure. They are better to be jointly designed. 						
			\\
			\hline
		\multirow{5}{1.5cm}{\textbf{Communi-cation-Centric (II.C)}}
			&
			Full-duplex technologies in the context of JCAS for co-located transmitter and receiver, an analogy to the mono-static radar \cite{8805161,8057284} (More detailed in Section \ref{sec-fullduplex}).
			&
			- Suppression of onboard leakage signals from the transmitter, particularly for MIMO systems.
			\\
			\cline{2-3}			
			&			
			Sensing with asynchronous transmitter and receiver, an analogy to the bi-static radar \cite{IndoTrack,ni2020uplink} (To be detailed in Section \ref{sec-resamb}).
			
			&
			- Resolution of sensing ambiguity in the estimation of propagation delay and Doppler frequency;
			
			- Removal of random phase shift across discontinuous temporal measurements to enable coherent processing and sensing parameter estimation. 
			\\
			\cline{2-3}
			
			&
			Sensing algorithms that can be adapted to communication signals and the propagation environment \cite{Kumari17,8057284,8827589} (To be detailed in Section \ref{sec-spe}).
			
			&			
			- Estimation of continuous sensing parameters  with potentially irregular measurements in time and frequency domains;
			
			- How to effectively use data symbols, in addition to pilots, for sensing;
			
			- How to remove unwanted multipath to improve sensing performance?
			
			\\
			\cline{2-3}
			
				&
			
			Joint signal design and optimization, such as beamforming optimization \cite{Liu18mumimo} and sensing assisted communications \cite{8422132, Liu20} (To be detailed in Section \ref{sec-waveform} and \ref{sec-sensingassit} ).
			&
			- How to optimize signals jointly for C\&S across one or more domains of spatial, time and frequency?
			
			- How to exploit environmental information to improve signal reception performance and the security of communications?
			
				\\
			\cline{2-3}
						
			&
			
			Information theories and technologies for sensing in a networked environment \cite{9186070,8847233} (To be detailed in Section \ref{sec-mi} and \ref{sec-networked}).
			
			&
			- Information theory for JCAS is very limited, particularly for a networked system, and almost needs to be started from a scratch;
			
			- Prior research on multi-static radar is very limited;
			
			- Cooperation and competition for sensing need to be jointly considered with the requirements for communications. 			
		
			\\
			
			\hline
		\multirow{2}{1.5cm}{\textbf{Joint Design (II.D)}}
		
			&
			
			High-frequency systems such as mmWave and sub-Terahertz-wave JCAS systems that can potentially achieve both high-data-rate communications and high-accuracy sensing \cite{8550811,8828030,elbir2021terahertzband}. 
			
			&
			- Design of effective signaling schemes that best exploit the large bandwidth;
			
			- Beamforming design to meet the conflictive requirements for high directivity for communications and wide-range of scanning for sensing;
			
			- Development of beamforming tracking techniques to support the communication and sensing of mobile nodes.
			
					\\
			\cline{2-3}			
			
			&			
			Multi-channel JCAS systems which can offer an overall large signal bandwidth for sensing, while without increasing instantaneous communication bandwidth \cite{9104263, 8423070}. 		
			
			&
			
			- Multi-channel stitching for sensing; 
			
			- Removal or reduction of the imperfections and distortions from the received signals across channels;
			
			- Smart signal design to ease channel stitching.
			\\
					
			\hline
		\end{tabular}
\end{table*}

\section{Framework for a PMN}\label{sec:intro2}

In this section, based on existing works, we present a framework of PMN that integrates radio sensing into the current communication-only mobile network, using JCAS technologies. In this framework, we describe the optional system architecture, introduce three types of unified sensing, and discuss communication signals that can be used for sensing. 

 \begin{figure*}[t]
	\centering
	\includegraphics[width=0.75\linewidth]{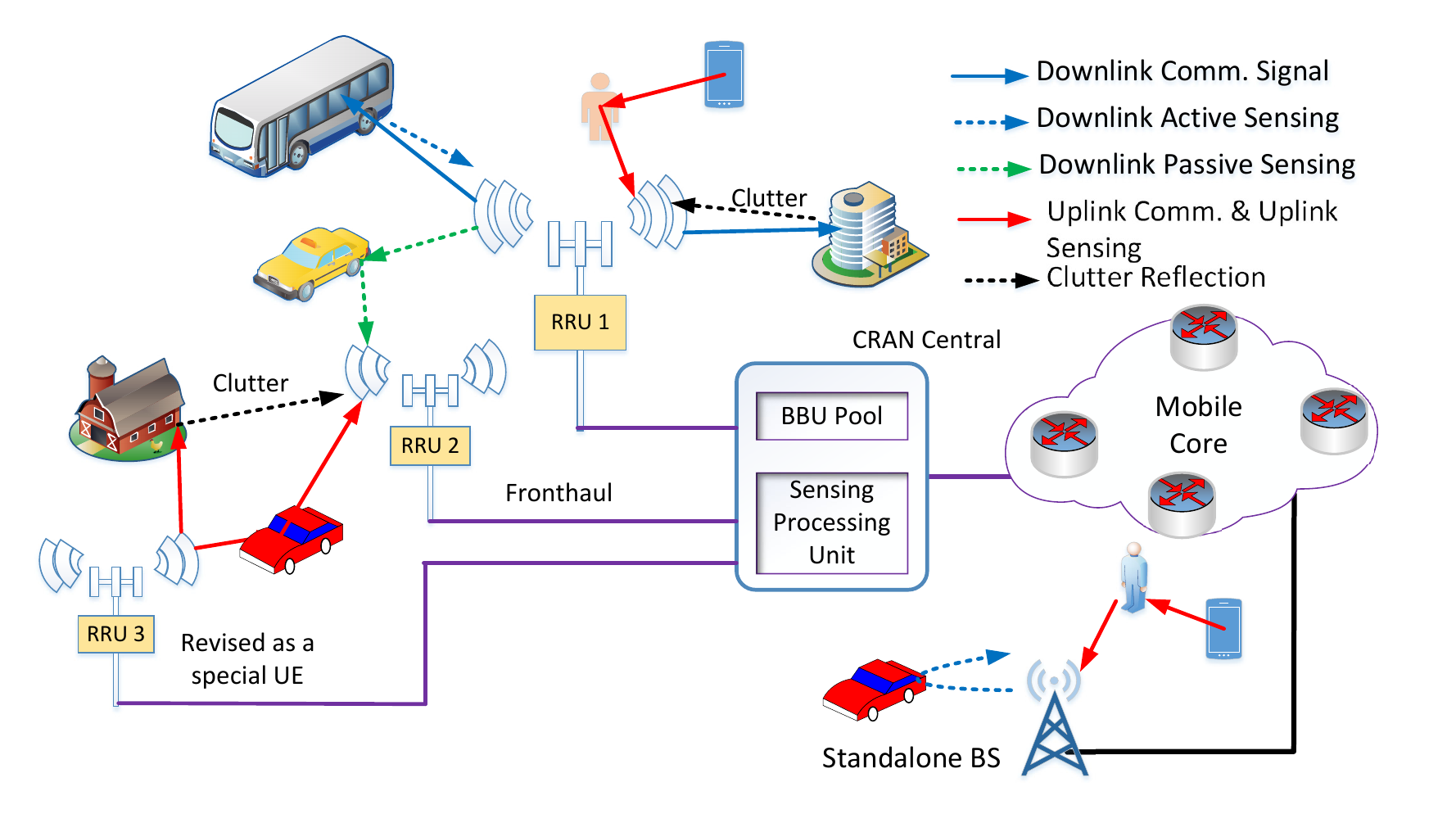}
	\caption{Illustration of sensing in a PMN with both standalone BS and CRAN topologies. RRU1 is a node supposed to have full-duplexing capability or equivalent. RRU3 is modified to be a special UE, transmitting uplink signals for uplink sensing in RRU2, with clock synchronization between them. RRU2 can also be modified as a receiver only, to do both uplink and downlink sensing, as well as communications (receiver only) .}
	\label{fig-system}
\end{figure*}

\subsection{System Platform and Infrastructure }\label{sec-problem}

The PMN can evolve from the current mobile network, with modification and enhancement to hardware, systems and algorithms. In principle, sensing can be realized in either the user equipment (UE) or base station (BS). Sensing in UE may motivate wider end-user applications. Compared to UE, BS has advantages of networked connection, flexible cooperation, large antenna array, powerful computation capability, and known and fixed locations to enable more reliable sensing results. Therefore, in the following, we mainly consider BS-side sensing. 

The evolution to PMN is not limited to a particular cellular standard. Hence we try to generalize the discussions by considering key components and technologies in modern mobile networks, such as antenna array, broadband, multi-user MIMO and orthogonal frequency-division multiple access (OFDMA), instead of a specific standard. When necessary, we will refer to the 5G new radio (NR) standard.

Depending on the network setup, we describe two types of topologies where JCAS can be implemented, that is, a cloud radio access network (CRAN) and a standalone BS. The realization of sensing in a PMN based on these two topologies is illustrated in Fig. \ref{fig-system}. Below we elaborate on the system and network setup for the two topologies. We will then discuss three types of sensing operations based on the topologies in subsection \ref{sec-updown}. Methods that modify the setup to enable sensing will be discussed in Section \ref{sec:structure}.

\subsubsection{CRAN}

A typical CRAN consists of a central unit and multiple distributed antenna units, which are called remote radio units (RRUs) \cite{7169508}. The RRUs are typically connected to the CRAN central via optical fibre. Either quantized radio frequency signals or baseband signals can be transmitted between RRUs and the central unit. As shown in Fig. \ref{fig-system}, in a CRAN PMN, the densely distributed RRUs, coordinated by the central unit, provide communication services to UEs. Their received signals, either from themselves, other RRUs, or from UEs, are collected and processed by the CRAN central, for both C\&S. The CRAN central unit hosts the original baseband unit (BBU) pool for processing communication functions and the new sensing processing unit for sensing. This setup aligns with the topologies for distributed radar systems \cite{Liang11,5703085}. 

A typical communication scenario is as follows: several RRUs work cooperatively to provide connections to UEs, using multiuser MIMO techniques over the same resource blocks (same time and frequency slots). In CRAN communication networks, power control is typically applied such that signals from one RRU may not reach other RRUs. While it is not necessary, we can relax this constraint and assume that cooperative RRUs are within the signal coverage area of each other. This assumption is reasonable when dense RRUs are deployed and used to support surrounding UEs via coordinated multipoint techniques. This is not necessary for some types of sensing as we are going to discuss in next subsection, but it increases the options of sensing \cite{8827589}. Technically, it is also feasible at the cost of increased transmission power, even if only for supporting sensing, as the downlink signals do not cause mutual communication interference to RRUs. 

Note that, in this configuration, all RRUs are typically synchronized using the timing clock from the GPS signals. This forms an excellent network with distributed nodes for sensing applications \cite{9186070}. Such CRAN-based PMN is investigated in \cite{RN9, RN30,8827589}, together with the sensing algorithms.

\subsubsection{Standalone BS}

The CRAN topology is not necessary for realizing sensing in PMNs. A standalone BS can also perform sensing using the received signals either from its own transmitted signals or from UEs.  This is actually the typical and simpler setup that has been widely considered in the literature.  This setup includes the small BS that may be deployed within a household, which pushes for the concepts of edge computing and sensing. Like WiFi sensing \cite{8067693}, such a small BS can be used to support indoor sensing applications such as fall detection and house surveillance. It also includes a roadside unit (RSU) that is part of the mobile network but specifically deployed to support vehicular communications \cite{liu2019joint, Yuan20Bayesian}.

From now on, our discussions will be referred to the CRAN topology, but most of the results are applicable to the standalone BS. Hence in the case without causing confusion, we will use CRAN and BS interchangeably.

\subsection{Three Types of Sensing Operations}\label{sec-updown}
There are three types of sensing that can be unified and implemented in PMNs, defined as \textit{uplink and downlink sensing}, to be consistent with uplink and downlink communications \cite{8827589}. In uplink sensing, signals received from UEs are used for sensing, while in downlink sensing, the sensing signals are from BSs.  The downlink sensing is further classified as \textit{Downlink Active Sensing} and \textit{Downlink Passive Sensing}, for the cases when an RRU collects the echoes from its own and other RRU-transmitted signals, respectively. The terms active and passive are used to differentiate the cases of sensing using self-transmitted signals and signals from other nodes. Below, we elaborate on each sensing operation. 

\subsubsection{Downlink Active Sensing}
In {downlink active sensing}, an RRU (or BS) uses the reflected/diffracted signals of its own transmitted downlink communication signals for sensing.  This is the typical case considered in systems where the sensing receiver is co-located with the transmitter \cite{Braun14, Liu18, Barneto21}, like a mono-static radar.  Downlink active sensing enables a BS to sense its surrounding environment. Since the transmitter and receiver are on the same platform, they can be readily synchronized at the clock-level, and the sensing results can be clearly interpreted by the node without external assistant.  However, this setup would require full-duplexing capability or equivalent \cite{Barneto21}.

\subsubsection{Downlink Passive Sensing}
Here, {downlink passive sensing} refers to the case where an RRU uses the received downlink communication signals from other RRUs for sensing. In terms of sensing, this corresponds to the setup of bi-static and multi-static radar, where the transmitter(s) and receiver are spatially separated, but their clocks may be synchronized \cite{9161314, 9180098}. Downlink passive sensing signals will be available to this RRU when the transmission power is sufficiently large. In this case, they will always be there together with the downlink active sensing signals, the reflection and refraction of the RRU's own transmitted signal. They may arrive at the sensing receiver slightly later than the downlink active sensing signals, due to longer propagation distances. When all RRUs cooperatively communicate with multiple UEs using SDMA, these two types of signals cannot be readily separated in time or frequency. Therefore sensing algorithms also need to consider downlink active sensing signals if downlink passive sensing is in operation. This setup and its sensing algorithms for this complicated scenario have been investigated in \cite{8827589}. In general, downlink passive sensing senses the environment between RRUs. 

\subsubsection{Uplink Sensing}
The {uplink sensing} conducted at the BS utilizes the received uplink communication signals from UE transmitters. It is similar to passive sensing \cite{8666692} in the sense that the transmitter and receiver are spatially separated and non-synchronized. The difference is that in uplink sensing, the receiver is fully aware of the system protocol and signal structure. Uplink sensing can be directly implemented without requiring change of hardware and network setup, and without requiring for full-duplex operation. However, it estimates the relative, instead of absolute, time delay and Doppler frequency since the clock/oscillator is typically not locked between spatially separated UE transmitters and BS receivers. This ambiguity may be resolved with special techniques, as will be discussed in Section \ref{sec-resamb}. Uplink sensing senses UEs and the environment between UEs and RRUs, and has been studied in \cite{ni2020uplink}.

\begin{table*}[b]
	
	\def\arraystretch{1.2}
	\centering
	\caption{Comparison of Three Types of Sensing Operations}
	\begin{tabular} {|m{2cm}|m{2.8cm}|m{2.4cm}|m{4.15cm}| m{4.0cm}|}
		
		\hline
		\textbf{Types} & \textbf{Signals} & \textbf{Action} & \textbf{Advantages} & \textbf{Disadvantages} \\
		\hline
		\textbf{Downlink Active Sensing}
		&
		Reflects from a RRU/BS’s own transmitted downlink communication signal
		&
		Sense surrounding environment of the RRU/BS.

		&
		
		All data symbols in the received signals can be used and are centrally known.
		
		&
		Generally require full-duplex operation and other network modifications. Devices can be specially deployed to resolve this problem.

		\\
		\cline{1-3}
		\textbf{Downlink Passive Sensing}
		&
		Received downlink communication signals from other RRUs
		
		&
		Sense environment between RRUs.
		&
		
		RRUs are synchronized.
		Privacy is less an issue because sensed results not directly linked to any UEs.

		&

		\\
		\hline
		\textbf{Uplink Sensing}
		&
		Uplink communication signals from UE transmitters
		
		&
		Sense UEs and environment between UEs and RRU. 
		&
		Require minimum modification to communication infrastructure. Does not require full-duplexing.
		
		&
		Timing and Doppler frequency measurement could be relative. Transmitted information signals are not directly known. Rapid channel variation when UEs are moving.
		
		\\
		
		\hline
		
	\end{tabular}
	
	\label{table:sensing}
	
\end{table*}

\subsubsection{Comparison}
Downlink sensing can potentially achieve more accurate sensing results than uplink sensing. This is because, in the downlink sensing case, RRUs generally have more advanced transmitters such as more antennas and higher transmission power, and the whole transmitted signals are centrally known. Additionally, as the sensed results in the downlink sensing are not directly linked to any UEs, the privacy issue is largely not a problem. Comparatively, uplink sensing may disclose the information of UE, causing privacy concerns.

Downlink and uplink sensing in PMNs are both feasible for practical applications in terms of sensing capabilities. According to the results in \cite{RN9} and \cite{RN30}, the downlink and uplink sensing with practical transmission power values (smaller than 25 dBm) can reliably detect objects more than 150 and 50 meters away, respectively, in a dense multipath propagation environment. Additionally, a distance resolution at a few meters can be achieved for a signal bandwidth of 100 MHz, an angle resolution of about 10 degrees for a uniform linear array of $16$ antennas, and a resolution of 5 m/s moving speed within channel coherence period. 

A comparison of the three types of sensing is provided in Table \ref{table:sensing}.

\subsection{Signals Usable from 5G NR for Radio Sensing}\label{sec-5gsignals}
For 5G NR, we can exploit the following signals for sensing: reference signals used for channel estimation, synchronization signal blocks (SSBs), and data payload, as was explored in, e.g., \cite{Zhangpmn20, 8805161}. The properties of these signals in terms of sensing are summarized in Table \ref{tab-property}. These communication signals may be further jointly optimized for C\&S, using methods in, e.g.,  \cite{Liu17ada,Liu18, Liu18mumimo}.
\begin{table*}[t]
	\centering
	\def\arraystretch{1.2}
	\caption{A summary of the properties of the signals that can be used for sensing in PMN, with reference to 5G NR.}
	\label{tab-property}
	\begin{tabular}{|m{2cm}|m{3cm}|m{3cm}|m{3cm}| m{4cm}|}
		\hline
		\textbf{Type of Signals}	& \textbf{Signal Pattern in Time Domain} & \textbf{Signal Pattern in Freq Domain} & \textbf{Signal Correlation} & \textbf{Signal Values}  \\
		\hline
		\textbf{Reference Signals}	& Irregular and varying length & May be on a regular comb structure & Typically orthogonal  & Known and fixed \\
		\hline
		\textbf{Synchronization Signal Blocks (SSBs)}	& Short and less frequent (every ~20 ms) & Sparsely distributed, using ~20 resource blocks & Orthogonal over smaller spatial layers  & Known and fixed \\
		\hline
		\textbf{Data Payload} & UE specific, irregular, and long & Allocation dependent, resource block based & Statistically independent & Known in downlink sensing, unknown in uplink sensing (become known after demodulation) \\
		\hline
	\end{tabular}
\end{table*}

\subsubsection{Reference Signals Used for Channel Estimation}
Deterministic signals specifically designed for channel estimations are available in many systems. The 5G NR \cite{3gp38}  includes the demodulation reference signals (DMRS) for both uplink (Physical uplink shared channel-PUSCH) and downlink (Physical downlink shared channel-PDSCH), sounding reference signals (SRS) for uplink, and channel state information– reference signals (CSI-RS) for downlink. Most of them are comb-type pilot signals, circularly shifted across OFDM symbols, and are orthogonal between different users. Especially, DMRS signals accompanying the shared channel are always transmitted with data payload and exhibit user-specific features. Therefore DMRS signals are random and irregular over time, which requires sensing algorithms that can deal with such irregularity. Comparatively, signals used for beam management in connected modes, like SRS and CSI-RS, can be either periodic or aperiodic, and hence they are more suitable for sensing algorithms based on conventional spectrum estimation techniques such as ESPRIT. Such training signals for channel estimation are the most widely exploited ones for sensing in the JCAS literature, e.g., in \cite{liu2019joint, yuan2020waveform,ni2020uplink}.

The number and position of DMRS OFDM symbols are known to BSs, and they can be adjusted and optimized across the resource grid, including slots and subcarriers (resource blocks). This implies good prospects for both channel estimation and sensing in different channel conditions. The allocation of resource grids can be optimized by considering requirements from both communications and sensing. With a given subcarrier spacing, the available radio resources in a sub-frame are treated as a resource grid composed of subcarriers in frequency and OFDM symbols in time. Accordingly, each resource element in the resource grid occupies one subcarrier in frequency and one OFDM symbol in time. A resource block consists of 12 consecutive subcarriers in the frequency domain. A single NR carrier is limited to 3300 active subcarriers as defined in Sections 7.3. and 7.4 of TS 38.211 in \cite{3gp38}. The number and pattern of the subcarriers that DMRS signals occupy have a significant impact on the sensing performance, as we will see in Section \ref{sec-spe}.

In \cite{8905229}, some simulation results for both uplink and downlink sensing using DMRS are provided. The signal is generated according to the Gold sequence as defined in \cite{3gp38} of \textit{3GPP TS 38.211}, for both \textit{PDSCH} and \textit{PUSCH}. The generated physical resource-block (PRB) is over a 3-D grid comprising a 14-symbol slot for the full subcarriers across the DMRS layers or ports. The interleaved DMRS subcarriers of PDSCH are used in downlink sensing, while groups of non-interleaved DMRS subcarriers of PUSCH are used in uplink sensing. The results demonstrate the feasibility of achieving excellent sensing performance with the use of the DMRS signals. However, a major problem of sensing ambiguity is also noted due to the interleaved pattern of the subcarriers. 

\subsubsection{Non-Channel Estimation Signals}
Several deterministic non-channel estimation signals such as the synchronization signal (SS) and the physical broadcast channel (PBCH), also called the SSB, can also be used for sensing. Such signals typically have regular patterns with a periodic appearance at an interval of several to tens of milliseconds. However, they only occupy a limited number of subcarriers, which may lead to limited identification of multipath delay values. Work on using this class of signals for JCAS has not been reported.

\subsubsection{Data Payload Signals}
In addition, we can also exploit the data payload signals in both the physical downlink shared channel (PDSCH) and physical uplink shared channel (PUSCH) for sensing, as has been investigated in, e.g., \cite{8827589, Liu20super}. In downlink sensing, the data symbols are known to the sensing receiver and hence can be directly used. In uplink sensing, symbols need to be used in a decision-directed mode. Since these data symbols are random and signals in different spatial streams are non-orthogonal, they are not ideal for sensing. If it is used for uplink sensing, the signals need to be demodulated first, which could also introduce demodulation error. However, they can significantly increase the number of available sensing signals, and hence improve the overall sensing performance at the cost of increased complexity. Precoders for these signals can be optimized by jointly considering the requirements from C\&S.

\subsection{Summary and Insights}

In this section, we have reviewed the system architecture of PMN, options for realizing sensing, and the communication signals that can be used for sensing in PMN networks. The key points and insights are summarized below.
\begin{itemize}
	\item PMN can be evolved from the current mobile network, with sensing being implemented at either the network side or UE side. Network-side sensing has the advantages of higher processing power and better information access, and hence is preferred. It can be realized in either a standalone BS, or multiple BSs or RRUs in a collaborative way;
	\item Three types of sensing can be realized in PMN: downlink active and passive sensings using downlink communication signals, and uplink sensing using uplink communication signals. A comparison of the three types of sensing operations is provided in Table \ref{table:sensing}. They can be realized individually or together;
	\item Almost all the communication signals can be used for sensing, with respective advantages and disadvantages. A comparison of three types of such signals is provided in Table \ref{tab-property}. Overall, the reference signals, such as the DMRS signals for channel estimation in communications, typically have the best properties for sensing. However, when more signals are needed, the data payload and synchronization signal blocks can also be used. Due to the different properties and therefore different performance impact on sensing, combined usage of these signals for sensing need to be carefully planned and optimized.
\end{itemize}

In the appendix, some mathematical introduction to the signal and channel models for JCAS in PMN is provided.

\section{Evolution: System Modifications to Enable Sensing}\label{sec:structure}

\begin{figure}[t]
	\centering
	\includegraphics[width=0.78\columnwidth]{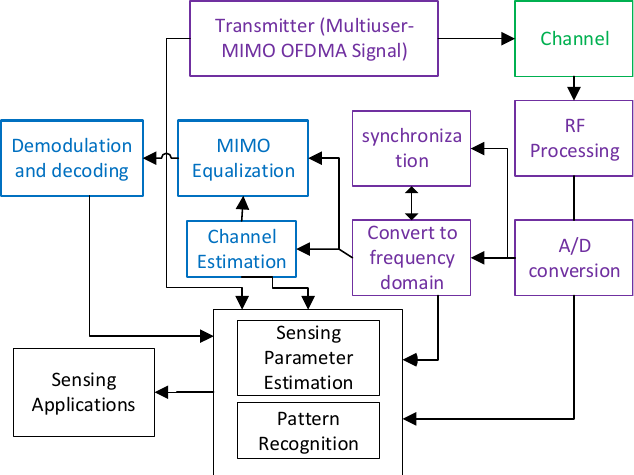}
	\caption{A block diagram of a transceiver showing the components that can be shared by C\&S. The blocks in purple are shared by C\&S; blocks in blue and black are for communications and sensing only, respectively. The sensing unit, including sensing parameter estimation and pattern recognition, can use signals outputs from multiple modules. A new module, sensing and communication cooperation, is not shown in the figure.}
	\label{fig-hardware}
\end{figure}

C\&S can share a number of processing modules in a MIMO-OFDM transceiver, as illustrated in Fig. \ref{fig-hardware}. The whole transmitter and many modules in the receivers that are shown in purple are shared by C\&S. The transmitted signal waveform can be optimized by jointly consider the requirements for C\&S, as will be detailed in Section \ref{sec-waveform}. Note that sensing parameter estimation can be done in both time domain and frequency domain. The sensing applications may demand either sensing parameter estimation or pattern recognition results, or both.

Despite the numerous modules shareable by C\&S, some modifications at hardware and network levels to existing mobile networks are necessary for realizing PMNs. As discussed in Section \ref{sec-diffcs}, communication signals can generally be directly used for estimating sensing parameters, but the communication system platform is not directly ready for sensing. On one hand, a communication node does not have the full-duplex capability at the moment, that is, transmitting and receiving signals of the same frequency at the same time. This makes mono-static radar sensing infeasible without modifying current communication infrastructure. On the other hand, for transmitter and receiver in two nodes spatially separated, there is typically no clock synchronization between them. This can cause ambiguity in ranging estimation, and makes processing signals across packets difficult. Thus bi-static radar techniques cannot be directly applied in this case. These are fundamental problems that need to be solved at the system level, to make sensing in primary communication systems feasible. 

We now describe the modifications of current hardware and systems that are required to evolve current communication only mobile networks to PMNs. The depicted changes focus on the fundamental reforms that allow the current mobile network to do radio sensing simultaneously with communication. In this section, we do not consider low-level changes such as joint waveform optimization \cite{Liu17ada,Liu18, Liu18mumimo}, joint antenna placement and sparsity optimization processing and power optimization \cite{Wang18}, but leave them to Section \ref{sec:wlk}. For the three types of sensing integrated in PMN, the realization of uplink sensing is relatively easy; the major challenges are with downlink sensing, where the leakage and reflected signals from the transmitter can cause significant interference to the received signals for both sensing and communications. We now review four options of system modifications to enable sensing in PMN.

\subsection{Dedicated Transmitter for Uplink Sensing}

Conventional uplink sensing can be realized in a similar way to passive sensing \cite{Hack14,Gogineni14, Abdullah2016}, with the difference that the receivers in PMN are part of the network and process signals for both communication and sensing. Uplink sensing confronts the same sensing ambiguity problem with passive sensing. Conventionally, the phase clock between UEs and BSs is not synchronized; hence, sensing ambiguity in time and Doppler frequency is present in uplink sensing. If the ambiguity can be tolerated, no change to hardware and system architectures of current mobile systems is required. Such sensing ambiguity may also be resolved using signal processing techniques \cite{ni2020uplink}, under some special situations, as will be detailed in Section \ref{sec-resamb}.  

To eliminate the ambiguity, dedicated (static) UEs that are clock-synchronized to BSs can be used. In terms of the required system modification, uplink sensing by static UE would be the most convenient way for achieving non-ambiguity sensing in the PMNs. This is shown as RRU3 in Fig. \ref{fig-system} for a CRAN, where RRU3 can be modified to operate as a UE, transmitting uplink signals. This option can also enable single-input multi-output (SIMO) sensing, where the transmitted signals can be processed jointly at multiple spatially separated BSs for collaborative sensing.

\subsection{Using Full-Duplex Radios for Downlink Sensing} \label{sec-fullduplex}
Resolving the signal leakage problem in downlink sensing would ideally require \textit{full-duplex} technologies \cite{7105651,8805161}. \bb{As has been reviewed in \cite{7105651}, {full-duplex} communications have been widely investigated. But it is still not very practical to be applied in an environment involving mobility and dynamics, particularly for TDD systems. In {full-duplex} communications, a device needs to be able to receive signals from other communication devices, while transmitting its signals using the same frequency channels. The receiver needs to recover the weak received communication signals, while suppressing the onboard leakage and echoes from the environment of the transmitted signal. The signal to be recovered can be several orders lower than the interference signal. Full duplex communications generally use a combination of antenna separation, RF suppression, and baseband suppression to mitigate the leaked transmit signal and its echoes from the environment at the receiver.} 
	
\begin{figure}[t]
	\centering
	\includegraphics[width=0.9\linewidth]{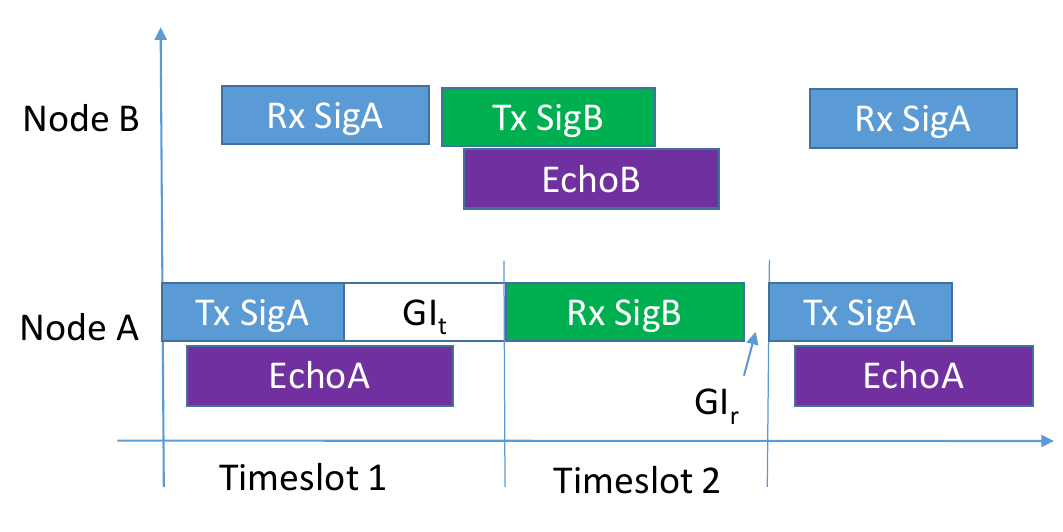}
	\caption{\bb{Illustration of the timeslot allocation in a full-duplex TDD JCAS system from the viewpoint of Node A. Full-duplex operation is only required during timeslot 1. GI stands for guarding interval, ``Sig'' is a shortened form of signal.}}
	\label{fig:tdd}
\end{figure}

\bb{We note that it may be easier to realize full-duplex for JCAS compared to for full-duplex communications, if we only require co-located sensing while transmitting, but not a simultaneous transmission from two communication nodes. This is because we only need to remove the impact of the directly leaked signal from the transmitter while keeping the echoes of the transmitted signals for sensing, without the presence of communication signals from other devices. There is generally no need to consider the interference between sensing and received communication signals. As illustrated in Fig. \ref{fig:tdd},  a sufficient guarding interval between communication transmission (sensing receiving) and communication reception, as existing in current TDD systems, shall be able to prevent mutual interference between communication and sensing. The guarding interval following transmission in Node A, $GI_t$, may only need to be slightly increased compared to conventional communication-only TDD systems, to accommodate possible longer echo sensing signals.}

Therefore, full-duplex operation is a potentially long-term solution to enable seamless integration of downlink sensing with communications. In \cite{8805161}, it is shown that moving targets are more robust to the leaked self-interference, whereas limited transmitter-receiver isolation is primarily a concern for detecting static objects. In \cite{Barneto21}, multibeam \cite{8550811, 9148935} techniques are investigated for full-duplex JCAS systems. Using the multibeam technology, where separated beams can be generated and optimized for C\&S, it is demonstrated that leakage and clutter signals can be significantly reduced.

Nevertheless, it is still very challenging to implement full-duplex JCAS, particularly for MIMO systems. The main reason is that in MIMO systems, a large number of leakage signals between pairs of transmitter and receiver antennas need to be handled simultaneously. Overall, full-duplex JCAS would be an ideal future technology for PMN, but it is immature and impractical for real implementations in the near future. 

\subsection{Dedicated Receiver for Downlink (and Uplink) Sensing}
For downlink sensing without requiring full-duplexing capability, one near-term option is to deploy a BS that only works on the receiving mode \cite{ni2020waveform}. It can be configured as a receiver either for downlink sensing only or for both communication and downlink sensing. 

To implement this near-term downlink sensing, changes to the hardware may be required. This is because the receiver in current BSs is conventionally designed to receive uplink communication signals only, and downlink sensing requires the receiving of downlink communication signals. The required change is insignificant for TDD systems since a TDD transceiver generally uses a switch to control the connection of antennas to the transmitter or receiver. Thus the change is only the adjustment of the transmitting and receiving period so that the switch is equivalently always connected to the receiver. For FDD systems, the BS receivers may be incapable of working on downlink frequency bands, and modification to the hardware is required. Therefore, it is more cost-effective to implement downlink sensing in TDD than in FDD systems. 

Alternatively, we can also deploy a dedicated receiving-only node for both downlink and uplink sensing, as well as communications if desired. This is particularly feasible for TDD systems. In TDD systems, downlink and uplink sensing signals can then be (largely) separated in time at the receiver. Of course, to remove the ambiguity in delay estimation, clock synchronization is required between the transmitters and this node. An example is shown as RRU2 in Fig. \ref{fig-system} for a CRAN, which can perform downlink and uplink sensing using received signals from RRU1 and RRU3, respectively. 

\subsection{BS with Spatially Widely Separated Transmitting and Receiving Antennas}

One possible solution for downlink sensing is to use well-separated transmitting and receiving antennas. The large separation will significantly reduce the leakage from transmitted signals. The receiver baseband also accepts feedback from the transmitter baseband, so that a baseband self-interference cancellation may be further applied. However, this spatially well-separated antenna structure requires extra antenna installation space and can increase the overall cost.  One option of minimizing the cost is to \textit{use a single spatially separated antenna for receiving sensing signals} in a conventional MIMO system, which could be the most cost effective solution. 

\begin{figure}[t]
\centering
\includegraphics[width=0.9\linewidth]{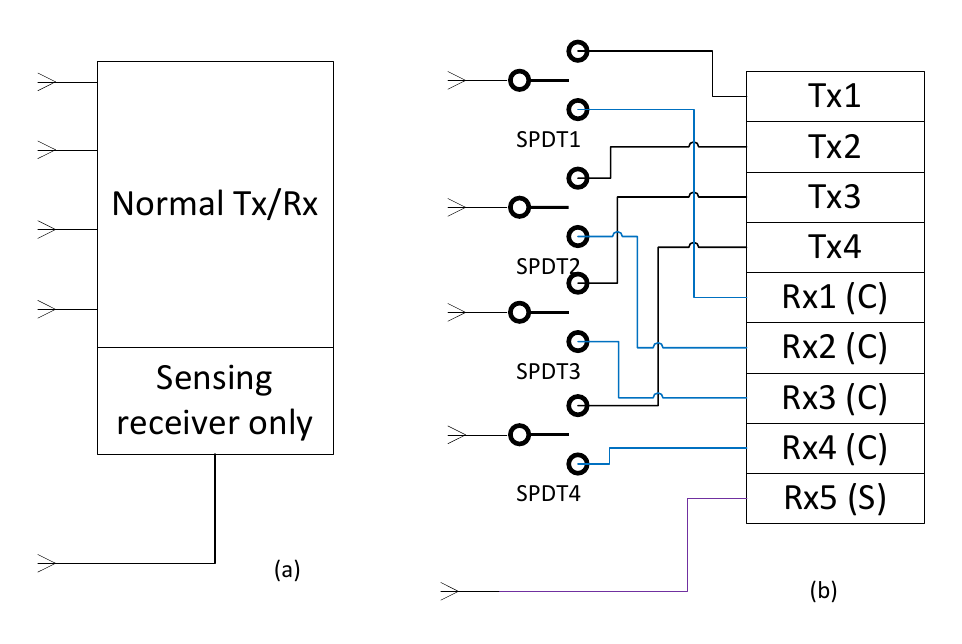}
  \caption{Simplified TDD transceiver model with a single receiving antenna dedicated to sensing. (a) A general concept; (b) Possible realization on existing hardware platform.}
\label{fig:trans}
\end{figure}

Fig. \ref{fig:trans} shows an example of this option in TDD systems. The system has a normal transceiver for communication with four antennas. A fifth antenna is installed at a position well separated from the four antennas, and it is connected to the receiver via a long cable. Signals from the fifth antenna are used for downlink active sensing.  Fig. \ref{fig:trans}-(a) plots the general concept, and Fig. \ref{fig:trans}-(b) shows a potential implementation in existing FDD systems. The switches (SPDT1-4) are operating normally for a TDD communication system. For the fifth antenna, it is always connected to the fifth receiver. Given that the onboard circuit leakage is small and the TDD switches can be separately controlled, this option can be conveniently realized in an existing TDD system that supports $5\times 5$ MIMO. 

Sensing using a single receiving antenna in this case can be realized by exploring the multiple transmitted spatial streams \cite{ni2020waveform}. Although sensing performance may be slightly degraded due to reduced number of receiving antennas for sensing,  the degradation is mainly on the received signal energy. All the sensing parameters except for AoA can still be correctly estimated. Exploiting the estimated  AoD and delay, the locations of targets can still be determined. 

If there are spaces for installing multiple dedicated receive antennas for sensing, a single sensing receiver (RF down-converter channel) can also be connected to these antennas in a time-interleaved (multiplexing) way, so that more spatial diversity gain can be achieved and AoA can also be estimated. This is due to the relatively low variation speed of channel parameters in sensing. The feasibility has been demonstrated in \cite{843078}, where a single receiver is connected to multiple antennas using fast antenna array multiplexing to identify the time-variant, directional structure of mobile radio channel impulse responses. Practical measurement results show that the delay, AoA, and Doppler can be effectively estimated using this setup.

\subsection{Summary and Insights}

 In this section, we reviewed several system deployment options that can enable sensing in mobile networks. Table \ref{tab-options} summarizes four of these options. Using full-duplex radios would be an ideal future option but is currently impractical. The other three are near-term suboptimal solutions, requiring only a few slight modifications on hardware and system to the existing network. The option of using a single spatially-separated receive antenna for sensing in a MIMO system seems to be the most cost-effective solution for downlink sensing.

\begin{table*}[t]
	\centering
	\def\arraystretch{1.2}
	\caption{Comparison of options of enabling sensing in mobile networks, in terms of system deployment.}
	\label{tab-options}
	\begin{tabular}{|m{3.5cm}|m{7cm}|m{6cm}|}
		
		\hline
		\textbf{Options} & \centering\textbf{Advantages} & \hspace{1.5cm}\textbf{Disadvantages}\\
		\hline
		\textbf{Using full-duplex radios (IV.B)} &  An ideal future option with best system integration and flexibility in system design and optimization. & Immature today; high system and signal processing complexity. \\
		\hline
		\textbf{Dedicated transmitter for uplink sensing (IV.A)}	&  Easy to realize. Still support transmission for communication. Enable SIMO sensing (similar to bi-static and multi-static radar). & Slightly increased cost. \\
		\hline
		\textbf{Dedicated receiver for downlink sensing (IV.C)}	& Easy to realize in TDD systems. Still support communication reception. Enable MISO sensing. & Not suitable for current FDD systems. Slightly increased cost. \\
		\hline
		\textbf{Using one spatially widely separated receive antenna for sensing (IV.D)}	& Co-located transmitter and receiver enables mono-static sensing for environment directly surrounding the sensing node. Lowest-cost solution for TDD systems. & Require separated control of RF chains. Slightly increased space requirement for installation.  \\
		\hline
	\end{tabular}
\end{table*}

In addition to hardware modifications, hardware calibration is also important for sensing. In \cite{843078, 6060881}, it is shown how an imperfect receiver and antenna array can impact high-resolution sensing parameter estimation in channel sounding experiments. Antenna array calibration techniques are further investigated in \cite{843078, 6060881} and shown to be able to mitigate the impact effectively.

\section{Major Research Challenges for PMN }\label{sec-challenge}

There exist a number of challenges in the research and development of PMN. These challenges are mainly associated with realizing sensing on the infrastructure of communication networks, joint design and optimization, exploring the mutual benefits of communication and sensing via the integration, and sensing in a networked environment. In this section, we will discuss several major research challenges from the signal processing aspect. In next section, we will then review detailed technologies and algorithms that have been developed to address these challenges, and present remaining open research problems and future directions.

\subsection{Sensing Parameter Extraction from Sophisticated Mobile Signals}\label{sec-challenge1}

The sophisticated signal structure of mobile networks makes sensing parameter estimation in PMNs challenging. A modern mobile network is a complex heterogeneous network, connecting diverse devices that occupy staggered resources interleaved and discontinued over time, frequency and space. Mobile signals are also very complicated because of multiuser access, diverse and fragmented resource allocation, and spatial multiplexing. The communication signals that are also used for sensing are randomly modulated using multiuser-MIMO and OFDMA technologies and can be fragmented for each user - discontinuous over time, frequency or space. This structure is detailed in our work in \cite{8827589}. Most existing sensing parameter estimation techniques are not directly applicable to the PMNs because of such signal structure. For example, active radar sensing technologies mostly transmit linear FM (LFM) chirp modulated transmitted signals \cite{4383615}; and most passive bi-static and multi-static radars consider simple single carrier and OFDM signals \cite{Hack14,Abdullah2016, Gogineni14, 7833233}. In addition, conventional spectrum analysis and array signal processing techniques, such as MUSIC \cite{RN32} and ESPRIT \cite{Braun14}, may not be directly applicable either, as ESPRIT requires continuous observations that do not constantly exist here, and MUSIC requires a large number of samples to separate signal and noise subspaces. As a result, specific sensing techniques need to be developed for estimating sensing parameters from complicated and fragmented signals. 

Sensing parameters describe the propagation of signals in the environment and the detailed composition of channels. They typically have continuous but not discrete values. Thus most existing channel estimation and localization algorithms are not directly applicable either. Existing channel estimation techniques developed for modern mobile networks principally emphasize on estimating composite channel coefficients at quantized discrete grids, and localization mainly focus on the line-of-sight path and determines the locations of signal emitting objects. However, some recent techniques developed for channel estimation in millimeter wave systems \cite{8720024,8377093} can potentially be extended and applied to sensing parameter estimation, as will be detailed in Section \ref{sec-spe}.  

\subsection{Joint Design and Optimization}
One key research problem in JCAS, as well as PMNs, is how to jointly design and optimize signals and systems for C\&S. A number of studies have investigated the impact of the waveform and basic signal parameters on the performance of a joint system, as will be detailed in Section \ref{sec-waveform}. Such waveform and system parameter optimisation can result in performance improvement in standalone systems, but it has less impact compared to those at high levels, i.e., system and network levels. 

C\&S have very different requirements at the system and network levels. For example, in a multiuser MIMO communication system, the transmitted signal is a mix of multi-users’ random symbols, while ideal MIMO-radar sensing signals are unmodulated and orthogonal \cite{Haim08}. When using an array, radar sensing focuses on optimizing the formation and structure of virtual subarrays to increase antenna aperture and then resolution \cite{5419124}, but communication emphasises beamforming gain and directivity. Such conflicting requirements can make joint design and optimisation very challenging. More research is required to exploit the commonalities and suppress the conflicts between the two functions. 

Another important issue is how C\&S can benefit more from each other via the integration. This is far from being well understood. Current research has been limited to propagation path optimization \cite{859,9162000,Liu20,Yuan20Bayesian} and secure communications \cite{Su20Secure}.

\subsection{Networked Sensing}

Integrating sensing into mobile communication networks provides great opportunities for radio sensing under a cellular structure. However, research on sensing under a cellular topology is still very limited. The cellular structure for communication is designed to greatly increase the frequency reuse factor and hence improve spectrum efficiency and communication capacity. A cellular sensing network intuitively also increases frequency reuse factor, and hence the overall ``sensing'' capacity. On one hand, there is almost no known performance bound for such cellular sensing networks yet, except for a limited number of slightly related works, such as performance analysis for coexisting radar and cellular communication systems \cite{Khawar15} and radar sensing using interfered OFDM signals \cite{Braun14}. On the other hand, although research exists on distributed radar and multi-static radar, sensing algorithms that consider and exploit the cellular structure, such as co-cell interference, node cooperation, and sensing-handover over base stations, are yet to be developed. The challenge lies in the way to address competition and cooperation between different base stations under the cellular topology, for both performance characterization and algorithm development of networked sensing.

\section{Detailed Technologies and Open Research Problems} 
\label{sec:wlk}

As a new platform and network, PMN is still in its very early stage of research and development. As described in the last section, there are a number of challenges to overcome to make it practical, which also imply great research opportunities. Here we review existing technologies and algorithms that have been developed to address these challenges, organized under eight topics. We also discuss open research problems for each topic. Since the major issue in PMN is how to achieve radio sensing without compromising the performance of existing communications, we focus on the issues in realizing radio sensing, leveraging the existing cellular communication infrastructure, and how communications may be affected and improved by integrating the sensing functions.

\subsection{Performance Bounds}\label{sec-mi}

There are two types of performance bounds that can be used to characterize the performance limits of sensing in PMNs. One is based on the mutual information (MI), and the other is based on the estimation accuracy of sensing parameters, such as the Cramer-Rao lower bounds (CRLBs).

MI \cite{ 7472362} can be used as a tool to measure both the radar and communication performance. To be specific, for communications the MI between wireless channels and the received communication signals can be employed as the waveform optimization criterion, while for sensing, the conditional MI between sensing channels and the sensing signals is used~\cite{Bell1993Information,Yang2007MIMO,Zhu2017Information}. So MI for sensing measures how much information about the channel, the propagation environment, is conveyed to the receiver. Maximizing MI hence is particularly useful for sensing applications that rely more on feature signal extraction than on sensing parameters estimation, for example, for target recognition. The usage of MI and capacity is well known to the communication community. The usage of MI for radar waveform design can also be traced back to the 1990s \cite{Bell1993Information}. MI has also been used to optimize the performance of coexisting radar communication systems, e.g., in \cite{8835671}.

Let us consider simplified signal models to illustrate the formulations of MI. Let $\Ht_s$, $\Ht_c$ denote the channels for sensing and communications, respectively; let $\Yf_s$ and $\Yf_c$ be the received signals for sensing and communications, respectively; and let $\Xt=f(\St)$ be the transmitted signals where $\St$ is the information symbols and $f(\St)$ denotes the function converting $\St$ to $\Xt$. The function $f(\St)$ can be linear or nonlinear, and can be across multiple domains, including time, frequency,and spatial domains. When a spatial precoding matrix $\Pc$ is used in place of $f(\cdot)$, the received signals are given by
\begin{align*}
\Yf_s=\Ht_s\Pc\St+\Zt;\notag\\
\Yf_c=\Ht_c\Pc\St+\Zt,
\end{align*}
for sensing and communications, respectively. The MI expressions for communication and sensing can be represented as
\begin{align}
&I(\Ht_s; \Yf_s|\Xt)=h(\Yf_s|\Xt)-h(\Zt), \ \text{for sensing;}\notag\\
&I(\St; \Yf_c|\Ht_c)=h(\Yf_c|\Ht_c)-h(\Zt), \ \text{for comm.},
\label{eq-mi}
\end{align}
where $I(\cdot)$ denotes the (conditional) MI, and $h(\cdot)$ denotes the entropy of a random variable. We can thus see that the MI for communications and sensing are defined as the metrics conditional on the channel and the transmitted signals, respectively. In each case, the signals $\Xt$, or more specifically the precoder $\Pc$, are optimized to maximize the MI. 
  
MI for JCAS systems has been studied and reported in a few publications. They are typically conducted by jointly optimizing the two MI expressions and their variations in \eqref{eq-mi}. The work in \cite{ XU2015102} formulates radar mutual information and the communication channel capacity for a JCAS system, and provides preliminary numerical results. In \cite{ 7472362}, radar waveform optimization is studied for a JCAS system by maximizing MI expressions. In \cite{RN15}, the estimation rate, defined as the MI within a unit time, is used for analyzing the radar performance, together with the capacity metric for communications. In \cite{ LIU2019317},  an optimized OFDM waveform is proposed by maximizing a weighted sum of the communication data rates and the conditional mutual information for radar detection in a JCAS system. %

These available results serve as good bases for studying the MI for PMNs. Some specific problems to PMNs can be considered to make the results more practical. 
\begin{itemize}
	\item Firstly, the MI formulations for uplink and downlink sensing are different, due to the different knowledge on the transmitted signals and the channel differences. In the downlink sensing, the symbols are known to the receiver, and the channels for C\&S are different but could be correlated. For uplink sensing, the symbols are unknown to the receiver, and the channels are the same for C\&S. Hence, the optimization objective functions and results can be quite different for uplink and downlink sensing.
	\item Secondly, the specific packet and signal structures in cellular networks can have a significant impact on MI for both C\&S. For example, a packet signal may include training sequence and data symbols which will lead to different MI formulations and results, as their statistical properties are different. In \cite{yuan2020waveform}, the MI is studied for PMN, considering the frame structure and estimation errors. The findings from \cite{yuan2020waveform} indicate that the optimal solution for one function (communication or sensing) is generally not optimal for the other, and some trade-off needs to be made, particularly when the requirements for C\&S are very different, for example, when the directions of sensing and communications deviate significantly. This implies the importance of sensing-motivated user scheduling, i.e., taking user scheduling into joint optimization of C\&S.   
\end{itemize}

CRLB is a more traditional metric that has been widely used in characterizing the lower bound of parameter estimation in radar \cite{4558509,5262999,6178073}. For PMN, the closest work on CRLB is reported in \cite{6620961}, in the scenario of passive sensing using UMTS (3G) narrowband mobile signals. However, the CRLB expressions are not always available in closed-form, particularly for MIMO-OFDM signals, primarily because the received signals are nonlinear functions of sensing parameters. Therefore, although they can be found and evaluated numerically, the CRLB metrics are not easy to be applied in analytical optimization. It is even harder to apply them in optimization, jointly with another cost function.

MI has also been combined with other metrics to study the performance of radar systems. For example, two criteria, namely, maximization of the conditional MI and minimization of the minimum mean-square error (MMSE), are studied in~\cite{Yang2007MIMO} to optimize the waveform design for MIMO radar by exploiting the covariance matrix of the extended target impulse response. In~\cite{Tang2010MIMO}, the optimal waveform design for MIMO radar in colored noise is also investigated by considering two criteria: by maximizing the MI and by maximizing the relative entropy between two hypotheses that the target exists or does not exist in the echoes. In \cite{ni2020waveform}, waveform optimization is studied with the application and comparison of multiple sensing performance metrics including MI, MMSE and CRLB. It is shown that there are close connections between MI-based and CRLB-based optimizations, and the MI-based method is more efficient and less complicated compared to the CRLB-based method. Overall, research for JCAS and PMNs based on these combined criteria is still very limited.

{A brief summary of recent analytic results of using MI in JCAS systems is given in Table \ref{tab-mutual}}. A very recent paper also provides a comprehensive review of the performance limits of JCAS systems \cite{liu2021survey}.

\begin{table*}[tb]
\caption{ Recent analytic results for usage of MI in JCAS and related systems.}
	\centering
	\bgroup
	\def\arraystretch{1.4}
	\begin{tabular}{|m{1.5cm}|m{7cm}|m{8.5cm}|} 
		\hline 
		\textbf{Reference} & \centering\textbf{Models} & \hspace{1.5cm}\textbf{Main contributions}  \\ 
		\hline \hline
		\cite{RN15} &  JCAS system consists of an active, mono-static, pulsed radar and a single user communications system. & MI  is used for analyzing the radar performance, together with the capacity metric for communications.
		   \\ 
		
		\hline 
	\cite{ 7472362} &  Radar and cellular systems  share the same spectrum for simultaneous operation.  & Radar waveform optimization is studied by maximizing MI expressions.  \\ 
		
		\hline 
	\cite{ XU2015102} & MIMO radar-based JCAS system. & Formulate radar MI and the communication channel capacity.   \\ 
		
		\hline 
	\cite{ LIU2019317} & Mono static radar transceiver is employed for target classification while simultaneously used as a communications transmitter  & Propose an OFDM waveform optimized by maximizing a weighted sum of the communication data rates and the conditional MI for radar detection.
		
		\\ 
		\hline 
	\cite{yuan2020waveform} & JCAS MIMO downlink system with a signal packet structure, including training sequence and information data symbols.  & MI is studied for PMN, considering the frame structure and estimation errors.
		
		\\ 
		\hline 	
	\end{tabular}
	\egroup
	 
	\label{tab-mutual}
\end{table*}

\subsection{Waveform Optimization}\label{sec-waveform}

\begin{table*}[tb]
	\caption{ Classification of waveform optimization techniques in PMN.}
	\centering
	\bgroup
	\def\arraystretch{1.9}
	\begin{tabular}{|m{1.8cm}|m{1.7cm}|m{2cm}|m{10cm}|} 
		\hline 
		\multirow{9}{1.8cm}{\textbf{Waveform Optimization} (Single carrier, OFDM, and new waveforms such as OTFS) (VI.B)}  
				&\multirow{5}{1.7cm}{Spatial Domain (VI.B.1)} & \multirow{3}{2cm}{Single spatial precoder} & Minimize waveform difference \cite{Liu18, Liu18mumimo}  \\ \cline{4-4}
				&														&																	& Weighted-optimal and pareto-optimal waveform design based on MI and/or CRLB \cite{LIU2017331,ni2020waveform, yuan2020waveform } \\ \cline{4-4}
				&														&																	& Multi-objective function optimization \cite{yliu17} \\\cline{3-4}
																																			
				&												    & \multirow{2}{2cm}{Decomposed spatial precoder} & Multibeam optimization where two pre-generated sub-beams are optimally combined, to support fast varying scanning directions and multiple targets \cite{ 8550811,Luo19jcas,9131843,Liu20joint}.\\ \cline{4-4}
				&													&    																				& {Additional spatial streams are introduced to either remove multiuser interference caused by scanning beam or enhance sensing performance}\cite{chen2020composite}. \\\cline{2-4}
				
				&\multirow{2}{1.7cm}{Time Domain (VI.B.2)} 	& \multicolumn{2}{c|}{Optimize the signal structure in terms of length of packet and interval between packets \cite{Kumari20}.} \\\cline{3-4}
				&													& \multicolumn{2}{c|}{Optimize power distribution between different parts of packets \cite{yuan2020waveform}.} \\	\cline{2-4}
				&\multirow{2}{1.7cm}{Frequency Domain (VI.B.2)} 	& \multicolumn{2}{c|}{Optimize subcarrier allocations, e.g., non-uniform subcarriers to achieve better delay estimation \cite{Hakobyan20}.} \\\cline{3-4}
				&															& \multicolumn{2}{c|}{Mimic traditional frequency-modulated radar signals in the preamble \cite{Cui18mat,liu2019joint}.} \\	\cline{2-4}
				\hline
			\end{tabular}		
			\egroup																								
	\label{tab-w-opt}
\end{table*}

For JCAS, joint waveform optimization is a key research problem as the single transmitted signal is used for both functions but the two functions have different requirements for the signal waveform. As discussed in Section \ref{sec-diffcs}, traditional radar and communication systems use very different waveforms, which are optimized for respective applications. For example, recall that radar uses orthogonal and unmodulated pulsed or FMCW signals, while in PMNs, typically the signals are random, with multicarrier modulation and multiuser access. However, the waveform for one function may be modified to accommodate the requirements of the other, under joint design and optimization. The work in \cite{RN32} is one of the earliest ones that investigate waveform design for JCAS systems. The waveform design and signal parameters can have a significant impact on the overall performance of a JCAS system. For example, the numerical analysis in \cite{8057284} demonstrates the close linkage between the sensing resolution capabilities and the signal parameters for both single carrier and multicarrier communication systems. 

\begin{figure}[t]
	\centering
	\includegraphics[width=0.95\linewidth]{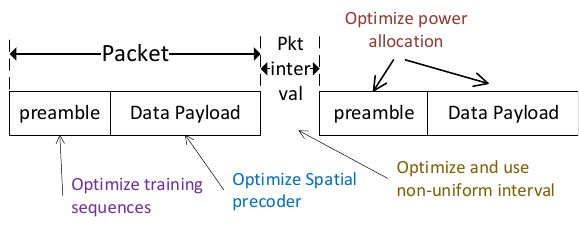}
	\caption{Waveform optimization depicted with respect to packets.}
	\label{fig-opt}
\end{figure}

In this subsection, we discuss three classes of waveform optimization techniques for PMN, classified in the spatial, time and frequency domains, as shown in Table \ref{tab-w-opt}. The first one is based on optimizing the spatial precoding matrix, and typically does not require the change of existing signals and can be seamlessly realized in current cellular networks. The second and third ones optimize signal parameters and resource allocation in the time and frequency domain, respectively, requiring slight modification. They are also depicted in Fig. \ref{fig-opt} with respect to signal packets. We also briefly review waveform optimization techniques that consider next-generation signaling schemes. 

\subsubsection{Spatial Optimization}
For spatial optimization with respect to the precoding matrix $\Pc$ in PMNs, apart from the MI-based waveform optimization as discussed in Section \ref{sec-mi}, there are two more practical methods. One method is optimizing the signal conversion function $f(\cdot)$, or specifically the precoding matrices $\Pc_{q,n,t}$ in \eqref{eq-yfnt0} when $f(\cdot)=\Pc$, to make the statistical properties of the transmitted signals $\Xt$ best suitable for both C\&S. Another method is to add the sensing waveform to the underlying communication waveform, while considering a coherent combination of the two waveforms for destination nodes. The two methods have respective advantages and disadvantages. We elaborate on them below.

In the first method,  when optimization is with respect to the spatial precoding matrix, it is designed to alter the statistical properties of the transmitted signal. This method is particularly suitable for global optimization of cost functions jointly formulated for C\&S. The basic optimization formulation is as follows:
\begin{align}
&\arg\max_\Pc \lambda(\Pc),\notag\\
&\text{subject to Constraints 1, 2} \cdots,
\label{eq-wfopt}
\end{align}
where $\lambda(\Pc)$ is the objective function. There could be various methods and combinations in defining the objective functions and the constraints. Each can be either for communication or sensing individually, or a weighted joint function. Some examples are as follows. In \cite{Liu18}, waveform optimization is realized via minimizing the difference between the generated signal and the desired sensing waveform under the restrictions of signal-to-interference-and-noise ratio (SINR) for multiuser MIMO downlink communications. A multi-objective function is further utilized to trade off the similarity between the generated waveform and the desired one \cite{Liu18mumimo}. In \cite{LIU2017331}, adaptive weighted-optimal and Pareto-optimal waveform design approaches are proposed to simultaneously improve the estimation accuracy of range and velocity and the channel capacity for communication. In \cite{yliu17}, the weighting vector for subcarriers in OFDM systems is optimized by considering a multi-objective function involving communication capacity and CRLBs for the estimates of sensing parameters. One main disadvantage of this method is that, the precoding matrix needs to be optimized or redesigned once the communication or sensing setup changes. 

In the second method, basic waveforms can be designed in advance for either or both C\&S, and the two waveforms are then added in a way to jointly optimize the performance of C\&S. The basic idea can be mathematically represented as follows:
\begin{align}
&\argmax _{\alpha, \Pc_c, \text{or } \Pc_s} \lambda(\Pc), \notag\\
&\text{subject to Constraints 1, 2} \cdots,
\label{eq-wfopt2}
\end{align}
where $\Pc_c$ and $\Pc_s$ are the precoding matrices primarily targeted for communication and sensing, respectively, and $\Pc$ is a function of $\Pc_c$ and $\Pc_s$. Examples are $\Pc=\alpha\Pc_c+(1-\alpha)\Pc_s$ and $\Pc=[\alpha\Pc_c, (1-\alpha)\Pc_s]$ where $\alpha$ is a complex scalar. Here, among $\{\alpha, \Pc_c, \Pc_s\}$, either $\Pc_c$, $\Pc_s$, or both of them can be pre-designed and fixed, while real-time optimization is with respect to the other parameters. Although the results may be suboptimal, this method provides great flexibility and can adapt quickly to changes on the requirements for C\&S. 

This method could be particularly useful for mmWave systems where directional beamforming is used. One example is available from \cite{ 8550811}, where a multibeam approach is proposed to flexibly generate communication and sensing sub-beams using analogue antenna arrays. Optimization of combining the two sub-beams is further investigated in \cite{Luo19jcas,9131843}. Of course, the efficiency of multibeam is related to the requirements of C\&S. According to \cite{WinNT}, getting the correct solutions of beam steering and beamwidth adaptation for JCAS operation highly depends on environmental context. Indeed, reflector position, blockage height, motion speed and other environmental context factors could have a significant impact on the efficiency of the multibeam method. 

For waveform optimization in PMNs, the following specific problem associated with multiuser access is yet to be considered, particularly for uplink sensing. For downlink sensing, multiuser access and multiuser interference only need to be considered for communications, because the transmitted signals are known to the sensing receiver and the environment to be sensed is common to multiuser signals. Thus waveform optimization only needs to consider the multiuser aspect for communication, as studied in \cite{Liu18mumimo}. However, for uplink, signals need to be specific to each user for both C\&S, because the signal propagation environments between different users and the BS could be different. But these environments could also be correlated. Thus waveform optimization in the uplink is a more challenging task.

{A brief summary of recent analytic results of waveform optimization in JCAS system is given in Table \ref{tab-waveformop}}.

\begin{table*}[tb]
\caption{ Recent analytic results of waveform optimization in JCAS.}
	\centering
	\bgroup
	\def\arraystretch{1.5}
	\begin{tabular}{|m{1.5cm}|m{5cm}|m{10cm}|} 
		\hline 
		\textbf{Reference} & \centering\textbf{Models} & \hspace{1.5cm} \textbf{Main contributions}  \\ 
		\hline \hline
		\cite{RN32} &  Joint implementation of digital beamforming radar and MIMO communications. & Investigate waveform design for JCAS systems.
		   \\ 
		
		\hline 
	\cite{Liu18} & Joint design of MIMO radar and multi-user MIMO communication & Waveform optimization is realized via minimizing the difference between the generated signal and the desired sensing waveform under the restrictions of SINR for multiuser MIMO downlink communications.  \\ 
	\hline 
	\cite{Luo19jcas} &   C\&S at different directions, using a common transmitted signal & Optimization of combining both of the communication and sensing sub-beams is investigated.
		\\ 
		\hline 
	\cite{yliu17} & OFDM JCAS system  & The weighting vector for subcarriers in OFDM systems is optimized by considering a multi-objective function involving communication capacity and CRLB for the estimates of sensing parameters.
		
		\\ 
		\hline 
	\cite{LIU2017331} & OFDM integrated radar and communication system  & Adaptive weighted-optimal and Pareto-optimal waveform design approaches are proposed to simultaneously improve the estimation accuracy of range and velocity and the channel capacity for communication.   \\

		\hline 
		\end{tabular}
	\egroup
	 
	\label{tab-waveformop}
\end{table*}

\subsubsection{Optimization in Time and Frequency Domains}

In addition to spatial optimization, communication signals can also be optimized across the time and frequency domains, by jointly considering the communication and sensing requirements. The optimization can be with respect to the frame structure, subcarrier occupation, power allocation, and pilot design and typically requires some slight changes of the mobile signals.

The preamble part in a frame can be optimized with respect to both the signal format and resource allocation. A typical communication frame consists of preamble and data payload, and in cellular networks, they are structured via logical channels. The preamble typically contains unmodulated and orthogonal signals, which can be directly used for sensing, as described in Section \ref{sec-5gsignals}. For a MIMO-OFDM signal, the format of the preamble signals can be designed to mimic the traditional radar waveform, while maintaining the properties required for communications. For example, in \cite{Cui18mat}, orthogonal linear frequency modulation (LFM) signals, which are commonly used in MIMO radar, are generated based on MIMO-OFDM JCAS. A similar LFM signal is also generated in \cite{liu2019joint}, for a JCAS system using a hybrid antenna array. The subcarrier occupation of the preamble can also be designed by incorporating the idea of non-equidistant subcarriers in MIMO-OFDM radar \cite{Hakobyan20} to balance the performance of C\&R.

The spatio-temporal power optimization between preamble and data payload is investigated for JCAS in \cite{yuan2020waveform}. It is shown that the length of the preamble and its power allocation have a much larger impact on communications than on sensing. Using a cost function as the weighted sum of the MIs for C\&R, a closed-form solution is obtained for optimal power allocation.

The interval of preambles or pilots can also be optimized to improve the sensing performance, while maintaining communication efficiency. This can also be realized via varying the length of the data payload over packets, if packets are continuously transmitted. In \cite{Kumari20}, for a single data-stream single-carrier JCR system based on 802.11ad, non-uniformly placed preambles are proposed to enhance velocity estimation accuracy. It is found that when preambles/pilots are equally spaced, the performance of radar or communications cannot be effectively improved without affecting the other. Comparatively, non-uniform preambles/pilots are found to achieve a better performance trade-off between C\&R, particularly at large radar distances. Although developed for a single carrier system, an extension to MIMO-OFDM in PMN is possible with the usage of non-uniform pilots in both frequency and time domains.   

\subsubsection{Optimization with Next-Generation Signaling Formats}
Most of communication-centric JCAS systems have been formulated on either signal carrier or OFDM(A) systems, which are consistent with those waveforms being used in radar. Joint JCAS waveform design may also be applied to next-generation communication signals such as orthogonal time-frequency space (OTFS) signaling and fast-than-Nyquist (FTN) modulation.

The OTFS signaling is developed to address the signal reception problem in both frequency and time selective channels, and is believed to be more effective than OFDM in such channels. The signal may be directly modulated in the so-called delay-Doppler domain, where the channels are shown to be sparse. Equalization of OTFS signals requires channel estimation, which can be efficiently represented by sparse parameters of delay, angles, and Doppler frequency. Thus the receive processing of OTFS can be naturally linked to sensing parameter estimation. In \cite{Gaudio20}, the effectiveness of using OTFS for JCAS is investigated. It is shown that using OTFS signals, like using OFDM, can generate as accurate sensing parameter estimation as using conventional radar waveform such as frequency modulated continuous wave (FMCW). \bb{In \cite{9473534}, an OTFS JCAS system is studied by explicitly taking into account the inter- symbol interference (ISI) and inter-carrier interference (ICI) effects. It is shown that ISI and ICI can be exploited to extend the maximum unambiguous detection limits in range and velocity. A generalized likelihood ratio test based detector/estimator considering the ISI and ICI effects is developed. To reduce the conventionally high complexity of OTFS sensing, an efficient Bayesian learning scheme is proposed in \cite{9414107}, together with the reduction of the measurement matrix's dimension by incorporating the prior knowledge on the motion parameter limit of the true targets. These works demonstrate the feasibility and potential efficiency of OTFS JCAS systems.} Although results on waveform optimization for OTFS JCAS is not available, we surmise that it can be conducted in a way similar to OFDM. In particular, the precoding may be more efficiently applied beyond the spatial domain, such as to the delay-Doppler domain.

\subsection{Antenna Array Design}

For radio sensing, each antenna with an independent RF chain is like a pixel in the camera. But a radio system allows more flexible control and processing of both transmitted and received signals. Therefore, there are more designs for antenna arrays in PMNs that we can do apart from the MIMO precoding for waveform optimization as discussed in the last subsection. A classification of these techniques is presented in Table \ref{tab-array}, including virtual array design, sparse array design, and spatial modulation. These techniques are elaborated on below. 

\begin{table*}[tb]
	\caption{ Classification of antenna array design techniques in PMN.}
	\centering
	\bgroup
	\def\arraystretch{1.5}
	\begin{tabular}{|p{2.5cm}|p{11cm}|p{3cm}|} 
		\hline
		\textbf{Techniques} & \centering{\textbf{Key Ideas}} & \textbf{Examples} \\
	\hline
	\textbf{Virtual MIMO and Antenna Grouping (VI.C.1)} & Group antennas and form virtual subarrays to achieve a balance between spatial multiplexing and diversity for communications, and spatial resolution and beamforming gain for sensing. & \cite{5419124,Qin20,8550811,liu2019joint}\\
	\hline
	\textbf{Sparse Array Design (VI.C.2)} & Optimize the number and placement of antenna elements in an array by jointly considering the sparsity in sensing and communication channels. & \cite{wang2018sparse,Wang18} \\
	\hline
	\textbf{Spatial Modulation (VI.C.3)} & Use the indexes and/or order of antennas to convey information bits. The randomness added to JCAS signals is also shown to improve sensing performance in some cases. & \cite{Hassanien_2019, Wuicc20,ma2020spatial}\\
	\hline
	\end{tabular}		
	\egroup																								
	\label{tab-array}
\end{table*}

\subsubsection{Virtual MIMO and Antenna Grouping}
There are many contradictory requirements for antenna array design between C\&S. Beamforming and antenna placement are two good examples. For beamforming, an array with varying beamforming and narrow beamwidth is typically required for sensing;  however, during at least a packet period, communications require fixed and accurately pointed beams to obtain non-time varying channels, and multibeam to support SDMA. For antenna placement, MIMO radar typically requires special antenna intervals to achieve increased virtual antenna aperture \cite{5419124}; while MIMO communications focuses on beamforming gain, spatial diversity and spatial multiplexing, therefore, low correlation among antennas is more important. These different and contradicting requirements demand new antenna design methods. 

\begin{figure}[t]
	\centering
	\includegraphics[width=0.75\linewidth]{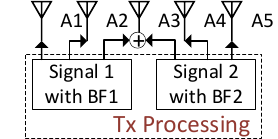}
	\caption{Two virtual subarrays with one overlapped antenna are formed: Virtual subarray 1 with antennas A1, A2 and A3; and subarray 2 with A3, A4 and A5. Each virtual subarray will transmit one data signal, and the signals between two virtual subarrays are orthogonal over time. Beamforming is applied in each virtual subarray. BF for beamforming.}
	\label{fig-var}
\end{figure}

One potential solution is to introduce the concept of antenna grouping and virtual subarrays \cite{5419124,Qin20}. By dividing existing antennas into two or more groups/virtual subarrays, we can designate tasks of C\&S and optimize the design across groups of antennas. There could be overlap between different groups of antennas, as shown in Fig. \ref{fig-var}. Virtual subarray introduces beamforming capability. Using orthogonal signals across virtual subarrays, we can maintain the orthogonality desired by MIMO radar, in order to achieve a larger aperture of an equivalent virtual array. Using overlapped antennas across neighbouring virtual subarrays can increase the spatial degree of freedom of the MIMO radar. Using virtual subarrays, we can also conveniently generate multibeams \cite{8550811} satisfying different beamforming requirements from C\&S. We can also virtually optimize the antenna placement, by antenna selection and grouping. Similar to the diversity and multiplexing trade-off in communications, there is a trade-off between processing gain and resolution in sensing, related to the number of independent spatial streams. 

The array structure as shown in Fig. \ref{fig-var} reminds us of the hybrid antenna array that has been widely studied in mmWave communications. Considering the benefits of antenna grouping for both C\&S, using hybrid antenna arrays \cite{RN36,8683591} will be an attractive low-cost option. This is particularly true for mmWave systems where propagation loss is high and beamforming gain is essential for achieving sufficiently high SNR for both C\&S. The research on hybrid array JCAS systems is still in its very early stage, with some limited results being reported in \cite{liu2019joint}.

\subsubsection{Sparse Array Design}
Besides antenna grouping, sparse array design is another method to exploit the degrees of freedom that can be achieved via configuring the locations of antennas when the total number of antennas is fixed.
 
The design of sparse arrays or thinned arrays \cite{9158394}, such as coprime arrays \cite{8472789}, is often cast as optimally placing a given number of antennas on a subset of a large number of (uniform) grid points \cite{wang2018sparse}. In this way, a small number of antennas can span a large array aperture with a high spatial resolution and low sidelobes. So far, the sparse array design-based JCAS has mainly been studied in integrating communication to radar systems, i.e., embedding information into radar waveforms to perform data communication \cite{wang2018sparse,Wang18}. In \cite{wang2018sparse}, antenna position and beamforming weights are optimized to design beams with mainlobe performing radar detection and sidelobe for communications through modulations like ASK or PSK. In \cite{Wang18}, the MIMO waveform orthogonality is further exploited to permute the waveform across selected antenna grids and hence convey extra information bits. 
	
Sparse array design is particularly suitable for massive MIMO arrays with tens to hundreds of antennas but a limited number of RF chains, i.e., switched arrays or hybrid arrays. This setup can provide more degrees of freedom and potential performance enhancement, with reduced cost, in PMNs. For example, the sparse array design can add index modulation to the communication part; while the sparse array design can provide better spatial resolution for radar detection. To this end, some interesting problems remain to be solved, such as how to formulate the problems with two goals satisfied and new trade-offs between C\&S.

\subsubsection{Spatial Modulation}
Spatial modulation uses the set of antenna indexes to modulate information bits and has been extensively investigated for communication systems \cite{8720024}. For multi-antenna JCAS systems, spatial modulation can also be potentially applied. In \cite{Hassanien_2019, Wuicc20}, a concept similar to spatial modulation is exploited to increase communication data rate in a frequency-hopping MIMO DFRC system. In \cite{ma2020spatial}, spatial modulation is applied to JCAS by allocating antenna elements based on the transmitted message, achieving increased communication rates by embedding additional data bits in the antenna selection. A prototype is developed in \cite{ma2020spatial} and demonstrates that the proposed scheme can improve the angular resolution and reduce the sidelobe level in the transmit beam pattern compared to using fixed antenna allocations.

Although these works are based on pulsed and continuous-wave radars, they can potentially be extended to PMN, by adding antenna selection to existing space-time modulations. In particular, the rich scattering environment in PMN provides a lower correlation between spatial channels, leading to potentially better performance. 

{A brief summary of recent analytic results of antenna array design in JCAS system is given in Table \ref{tab-antennarray}}.

\subsubsection{\bb{Reconfigurable Intelligent Surface Assisted JCAS}}

\bb{Reconfigurable Intelligent Surface (RIS) \cite{9140329}, also known as reconfigurable intelligent meta-surfaces or intelligent reflective surface, can be treated as a special type of relaying ``antenna'' arrays that are deployed to influence the signal propagation. The environmental objects are coated with artificial thin films of electromagnetic and reconfigurable meta-surfaces, which can be controlled to shape radio propagation. RIS provides a large volume of spatial degrees of freedom, which can be typically modeled as adjustable phase shifts. Using RIS can significantly improve communication performance by increasing beamforming gain, reducing interference, and reducing fading; it can also have a notable impact on sensing via generating location-dependent radio fingerprints and directed sensing. Therefore, RIS has been extensively studied for improving the performance of both radar \cite{9133157,buzzi2021foundations} and communications \cite{9122596}, separately. Although limited, the research on RIS-assisted JCAS is emerging.}

\bb{One direct application of RIS to JCAS is to treat the phase shifts in RIS as increased degrees-of-freedom in signal optimization. In \cite{9364358}, the precoding matrix of the transmitter and the phase shift matrix of the RIS are jointly optimized, with a problem formulated for maximizing the SNR at the radar receiver under the constrained SNR for communications. In \cite{9534682}, RIS is introduced as additional configurable channel parameters to assist the realization of a JCAS system based on a novel sparse code multiple access scheme.}  

\bb{Another important application of RIS to JCAS is to reduce the potential interference that may be caused by accommodating the sensing requirement, particularly for SDMA communications. In \cite{Wangtvt21}, the RIS technology is introduced to mitigate the multiuser interference, which may be increased when optimizing the beams by jointly considering the beamforming requirements for C\&S in a JCAS system. The JCAS waveform and the RIS phase shift matrix are jointly optimized, and the trade-off between C\&R performance is investigated. The work demonstrates that RIS can significantly improve the system throughput of communications, while without distorting the desired beampattern for sensing. In \cite{Wang21}, it is also exploited to deal with the interference in a coexisting MIMO radar and SDMA communication system.}
   
\bb{Given the extensive research of RIS in C\&S separately, we can expect rapidly growing research outputs on RIS-assisted JCAS. It would be interesting to see more diverse applications of RIS in JCAS, for example, how to use RIS to generate location-dependent radio fingerprints to enable simpler and more accurate sensing, while simultaneously improving the communication performance? Such work can better explore RIS's potential in JCAS, making it a game-changer in JCAS design, rather than just offering more adjustable phase values in an optimization problem.
}

\begin{table*}[tb]
\caption{ Recent analytic results for antenna array design in JCAS.}
	\centering
	\bgroup
	\def\arraystretch{1.5}
	\begin{tabular}{|m{1.5cm}|m{6.5cm}|m{8.5cm}|} 
		\hline 
		\textbf{Reference} & \centering\textbf{Models} & \hspace{1.5cm}\textbf{Main contributions}  \\ 
		\hline 
		\hline
		\cite{8550811} & C\&S at different directions, using a common transmitted signal.  & Generate multibeams satisfying different beamforming requirements from C\&S.
		   \\ 
		\hline 
\cite{Wang18} & Joint MIMO-radar communications platform equipped with a reconfigurable transmit antenna array.   & MIMO waveform orthogonality is exploited to permute the waveform across selected antenna grids and hence convey extra information bits.  \\

		\hline 
	\cite{wang2018sparse}& Joint radar communications platform equipped with a re-configurable transmit antenna array through an antenna selection network. 
	& Antenna position and beamforming weights are optimized to design beams with mainlobe performing radar detection and sidelobe for communications through modulations like ASK or PSK.  \\ 
		
		\hline 
	\cite{ma2020spatial} &  Joint system equipped with a phased array antenna implementing active radar sensing while communicating with a remote receiver.  & Spatial modulation is applied to JCAS by allocating antenna elements based on the transmitted message, achieving increased communication rates by embedding additional data bits in the antenna selection.
		
		\\ 
		\hline 
		\end{tabular}
	\egroup
	 
	\label{tab-antennarray}
\end{table*}

\subsection{Clutter Suppression Techniques}

Rich multipath in mobile networks creates another challenge for sensing parameter estimation in PMNs. In a typical environment, BSs receive many multipath signals that are originated from permanent or long-period static objects. These signals are useful for communications, but for a fixed BS, they are generally not of interest for continuous sensing because they bear little new information. Such undesirable multipath signals are known as \textit{clutter} in the traditional radar literature. In PMNs, we treat multipath signals as clutter if they remain largely unchanged and have near-zero Doppler frequencies over a period of interest. A lot of clutter could be present in the received signals because of the rich multipath environment of mobile networks. Clutter contains little information and is better to be removed from the signals sent to the sensing parameter estimator. In \cite{7421368}, the inner bounds of the impact of clutter on the performance of JCAS is evaluated. It is shown that clutter originating from motion objects can significantly degrade the inner bounds of the performance. 

\begin{figure}[t]
	\centering
	\includegraphics[width=0.95\linewidth]{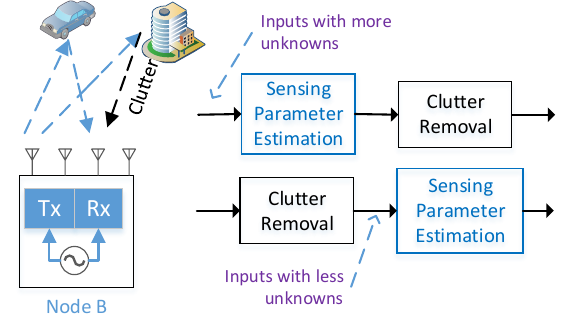}
	\caption{Clutter and two ways of clutter suppression.}
	\label{fig-clutter}
\end{figure}

There may be two ways of clutter suppression, as shown in Fig. \ref{fig-clutter}: doing suppression after or before estimating the sensing parameters. The former does not introduce signal distortion for sensing parameter estimation, and the latter can reduce the unknown sensing parameters to be estimated. High-end military/domestic radar can simultaneously detect and track hundreds of objects, and the capability is built on advanced hardware such as huge antenna arrays of hundreds and thousands of antenna elements. Thus both ways can be applied. For a PMN BS with only tens of antennas and limited signal bandwidth, the sensing capability largely depends on the sensing algorithm, which is closely related to the number of unknown parameters. Most existing parameter estimation algorithms require more measurements than unknown parameters and the estimation performance typically degrades with the number of unknown parameters increasing.  Therefore, it is crucial to identify and remove non-information-bearing clutter signals before applying sensing parameter estimation.  

Clutter suppression techniques for conventional radars are not directly applicable here because the signals and working environment for the two systems are very different. Typical radar systems are optimized for sensing a limited number of objects in open spaces using narrow beamforming, and clutter is typically from ground, sea, rain, etc. and has notable distinct features \cite{6159475,6843479, RN32,RN54}.  The well-known algorithms in radar systems, such as space-time adaptive processing \cite{6159475,1706131}, independent component analysis \cite{6843479}, singular value decomposition \cite{LIU2017125} and Doppler focusing \cite{6883164}, are adapted to such scenarios. These techniques also need to exploit different features of desired and unwanted echoes, such as low correlation between them. These different features may not always be available in mobile networks, because the desired multipath and clutter can come from the same classes of reflectors. For communications, narrow beamforming may occur in emerging millimetre wave systems, but not in more general microwave radio systems due to the limited number of antennas and the use of multibeam technology to support multiuser MIMO. The signal propagation environment in PMNs can also be very complex and different from a typical radar working environment. Therefore, existing clutter suppression methods developed for radar systems, e.g., those in \cite{RN54, RN50, RN52}, may not be directly suitable for clutter reduction in PMNs. 

Alternative approaches exploit the correlation in time, frequency and space domains, and use recursive averaging or differential operation to construct or remove clutter signals \cite{7944212,8827589, RN62,RN31}. These approaches could be more viable for perceptive mobile networks. They have similarities to \textit{background subtraction} in image processing \cite{SOBRAL20144}. However, there are two major differences:
\begin{itemize}
	\item  In image processing, the difference between two images is exhibited via pixel variation. In radio sensing, both Doppler shifts and variation in sensing parameters cause differences in received sensing signals at different time;
	\item  In an image, the background is overlapped/covered by foreground. In radio sensing, clutter and desired multipath signals are typically additive, and coexist in the received signals. 
\end{itemize}
Nevertheless, the many background subtraction methods developed for image processing can be revised and applied for radio sensing in PMNs. Below we review two types of typical background subtraction techniques that can be used in PMNs: \textit{recursive moving averaging (RMA) and Gaussian mixture model (GMM).  }

\subsubsection{Recursive Moving Averaging (RMA)}

Assume sensing parameters are fixed over the coherence time period, then ideally the received signals for each path at two different times will only have a phase difference caused by the Doppler phase shift. If the Doppler frequency is near zero, then the two signals are nearly identical. Based on this assumption, we can use an RMA method \cite{8827589} to estimate the clutter and then remove it from the received signal.

The received signals as shown in, e.g.,\eqref{eq-yf1} and \eqref{eq-yfnt0}, cannot be directly used in RMA, due to the transmitted signals $\xt$ and possibly the unknown timing offset $\tau_o$ which introduce a time-varying phase shift across discontinuous packets. They need to be removed or compensated before RMA can be applied. To demonstrate the idea of RMA, we refer to the signal stripping approach \cite{8827589}. We assume that $\tau_o=0$ and an estimate of the channel matrix $\Ht_n$ is available, given by $\hat{\Ht}_{n}=\Ht_{n}+\bm{\Delta}_{n}$, where $\bm{\Delta}_{n}$ is the reconstruction error. 

The RMA method uses a small forgetting vector to recursively average the received signal over a window, with a length sufficiently large to allow suppressing time-varying signals of non-static paths, but smaller than the coherent time. Mathematically, it can be represented as a recursive equation
\begin{align}
\bar{\Ht}_n(i)=\alpha\bar{\Ht}_n(i-1) +(1-\alpha)\hat{\Ht}_{n}(i),
\label{sec-recur}
\end{align} 
where $\bar{\Ht}_n(i)$ is the output at the $i$-th iteration, $\alpha$ is the forgetting factor (learning rate), and the initial $\hat{\Ht}_{n}(1)$ can be either zero or computed as the average of several initial $\hat{\Ht}_{n}(i)$s. In RMA, the window length can be adapted to the variation speed of the channels. The time interval between the inputs to the averaging determines how signals with different Doppler frequencies are added, either constructively or destructively. Hence it has a significant impact on suppressing signals of different Doppler frequencies. The forgetting factor and the window length determine the suppression power ratio. Although experimental results have been reported in \cite{8827589} for the relationship between these parameters and the effect of clutter estimation and suppression, optimal combinations of these parameters, in consideration of channel statistical properties, are yet to be studied. 

Although the RMA method works well in principle, it may become inefficient due to practical issues, such as timing and frequency offset commonly existing in actual systems. These signal imperfectness needs to be well compensated before the RMA method can work effectively.

\subsubsection{Gaussian Mixture Model}

 GMM has been widely used for analyzing and separating moving objects from the background in image and video analysis \cite{SOBRAL20144}, target identification and classification in radar system \cite{7995021}, and positioning solutions \cite{7233072}. The statistical learning of the GMM model with respect to the mean and variance in background subtraction is used to determine the state of each pixel whether a pixel is background or foreground.  It has also been applied recently to extract static channel state information from channel measurement in \cite{7752732}. Different from GMM in video analysis, where background and foreground overlap each other, clutter and multipath of interest in PMNs are additive and can coexist. {Therefore, it is infeasible for PMNs to place foreground (dynamic signals) and background (static signals) into two different sets by classical clustering approaches that happened in image or video signal processing.} 

GMM's working principle for clutter suppression in PMNs is as follows. Wireless channels can be modeled and estimated by a mixture of Gaussian distributions since each density represents the distribution of paths in the channel \cite{7752732,Rahmanclutter21}. Static and dynamic paths can be represented by Gaussian distributions with very different parameters over the time domain. This is because over a short time period, static paths change little and dynamic paths may vary significantly. It is also quite common that static paths typically have larger mean power than dynamic ones. Hence, in terms of their distributions, static paths have near-zero variances, which are much smaller than those of the dynamic ones. Therefore, by learning the mean values of the distribution, static paths can be identified and separated via comparing the variance.

The main advantage of GMM for clutter estimation in PMM is that much less samples are required to achieve a given accuracy, compared to the matched filtering and RMA methods. However, the estimation usually needs to be realized by high-complexity algorithms such as expectation maximization. Low-complexity estimation based on the GMM formulation is a key research problem here.

\subsection{Sensing Parameter Estimation}\label{sec-spe}
The tasks of sensing in PMNs include both explicit estimation of sensing parameters for locating objects and estimating their moving speeds, and application-oriented pattern recognition such as object and behavior recognition and classification. In this subsection, we review research on sensing parameter estimation, considering typical multiuser-MIMO OFDM signals used in modern mobile networks. We will review work on pattern recognition in subsection \ref{sec-patternana}.

\begin{figure}[t]
	\centering
	\includegraphics[width=0.95\linewidth]{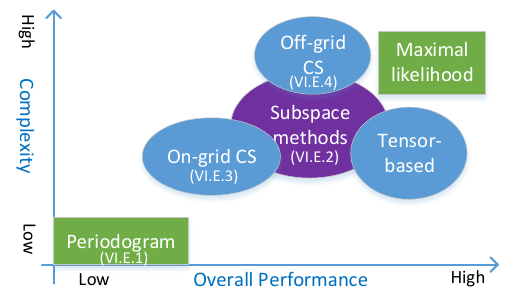}
	\caption{Sensing parameter estimation algorithms: Overall performance versus complexity. The overall performance considers estimation accuracy and application limitations. }
	\label{fig-sensing}
\end{figure}

We note that sensing parameter estimation is a non-linear problem, and hence most classical linear estimators, which have been widely used in channel estimation in communications, cannot be applied. Typical sensing parameter estimation techniques can be classified as follows: periodogram such as 2D DFT, subspace-based spectrum analysis techniques, on-grid compressive sensing  (CS) algorithms, off-grid CS algorithms and grid densification, and Tensor tools. Most of these techniques have higher complexity than classical channel estimation algorithms. Since the required sensing rate is typically in the order of milliseconds to seconds, such high computational complexity is affordable at BSs. Comparison of these techniques for sensing parameter estimation in PMNs is summarized in Table \ref{table:method} and illustrated in Fig. \ref{fig-sensing} in terms of the overall performance and complexity. Details of the research are elaborated below, together with additional techniques for sensing in clustered channels and resolution of sensing ambiguity.

\begin{table*}[t]
	
	\centering
		\def\arraystretch{1.4}
	\caption{Classification and comparison of Sensing Parameter Estimation Algorithms}
	\begin{tabular} {|m{3cm}|m{7cm}|m{6.5cm}|}
		
		\hline
		\textbf{Algorithms} & \centering\textbf{Properties} &\hspace{1.5cm}\textbf{Suitability and main limitation} \\
		\hline
		Periodogram such as \textbf{2D DFT and other Fourier transform based techniques} \cite{8827589}
		&
		Traditional techniques. Simple to implement. May be used as the starting point for other algorithms.
		&
		Low resolution. Generally, requires a full set of continuous samples 
		in all domains, which may not always be satisfied. 
		
		\\
		\hline
		\textbf{Subspace methods} such as MUSIC, ESPRIT and Matrix Pencil \cite{7935499,Liu20super,7448951,8812953,9529026}
		&
		High resolution and can do off-grid estimation. High complexity. Work with a small number of measurements. 
		
		&
		Typically require a large segment of consecutive samples, with the exclusion of the MUSIC algorithm, which may not always be satisfied.
		\\
		\hline
		\textbf{Compressive sensing (On-grid)} \cite{8827589,Rahmancs20,8918315,8788549}
		&
		Flexible. Does not require consecutive samples. Many options of problem formulations that can be tailored to signal models. Various recovery algorithms can be selected to adapt to complexity and performance requirements. 
		
		&
		Works well even for estimating a small amount of off-grid parameters. Performance can degrade significantly with many paths of continuous parameter values.
		
		\\
		\hline
		\textbf{Compressive Sensing (Off-grid) }such as atomic norm minimization \cite{RN45, Chi20cs, 8720024}
		&
		Flexible and do not require consecutive samples. Capable of estimating off-grid values.
		
		&
		Limitation in real time operation due to very high complexity. Still require sufficient separation between parameter values.
		\\
		\hline
		\textbf{Tensor-based algorithms} \cite{Nion10,9195839}
		&
		High-order formulation using the Tensor tools
		such as 3D Tensor CS simplifies computational complexity and provides capability
		in resolving multipath with repeated parameter values. Improve SNR by combining multiple measurements coherently.
		
		&
		Need to be combined with other algorithms such as ESPRIT and CS. Thus they face the inherent problems of these algorithms.
		\\
		\hline
		
	\end{tabular}
	
	\label{table:method}
	
\end{table*}

\subsubsection{Periodogram such as 2D DFT}
The classical 2D DFT method is a periodogram method  being widely used in radar. It can be used to coarsely estimate sensing parameters by combining two of the following three transformations: converting the time-domain samples to the frequency domain, spatial-domain samples to angle domain, and phase shifting samples to Doppler frequency domain. A 3D DFT may also be used. But due to the complexity, it is generally replaced by two or three 2D DFTs. The resolution of this method is low because of the long tail of the inherent sinc function in the DFT. A windowing operation can be applied to slightly improve the resolution. This method typically requires a full set of continuous measurements in time or frequency domain, which can limit its application in PMNs due to the discontinuous samples.

\subsubsection{Subspace-Based Spectrum Analysis Techniques}
 Classical subspace-based spectrum analysis techniques such as MUSIC and ESPRIT can estimate parameters of continuous values with high resolution \cite{7935499,Liu20super}. However, the sensing accuracy of MUSIC depends on the searching granularity, and ESPRIT typically requires samples of equal intervals. Techniques that can deal with non-uniform sampling have been proposed, e.g., the coupled canonical polyadic decomposition approach in \cite{7448951} and the generalized array manifold separation approach in \cite{8812953}, but they have very high computational complexity. To achieve high resolutions, MUSIC and ESPRIT typically also require a large number of samples so that the signal subspace and noise subspace can be well separated via computing the signal correlation matrix. This may not always be available in some domains, such as the spatial domain, which would require a large number of antennas. However, it may be a good option to combine them with other techniques for sensing parameter estimation, by exploiting their capabilities of high resolution and estimating parameters of continuous values. \bb{One good example can be seen from \cite{9529026}, where MUSIC is first used to obtain angle estimates via the spatial correlation matrix, and the delay and Doppler are then estimated by forming a maximum likelihood estimation problem.}
 
\subsubsection{On-Grid Compressive Sensing Algorithms}
Compressive Sensing (CS) techniques \cite{RN25} have been widely used in communication systems for channel estimation \cite{RN34,RN55, RN39} and in radar systems \cite{Hadi2015CompressiveSA}. CS techniques formulate parameter estimation as a sparse signal recovery problem, which can be solved by many algorithms such as $l_1$ recovery (convex relaxation), greedy algorithms and probabilistic inference \cite{RN25}. At the least, only twice the number of samples are required to accurately recover a certain number of unknown parameters, in the noise-free case. Typical CS techniques use on-grid quantized dictionaries, and hence errors are caused due to quantization when the original parameters have continuous values. One main advantage of CS for sensing parameter estimation in PMNs is that it does not require consecutive samples. Actually, higher randomness of samples in time, frequency and spatial domains can generally lead to better estimation performance. 

The sensing parameters to be estimated in PMNs include delay, AoA and Doppler in three different domains. Sometimes, the AoD and magnitude of a path are also of interest, which are not considered here. Since the signals are relatively independent in the three domains, they can be formulated in a high-dimension (3D here) vector Kronecker product form or even Tensor form. Therefore, we can apply 1D to 3D CS techniques to estimate these sensing parameters \cite{Rahmancs20}. The following two problems need to be considered when selecting CS techniques of different dimensions. 

\begin{itemize}
	\item \textit{Quantization error and number of available samples:} Although high-dimensional on-grid CS algorithms such as the Kronecker CS \cite{Caiafa13} could offer better performance, they require more samples than unknown variables in each dimension. In a typical BS, we can get a sufficient number of observations for the delay (linked to subcarriers), a reasonable number of samples in the Doppler frequency domain (linked to intermittent packets over a segment of channel coherent period), and a limited number of AoA observations (linked to antennas). 
	\item \textit{Complexity:} Exploiting the Kronecker CS property, the computational complexity is in the order of the product of the complexity in each domain, which is typically proportional to the cube of the number of samples.  
\end{itemize}

Therefore, a high dimensional CS algorithm is not always the viable option, particularly for the Doppler frequency and AoA estimation due to the limited number of samples. Comparatively, mobile signals generally have tens to thousands of subcarriers, which provide numerous samples for delay estimation. Thus, we can formulate two multi-measurement vector (MMV) CS problems, by stacking spatial-domain and Doppler-frequency domain signals, respectively with frequency domain signals. From the MMV-CS amplitude estimates, we can then estimate the AoA and Doppler frequencies \cite{8827589,8905229}. The details of CS algorithms from 1D to 3D and their performance are presented in \cite{8905229,Rahmancs20}. One common problem associated with using lower dimension CS is that parameters with overlapped values in one or more dimensions cannot be separately estimated. In this case, techniques such as the one proposed in \cite{8790748} can be used, by taking advantage of  the capability of model-based algorithms, for example, modified matrix enhancement and matrix pencil. 

For multiuser-MIMO signals, for example, signals received at an RRU from multiple RRUs in downlink passive sensing, we can use two methods to  formulate the CS problems \cite{8827589}. The first, \textit{direct sensing} method, directly uses the received signals as inputs to CS sensing algorithms. Since the receiver knows the transmitted information data symbols, the problem can be formulated as a block CS model  \cite{6415293, 6338280,8827589}, without decorrelating signals from multiusers. Correlation between the parameters can also be exploited in this model, via introducing intra-block correlation coefficients. The second, \textit{indirect sensing}, is based on \textit{signal stripping} that decorrelates signals between users \cite{8827589,RN30}. Then the sensing parameters can be estimated for each individual user by conventional CS algorithms. Direct sensing can achieve better performance than indirect sensing, as the decorrelation process introduces noise enhancement, at the cost of higher complexity. If the data symbols are unknown, e.g., in uplink sensing, decorrelating and demodulation errors also exist. Such errors may be explicitly considered and removed in the estimation \cite{8918315}. In \cite{8918315}, a passive sensing algorithm for multiple objects is proposed by using demodulated signals. The delay-Doppler values are estimated by exploiting the sparsity of the demodulation errors and numbers of objects. The positions and velocities of objects are then estimated based on the estimated delay-Doppler, using neural network techniques.

Overall, on-grid CS algorithms are promising for sensing parameter estimation in PMNs. However, the quantization error is a major problem as true sensing parameters have continuous values. For parameters of continuous values, there exist a mismatch between the assumed and actual dictionaries, generally known as ``dictionary mismatch'', which can cause significant performance degradation \cite{5710590}.  The degradation is severer when the number of unknown variables is larger. Therefore, resolving the quantization error and dictionary mismatch is a major challenge here.

\subsubsection{Grid Densification and Off-Grid CS Algorithms}
There are mainly two types of techniques that have been developed to tackle the quantization error problem in CS: grid densification and off-grid CS algorithms \cite{RN45,RN37}. Both techniques have higher complexity than conventional on-grid CS algorithms.

Grid densification uses denser dictionaries to reduce quantization error. The discretization of the physical space is unavoidable since CS has been focused on the signals that can be represented under a finite dictionary by reconstruction. It is intuitively reasonable that both dictionary mismatch and parameter estimation error can be reduced with a dense grid. Therefore, the question comes whether a denser grid leads to more accurate sparse signal recovery or not. In fact, according to the CS theory, the sampled grids should not be too dense. As in densely sampled grids, the dictionaries have a high inter-column correlation. The high correlation of dictionary items violates the restricted isometry property (RIP) condition of CS \cite{RN34}. This is particularly of concern when the SNR is not very high. Therefore, there is a trade-off in dictionary mismatch and estimation accuracy while constructing a densified dictionary. Dynamic dictionaries with multi-resolution capability are proposed to resolve this problem. For example, in \cite{8788549}, a dynamic dictionary-based re-focused DOA estimation method is developed with the number of extremely sparse grids refined to the number of detected sources. 

There are extensive research interests in extending CS to off-grid models, via, e.g., the perturbation method \cite{0266-5611-29-5-054008}, CS plus maximal likelihood \cite{8668429}, and the atomic norm minimization (ANM) method \cite{RN45, Chi20cs, 8720024}. The ANM method \cite{RN45, Chi20cs} can handle a continuous dictionary and recover unknown variables with a reasonable number of samples at a high probability via a semi-definite program. It has been widely applied for channel estimation in, e.g., generalized spatial modulation systems \cite{8720024}, MIMO radar via MMV models \cite{8708914}, and mmWave MIMO systems with planar arrays \cite{8432470}. However, the ANM method still requires that the variables such as delays have well-separated values. This may not always be satisfied in PMNs as an object may not always be approximated as a point reflector/scatter and reflected/scattered signals may come in clusters due to the limited distance among the transmitter, the object and the receiver. Enhancing the ANM method and making it capable of handling such signals are important for its practical application in PMNs.

\subsubsection{Sensing with Clustered Multipath Channels}\label{sec-challenge3}
In cluster sparsity patterns, non-zero taps of sparse signal appear in clusters rather than being arbitrarily spread over the vector, which means that sparse signal exhibits a structure in the form of non-zero coefficients occurring in clusters. In practice, multipath signals in mobile systems often arrive in clusters \cite{1287404}, and paths from one cluster typically come from the same scatter(s) and have similar parameter values. The situation becomes complex once the clusters originated in a propagation scene have correlation among other clusters of the same user and across different users. Eventually, getting sensing parameters from delay or spatial domain without acknowledging the channel cluster structure can create accuracy problems.  

We can find several research results on reconstructing cluster sparse signals in general, for example, through periodic compressive support \cite{7032149}, model-based CS \cite{Yu:2015:MBB:2735593.2739194}, variational Bayes approach \cite{7558378}, and block Bayesian method \cite{6288632}. The exploitation of the cluster property in multipath channels for sensing parameter estimation in PMNs is possible through creating a prior probability distribution. In particular, a cluster prior probability density function needs to be introduced in the CS reconstruction algorithm in order to efficiently detect the coarse locations of the clusters, leading to more accurate sparse reconstruction performance when CS algorithms are applied \cite{Rahmancs20}. Detailed technology on how cluster sparsity can be exploited in JCAS systems such as PMNs that involve OFDMA and multi-user MIMO is yet to be developed. 

\subsection{Resolution of Sensing Ambiguity}\label{sec-resamb}

\begin{figure}[t]
	\centering
	\includegraphics[width=0.9\linewidth]{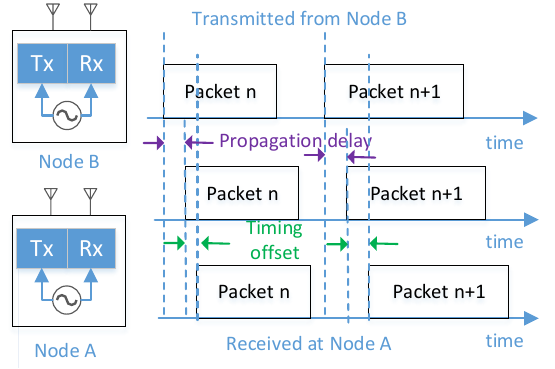}
	\caption{Illustration of the propagation delay $\tau_\ell$ and timing offset $\tau_o(t)$. $\tau_o(t)$ is time varying due to instability of the oscillators' clocks.}
	\label{fig-sync}
\end{figure}

As discussed in Section \ref{sec:structure}, there is typically no clock-level synchronization between a sensing receiver and the transmitter in PMNs, particularly in uplink sensing. In this case, there exist both timing and carrier frequency offsets in the received signals. The timing offset is illustrated in Fig. \ref{fig-sync}. Both of them, as shown in \eqref{eq-hfbasic}, are typically time-varying due to oscillator stability. In communications, timing offset can be absorbed into channel estimation and CFO can be estimated and compensated. Their residuals become sufficiently small and can be ignored. Differently, in sensing they cause measurement ambiguity and accuracy degradation. Timing offset can directly cause timing ambiguity and then ranging ambiguity, and CFO can cause Doppler estimation ambiguity and then speed ambiguity. They also prevent aggregating signals from discontinuous packets for joint processing, as they cause unknown and random phase shifting across packets or CSI measurements. Thus it is very important to resolve the clock timing offset problem. Should it be solved, uplink sensing can be efficiently realized, requiring little changes of network infrastructure.

There have been a limited number of works that address this problem in passive sensing \cite{passive10, IndoTrack, widar2.0}. A cross-antenna cross-correlation (CACC) method is applied to passive WiFi-sensing, to resolve the sensing ambiguity issues. The basic assumption is that timing offsets and CFO across multiple antennas in the receiver are the same, because the common oscillator clock is used in the RF circuits for all antennas. Hence they can be removed by computing the cross-correlation between signals from multiple receiving antennas. However, cross-correlation causes increased terms and unknown parameters. The sensing parameters also become relative ones. 

To proceed with the estimation of sensing parameters, it is assumed that line-of-sight (LOS) path exists and has much larger magnitude than the non-LOS (NLOS) paths. The UE and BS are assumed to be static and the relative location of the UE is also assumed to be known to the BS. In this case, terms multiplied with the LOS path are much larger than the others. The cross-product between LOS paths contain static multipath only and is invariant over the channel coherent time. Thus it can be removed by passing the correlation output through a high pass filter. The cross-terms between NLOS and LOS paths thus dominate in the output of the filter. The sensing parameters can then be estimated, with respect to the known parameters of the LOS path.  
   
The CACC method has been successfully used in WiFi-sensing. In \cite{widar2.0}, CACC is used to obtain estimates for ranges and velocities of targets. In \cite{IndoTrack}, CACC is adopted to get the \textit{AoA spectrum}, which represents the probabilities of the direction or angle of target. However, the outputs after CACC contain cross-product terms, which doubles the number of unknown parameters to be estimated. In particular, the relative delays and Doppler frequencies have values symmetric to zero. These increased symmetric terms are known as image components, which causes sensing ambiguity and degrades the performance of sensing algorithms. The authors in \cite{IndoTrack} proposed a method to suppress image signals, by adding one constant value to and subtracting another from to signals between the cross-correlation operation. The method works to some extent, however, it is found to be susceptible to the number and power distribution of static and dynamic signal propagation paths. Therefore, although the idea of CACC looks attractive in resolving the sensing ambiguity problem, more advanced techniques need to be developed to handle the output signals from CACC. 

The required setup to enable CACC working is practical in PMNs. For example, the fixed nodes that receive fixed broadband service are ideal options. Those with LOS-path links can also be found, particularly when mmWave communication is deployed. It is generally more challenging to extend this method to more complicated signal formats. In PMNs, the transmitted signals may also be optimized to enable better implementation of the CACC method. In \cite{ni2020uplink}, a mirrored MUSIC algorithm is proposed to handle the image components in the CACC output. Noticing the symmetry of unknown parameters, new signals and basis vectors are constructed by adding the original ones with their sample-reversed versions. This equivalently reduces the unknown parameters by half, and improve the estimation performance.

\bb{Another method for removing the random phase shifts due to clock asynchronism is using the ratio of CSI measurements across antennas \cite{Zengwifi19, Zengwifi2020}. The advantages of doing this, compared to the CACC method, are that (1) the measurement noise can be largely suppressed \cite{Zengwifi19}; and (2) the ratio may be better used as more information can be maintained compared to the cross correlation \cite{Zengwifi19,Zengwifi2020}. In \cite{Zengwifi19}, a close relationship is established between the ratio of CSI measurements and target movement, which enables the determination of movement direction and distance via the changes of the ratio. This method may be widely used in PMN sensing in the presence of clock asynchronism. For the particular respiration sensing application in \cite{Zengwifi19}, it is shown that the sensing range can be significantly extended with very high accuracy. The work is extended in \cite{Zengwifi2020}, where respiration sensing for multiple objects is studied. To be able to separate the measurement CSI signals for multiple persons, a blind source separation technique based on the independent component analysis (ICA) is applied. The ICA technique requires that the sources are mixed linearly. The direct ratio of CSI signals possesses non-linearity and needs to be modified. In \cite{Zengwifi2020}, a filter is designed through the genetic algorithm to nullify respiration signals with only static background signals left. Since the timing offset appears as a common phase shift in both the originally received and background signals, the filtered signal can then be used as the denominator in the CSI ratio to suppress the timing offset in the respiration signals. In this way, the modified ratios form linear combinations in the total observations, and ICA can be applied. These techniques may find good applications in PMN, particularly when the main goal is to coherently use CSI measurements for sensing, and/or for extracting the Doppler frequencies.}

\subsection{Pattern Analysis}\label{sec-patternana}

Using radio signals, high-level application-oriented object, behavior and event recognition and classification can be achieved by combining machine learning and signal processing techniques. They can be realized with or without using the sensing parameter estimation results, which provide location and velocity information. 

The feasibility and benefits of applying machine learning technologies to  communication systems have been well demonstrated, for example, fast beamforming design via deep learning \cite{8880526}, behavioral modeling and linearization of wideband RF power amplifiers in 5G system\cite{8750814}, vehicular network modeling in 6G by machine learning \cite{8926369}, route computation for software-defined communication systems by deep learning strategy \cite{ 8642376}, and heterogeneous network traffic control by deep learning \cite{ 7792369}. 

Although the work on pattern analysis using mobile signals is still in its infancy stage, we have seen some interesting examples, such as  \cite{RN48, RN43,doi:10.1002/wat2.1289}. We can foresee its booming in the near future, as we have been observing from many successful WiFi sensing applications. Using WiFi signals for object and behavior recognition and classification has been well demonstrated \cite{7875148,7384744,8067693}. Mobile signals are more complicated than WiFi signals, and the outdoor propagation environment is also more challenging. However, the PMNs have more advanced infrastructure than WiFi systems, including larger antenna arrays, more powerful signal processing capability, and distributed and cooperative nodes. Using massive MIMO, a PMN BS equivalently possesses a massive number of ``pixels'' for sensing. It is able to resolve numerous objects at a time and achieve imaging results with better field-of-view and resolution, like optical cameras. 

\begin{figure}[t]
\centering
\includegraphics[width=1\linewidth]{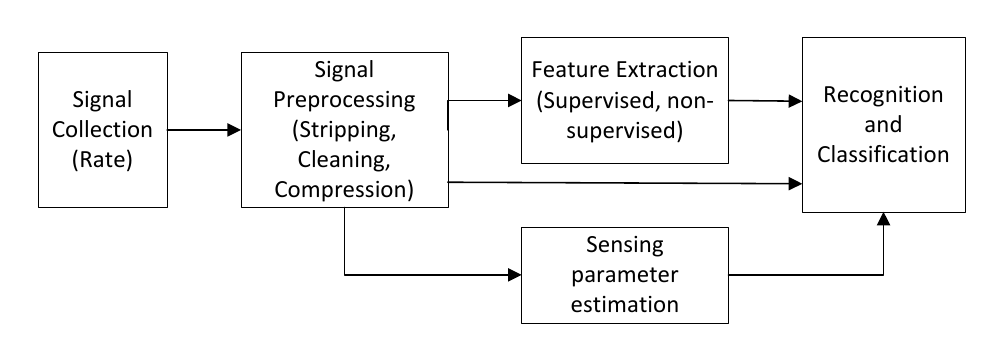}
\caption{Block diagram showing the procedure for pattern recognition.}
\label{fig-patr}
\end{figure}

Based on the various approaches developed for WiFi sensing, we can deduce the procedures of applying pattern analysis to mobile signals, as shown in Fig. \ref{fig-patr}. They typically involve four steps: signal collection, signal preprocessing, feature extraction, and recognition and classification. In the signal collection step, the signals are collected at the receiver according to the desired rate. In the signal preprocessing step, the collected signals may be stripped, cleaned and compressed. Signal stripping removes the modulated symbols from the received signal, and hence the pure channel state information (CSI) is obtained. Multiuser signals may also be decorrelated here. Signal cleaning removes signal distortions associated with, e.g., timing, CFO and phase noise, and suppresses clutter signals. The purpose is to keep mostly information-carrying signals. Many of the algorithms described before can be applied for this purpose. If signals arrive irregularly in the first step, the CSI can also be interpolated here if desired. Signal compression makes the signal condense, so that the useful information can be enhanced and the processing complexity in the following steps can be reduced. Common compression techniques include principal component analysis and correlation \cite{8761445}. Feature signals are then extracted from preprocessed signals, using machine learning techniques such as supervised and non-supervised deep learning. Finally, recognition and classification are conducted, with inputs from the extracted feature signals, the preprocessed signals, and estimated sensing parameters. 

\subsection{Networked Sensing under Cellular Topology} \label{sec-networked}

PMNs provide great opportunities for radio sensing under a cellular structure, which could be well beyond the scale and complexity of distributed radar systems.  The main challenge for networked sensing under a cellular topology remains in the way to address competition and cooperation between different nodes for sensing performance characterization and algorithm development. The research in this area is almost blank at the moment. Here, we envision two potential research directions.

 \subsubsection{Fundamental Theories and Performance Bounds for ``Cellular Sensing Networks''}
This is about investigating the potentials of the cellular structure on improving the spectral efficiency and performance of sensing, and developing fundamental theories and performance bounds for such improvement. Similar to communications, a cellular network intuitively also increases frequency reuse factor and hence the overall “capacity” for sensing. Stochastic geometry model may be an excellent tool for analyzing the dynamics in the sensing network, as have been applied to characterize the aggregated radar interference in an autonomous vehicular network in \cite{Al-Hourani, 8967012}. New metrics such as the \textit{radar detection coverage probability} \cite{9114637, ram2021optimization} can also be introduced to characterize the sensing performance with the application of stochastic geometry models. Both intra-cell and inter-cell interference would then be taken into consideration in deriving the mutual information for networked sensing. 

\subsubsection{Distributed Sensing with Node Grouping and Cooperation} 
One way of exploiting networked sensing is to develop distributed and cooperative sensing techniques by scheduling and grouping UEs and enable cooperation between RRUs. On one hand, existing research has shown that distributed radar techniques can improve location resolution and moving target detection by providing large spatial diversity and wide angular observation \cite{Liang11}. Such diversity can be maximized by optimizing both waveform design and the placement of radar nodes. In PMNs, we can group multiple UEs’ sensing results to improve uplink sensing. On the other hand, distributed radar can enable high-resolution localization, exploiting coherent phase differences of carrier signals from different distributed nodes \cite{Haim08}. This requires phase synchronization among radar nodes, and can only be potentially achieved in downlink sensing by grouping RRUs. For both cases, we may develop distributed sensing techniques, leveraging extensive research works on distributed beamforming and cooperative communications. One example of cooperative detection and localization/sensing is depicted in \cite{9186070}, where  a cooperative target detection scheme based on a generalized likelihood ratio test detector is proposed. The impact of network dynamics, including position uncertainty and the number of collaborative nodes, on the detection performance is also evaluated. Another example can be seen from \cite{8847233}, where cooperative passive coherent location is explored in the 5G radio network. The timing-ambiguity problem, as mentioned in Section \ref{sec-resamb}, may also be efficiently resolved via cooperative sensing, in a way analogous to the triangulation method for the removal of timing asynchronism in current localization systems \cite{9113730}.


\subsection{Sensing-Assisted Communications} \label{sec-sensingassit}

When communication and sensing are integrated, it is important to understand how they can mutually benefit from each other. In the context of PMNs, at least the following techniques can be exploited to improve communications using sensing results: Sensing assisted beamforming and secure communications.

\subsubsection{Sensing Assisted Beamforming}

Beamforming is an important technique used for concentrating transmission at certain directions to achieve high antenna array gain, and is critical for mmWave systems. However, due to the narrow beam width, it is generally time-consuming in mmWave communication systems to find the right beamforming directions and update the pointing directions once the LOS propagation channel is blocked. Techniques exploiting the propagation information of sub-6GHz signals have been proposed for improving beamforming speed for mmWave communications in mobile networks \cite{8198818}. These techniques exploit the spatial correlation between channels for the two frequency bands, which, however, is site-specific and needs to be updated in real time because of the environmental dynamics. The translation may also be inaccurate because of the large difference in the signal wavelength between the two bands. Comparatively, using radar operating in a similar mmWave band can potentially provide detailed propagation information, which is ideal for beamforming update and tracking.

In \cite{859,Ali20}, radar aided beam alignment is investigated for mmWave communications, where a mmWave radar is installed on a BS, co-locating with the communication system. A compressive covariance estimation algorithm is proposed to estimate the covariance output at the output of the combiner of a mmWave hybrid receiver, from the covariance matrix of the signals received from radar. Simulation results demonstrate the feasibility of the proposed scheme, while there is a relatively large gap to the upper bound performance. In \cite{9162000}, it is further shown that using location information obtained via automotive radar can largely reduce the beam training overhead in mmWave communications. Since different frequency bands and two separate systems are used, there are always some limits in the similarity of the channels, and hence the performance is not ideal.

\begin{figure}[t]
	\centering
	\includegraphics[width=1\linewidth]{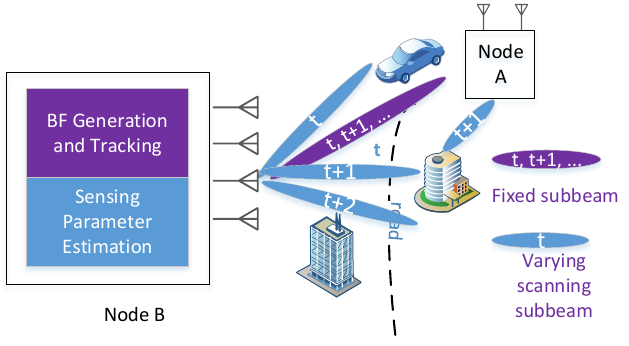}
	\caption{Sensing assisted communication using multibeam where a fixed sub-beam points at the communication receiver, and a scanning sub-beam varies pointing directions to sense the environment. The multibeam can be realized in all of analog, hybrid, and full-digital arrays. The sensing parameter estimation module in Node B can use the echo signals from both the fixed and the scanning sub-beams to establish a radio propagation map.}
	\label{fig-bf}
\end{figure}

Such limits can be removed by using the mmWave JCAS technology, as both radar sensing and communications are integrated to the same device and use the common signal now. JCAS enables a BS to construct a radio propagation map that limits to one hop in general. This map can provide detailed information for generating initial beamforming and updating it when either the BS or the UE moves, using downlink and uplink sensing \cite{8422132}. In particular, the multibeam scheme in \cite{8550811,9131843} introduces protocols and algorithms to enable communication and sensing in different directions at the same time even with an analog antenna array, as shown in Fig. \ref{fig-bf}. This makes it possible for a JCAS transmitter or a receiver to scan the surrounding environment and update the propagation map, while maintaining the communications. The basic idea of the multibeam scheme is to generate beamforming with multiple sub-beams, consisting of fixed sub-beams primarily for communications and packet-varying sub-beams primarily for scanning and sensing. Note that the same data signals are transmitted at the different sub-beams here. The basic sub-beams can be pre-generated, and are combined to readily generate the multibeam in real-time by optimizing the combination coefficients, using the optimization techniques described in \ref{sec-waveform}. The idea can also be extended to hybrid and full-digital arrays. 

In addition to achieving environmental awareness, radar sensing in PMNs can also be used for predictive beamforming and tracking. In \cite{Liu20}, a predictive beamforming design for a multi-user vehicular-to-infrastructure (V2I) network is proposed. The design introduces a different frame structure, by removing downlink pilots and uplink feedback parts in conventional training/tracking frames, because the BS (road-side unit) can actively sense the locations of the vehicles. An extended Kalman filtering (EKF) method is proposed for the tracking and predicting of the state of vehicles based on the sensing results. Thus instead of conventional beamforming prediction and tracking methods that are purely based on received signals, sensing information on vehicle location and signal propagation paths enables more efficient and direct prediction. In \cite{Yuan20Bayesian}, an advanced Bayesian predictive beamforming scheme exploiting JCAS is presented, using a message passing algorithm based on a factor graph. The scheme is shown to achieve near-optimal performance by the maximum a posteriori estimation. 

Despite that the works above demonstrate the feasibility and potential of sensing assisted beamforming in PMN, some major problems are yet to be tackled to make it practical. One problem is how to translate the sensing results to the beamforming design. In particular, the downlink sensing results are associated with the object, while the communication channels are associated with the object's antennas. Given the size of a vehicle, there may be a large offset between them. Another problem is how to deal with multiple reflections. When there are multiple reflections coming from different directions, the phase of signals will also play an important role in beamforming, but they cannot typically be accurately estimated in radar sensing. One potential solution to both problems is combining uplink and downlink sensing in PMN. Downlink sensing can provide quick and coarse information for initial beamforming, while uplink sensing can offer more detailed and accurate information in a complicated propagation environment.

\subsubsection{Sensing Assisted Secure Communications}
Radio sensing offers informative channel compositions for both active transmitters and passive objects in the surrounding environment. Such detailed channel information can be exploited for secure wireless communications, for which the research is still in its infancy.

One important application of the detailed channel composition information is in physical layer security techniques. Current physical layer security studies are mainly based on channel state information \cite{7467419}, using, e.g., artificial-noise-aided, security-oriented beamforming and physical-layer key generation approaches. This has been attempted in \cite{Su20Secure}, where the artificial noise method is employed to realize secure beamforming in a JCAS system, to combat eavesdroppers. Physical-layer key generation based on CSI has been widely investigated and is shown to be very effective for secure communications. However, the secrecy capacity of CSI is generally limited as it hides the channel propagation details and is the sum of many propagation paths. Comparatively, the sensing results contain more essential information about the environment between a pair of transmitter and receiver. They can motivate more informative secret-key generating methods and agreement in cellular communication networks. As a start, we can characterize the secrecy capacity of PMNs, and develop practical secret-key generating methods for information encryption.

\section{Concluding Remarks}\label{sec:conclu}
We have provided a comprehensive review on PMN, which integrates radio sensing into the current communication-only mobile network, using the JCAS techniques. Referring to the 5G NR standard, we have illustrated that uplink and downlink sensing can be realized with different degrees of modifications and enhancement to the current mobile network infrastructure. We have provided a detailed review of major research challenges, potential solutions and diverse research opportunities within the context of PMN.

In Table \ref{tab-marks}, we present our evaluation on the technology maturity and research difficulty, and highlight selected key open research problems for major techniques that have been reviewed in Section \ref{sec:wlk}. The scores for maturity and difficulty are based on our own expertise and experience, and are indicative only. As can be seen from the table, significant technical advancement for PMN is demanded in many areas, particularly in networked sensing, sensing-assisted communication, and application-driven pattern analysis. They also represent the most breakthrough that we may expect from the integration. With the integration, large-scale networked sensing becomes feasible at very low costs. Like networked communications based on the key idea of frequency reuse, it will motivate brand new sensing theories and technologies. Sensing-assisted communication will answer why we need sensing in communication networks, which would be the inspiration for communication equipment manufactures and operators to adopt the technology. With the combined use of machine learning, pattern analysis, and signal processing techniques, PMN with both communication and sensing capabilities will certainly motivate numerous new applications, which promises great opportunities for both research and commercial development.  

In addition to reviewing and evaluating these important technical problems in PMN, here we would also like to share some important lessons and tips that we have learned in the research and development of PMN systems, listed as follows.
\begin{itemize}
	\item It is important to cast the signal design and optimization problem within the framework of mobile networks. That is, consider the actual network structure, protocols, resource allocation, and organization of logical channels, so that the work can be really meaningful and be more creative. This is because in this case the sensing problem will be out of the comfort zone of conventional radar;	
	\item Problem formulation in PMN is more challenging compared to that in communications, because of the involvement of both C\&S functions, various parameters, and performance metrics. It is important to consider significant application scenarios and select the proper performance metrics in the research; and
	\item Sensing can typically be implemented and updated at a much slower speed compared to communications, which should be an important consideration in the joint design. 
\end{itemize}

To conclude, the PMN is expected to deliver a revolutionary ubiquitous radio sensing network that can significantly drive smart initiatives such as smart cities and smart transportation, integrated with enriched mobile communication. In relation to the (stereo) optical vision in camera sensing, the PMN is expected to realize 3D+ radio vision, including 3D location + speed + features for objects surrounding the radio transceivers, with additional attractive features such as day-and-night availability, fog/leaf-penetration, and continuous tracking. It will enable many new applications for which current sensing solutions are impractical or too costly. While there are significant challenges and a long way ahead to make the PMN fully operational, our survey here is a solid presentation, indicating the feasibility and providing the potential directions to pursue.

\begin{table*}
	\def\arraystretch{1.2}
	
	\caption{Technology matureness, research difficulty and selected key open research problems. Higher scores stands for more mature, and more difficult.}
	\centering
	\begin{tabular}{|m{1.7cm}|m{1.5cm}|m{1.5cm}|m{11.5cm}|}
		\hline 
		\textbf{Research Topics} & \textbf{Technology Matureness (1 to 10)} & \textbf{Research Difficulty (1 to 10)} & \textbf{Selected Key Open Research Problems} \\ 
		
		\hline 
		\textbf{Performance Bounds} & 5  &  7 & \begin{itemize}
			\item MI formulation specific to PMNs by considering uplink and downlink sensing, and actual signal and packet structure;
			\item Closed-form expressions of CRLB of sensing parameter estimation for broadband mobile signals;
			\item Combine MI and other metrics such as CRLB of estimates for performance characterization.
		\end{itemize}\\
		\hline 
		\textbf{Joint waveform optimization} & 6 & 5 &  \begin{itemize}
			\item Waveform optimization for hybrid antenna arrays;
			\item Low-complexity optimization schemes that can be quickly adapt to channel variation in both C\&S;
			\item Multiuser correlation in waveform optimization for uplink sensing.
		\end{itemize}
		\\ 
		\hline 
		\textbf{Antenna array design} & 3 & 7 & \begin{itemize}
			\item Using virtual array and antenna grouping techniques to achieve a balance between processing gain and resolution in sensing, and diversity and multiplexing in communications;
			\item Sparse array design and signal processing in PMNs. 
		\end{itemize}\\ 
		\hline 
		\textbf{Clutter suppression} & 7 & 5 & \begin{itemize}
			\item Parameter optimization in the recursive moving averaging method;
			\item Low-complexity algorithms for parameter estimation in Gaussian mixed model.
		\end{itemize}\\ 
		\hline 
		\textbf{Sensing parameter estimation} & 5 & 8 & \begin{itemize}
			\item Off-grid compressive sensing with discontinuous samples; 
			\item Off-grid Tensor signal processing algorithms;
			\item Sensing parameter estimation with clustered multipath channels;
			\item Resolution of sensing ambiguity with asynchronous nodes.
		\end{itemize}\\ 
		\hline 
		\textbf{Pattern analysis} & 2 & 5 & \begin{itemize}
			\item Application-driven problem formulation and pattern analysis;
			\item Environment robust algorithms.
		\end{itemize}\\ 
		\hline 
		\textbf{Networked sensing} & 1 & 8 & \begin{itemize}
			\item Fundamental theories and performance bounds for cellular sensing networks;
			\item Distributed sensing with node grouping and cooperation.
		\end{itemize}\\ 
		
		\hline 
		\textbf{Sensing-assisted  communication} & 2 & 6 & \begin{itemize}
			\item Sensing-assisted beamforming generation and tracking;
			\item Characterize the Secrecy capacity and develop practical code design methods for information encryption using sensing results.
		\end{itemize}
		\\ 
		\hline 
	\end{tabular} 
	\label{tab-marks}
\end{table*}

\appendix

\section*{Basic Signal and Channel Models}

We describe the basic signal and channel models here that involve some mathematics but may be helpful for the readers to understand the technologies better. We consider a CRAN system with $Q$ RRUs. Each RRU has a uniform linear array (ULA) with $M$ antenna elements spaced at half wavelength. These RRUs cooperate and provide links to $K$ users using multiuser-MIMO OFDMA. Each user has a ULA of $M_T$ elements, and may occupy and share a part of the total subcarriers with other users. The data symbols are first spatially precoded in the frequency domain, and an IFFT is then applied to each spatial stream. The time-domain signals are then assigned to the corresponding RRUs. Let $N$ denote the number of total subcarriers and $B$ the total bandwidth. Then the subcarrier interval is $f_0=B/N$ and the OFDM symbol period is $T_s=N/B+T_p$ where $T_p$ is the period of the cyclic prefix. 

\subsection{A General Channel Model}
The array response vector of a size-$M$ ULA is given by
\begin{align}
\av(M,\theta)=[1,e^{j\pi \sin(\theta)},\cdots, e^{j\pi (M-1)\sin(\theta)}]^H,
\end{align}
where $\theta$ is either AoD or AoA. Let the AoD and AoA of a multipath be $\theta_{\ell}$ and $\phi_{\ell}$, $\ell\in[1,L]$, respectively.  For $M_1$ transmitting and $M_2$ receiving antennas, the $M_2\times M_1$ frequency-domain baseband channel matrix at the $n$-th subcarrier in the $t$-th OFDM block is given by
\begin{align}
\Ht_{n}=&\sum_{\ell=1}^L b_\ell e^{-j2\pi n(\tau_{\ell}+\tau_o(t))f_0}e^{j2\pi t (f_{D,\ell}+f_o)T_s}\av(M_2,\phi_{\ell})\av^T(M_1,\theta_{\ell}),
\label{eq-hfbasic}
\end{align}
%
where for the $\ell$-th multipath, {$b_{\ell}$ is its amplitude of complex value accounting for both signal attenuation and initial phase difference},  $\tau_{\ell}$ is the propagation delay, and $f_{D,\ell}$ is the associated Doppler frequency, $\tau_o(t)$ denotes the potential timing offset due to asynchronous clocks between transmitter and receiver, and $f_o(t)$ denotes the carrier frequency offset (CFO). Note that in \eqref{eq-hfbasic}, we have approximated the Doppler phase changes over the samples in one OFDM block as a single value.   

Equation (\ref{eq-hfbasic}) represents a general channel model that can be used for both communication and sensing. Note that for communications, we generally only need to know the composited values of the matrix $\Ht_(n)$, which are typically obtained by directly estimating channel coefficients at some subcarriers and obtaining the rest via interpolation. For radio sensing, however, the system needs to resolve the detailed channel structure and estimate the \textit{sensing parameters} $\{\tau_{\ell},f_{D,\ell},\phi_{\ell}, \theta_{\ell}, b_{\ell}\}$. We define a \textit{coherent processing interval} (CPI) when all these parameters maintain almost unchanged, which is typically a few milliseconds (equivalent to the length of hundreds of OFDM symbols). When the oscillator clocks of the transmitter and receiver are not locked, the timing offset $\tau_o(t)$ and CFO $f_o(t)$ are nonzero. They may be unknown to the receiver and can cause ambiguity in range and speed estimation if not being removed in sensing. Their values could be time varying due to the instability of the oscillator clock. Both $\tau_o(t)$ and CFO $f_o(t)$ can be easily handled in communications; however, handling them for radar sensing is challenging, as will be detailed in Section \ref{sec-resamb}. 

\subsection{Formulation for Downlink Sensing}
For downlink sensing, each RRU sees reflected downlink signals from itself and the other $Q-1$ RRUs. For any RRU, a general expression of its received signal at the $n$-th subcarrier and the $t$-th OFDM block is given by
\begin{align}
\yf_{n,t}=&\sum_{q=1}^Q\sum_{\ell=1}^{L_q} b_{q,\ell} e^{-j2\pi n (\tau_{q,\ell}+\tau_{q,o})f_0}e^{j2\pi t f_{D,q,\ell} T_s}\cdot\notag\\
& \quad\av(M,\phi_{q,\ell})\av^T(M,\theta_{q,\ell})\xf_{q,n,t}  +\mathbf{z}_{n,t},
\label{eq-yf1}
\end{align}
where variables with subscript $q$ are for the $q$-th RRU, $\xf_{q,n,t}$ are the transmitted signals at subcarrier $n$ from the $q$-th RRU, and $\mathbf{z}_{n,t}$ is the noise vector. There is typically a common clock between the transmitter and receiver in one RRU, and between RRUs, hence the timing offset $\tau_{q,o}=0$. Note that the signals here are specifically collected and used for sensing, not required for communications. However, the sensing results can be used for improving communication performance, as will be discussed in Section \ref{sec:wlk}.

The signals $\xf_{q,n,t}$ are typically the precoded output. Let the precoder at the $n$-th subcarrier, $t$-th OFDM block of the $q$-th RRU be $\Pc_{q,n,t}$, and the data symbols be $\st_{q,n,t}$. Then 
\begin{align}
\xf_{q,n,t}=\Pc_{q,n,t}\st_{q,n,t}.
\end{align}
When joint precoding is applied across RRUs, the precoding matrices $\Pc_{q,n,t}, q=1,\cdots, Q$ are jointly designed. 

According to (\ref{eq-yf1}), we can see that packing $\yf_{n,t}$ from multiple RRUs can increase its length, but the unknown parameters are similarly increased. Hence sensing does not directly benefit from jointly processing, although there may be some correlation between sensing parameters for different RRUs. However, due to channel reciprocity, parameters for signal propagation between RRUs could be similar. Such a property can be exploited for joint processing across RRUs.

When RRUs' signals do not reach each other, then $Q=1$ in \eqref{eq-yf1}, and each RRU only sees its own reflected transmitted signals.

\subsection{Formulation for Uplink Sensing}\label{sec-direct}
For uplink signal transmitted from one or more UEs, the received signal at an RRU at the $n$-th subcarrier and the $t$-th OFDM block can be represented as
\begin{align}
\yf_{n,t}&=\sum_{k=1}^K\sum_{\ell=1}^{L_k} b_{k,\ell} e^{-j2\pi n (\tau_{k,\ell}+\tau_{k,o})f_0}e^{j2\pi t f_{D,k,\ell} T_s}\cdot\notag\\
& \quad\av(M,\phi_{k,\ell})\av^T(M_T,\theta_{k,\ell}) \xf_{k,n,t}  +\mathbf{z}_{n,t},
\label{eq-yfnt0}
\end{align}
When a single UE exclusively occupies the $n$-th subcarrier, $K=1$ in \eqref{eq-yfnt0}. For uplink sensing, the received signals for sensing and communications are identical.

Comparing \eqref{eq-yfnt0} and \eqref{eq-yf1}, we can see that their expressions are quite similar, except that the variables have different values. Therefore, many signal processing techniques as will be discussed in this paper are applicable to both downlink and uplink sensing.

\bibliographystyle{IEEEtran}
\bibliography{IEEEabrv,references}

\begin{IEEEbiography}
	[{\includegraphics[width=1in,height=1.25in,clip,keepaspectratio]{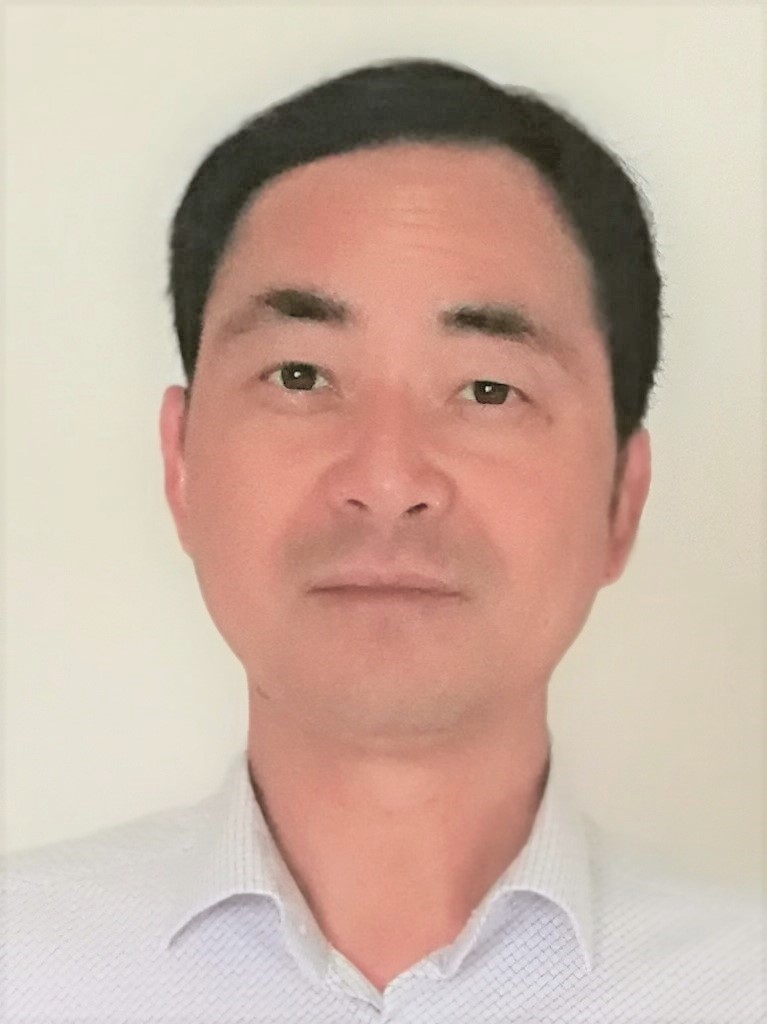}}]{J. Andrew Zhang} (M'04-SM'11) received B.Sc. degree from Xi'an JiaoTong University, China, in 1996, M.Sc. degree from Nanjing University of Posts and Telecommunications, China, in 1999, and Ph.D. degree from the Australian National University, in 2004. Currently, He is an Associate Professor in the School of Electrical and Data Engineering and the director of the Radio Sensing and Pattern Analysis (RaSPA) laboratory, University of Technology Sydney, Australia. He was a researcher with Data61, CSIRO, Australia from 2010 to 2016, the Networked Systems, NICTA, Australia from 2004 to 2010, and ZTE Corp., Nanjing, China from 1999 to 2001.
	
	Dr. Zhang's research interests are in the area of signal processing for wireless communications and sensing, with current focuses on joint communications and radio/radar sensing, and autonomous vehicular networks. He has published over 200 journal and conference papers, and has won 5 best paper awards including in IEEE ICC 2013. He is a recipient of CSIRO Chairman's Medal and the Australian Engineering Innovation Award in 2012 for exceptional research achievements in multi-gigabit wireless communications. He is serving as an Editor for IEEE Trans. on Communications, and the publication co-chair of the IEEE Communication Society's Integrated Sensing and Communication Emerging Technology Initiative (ISAC-ETI).
\end{IEEEbiography}

\begin{IEEEbiography}
[{\includegraphics[width=1in,height=1.25in,clip,keepaspectratio]{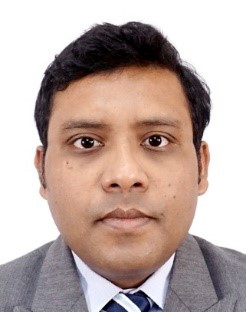}}]{Md Lushanur Rahman} received the Ph.D. degree in electrical engineering from University of Technology Sydney (UTS), Australia in 2020, the  M.Sc. degree in electrical engineering (with distinction) from Tampere University of Technology (TUT), Finland in 2014 and the B.Sc. degree in electrical and electronic engineering from Islamic University of Technology (IUT), Bangladesh in 2008.  His research interests are in the areas of signal processing for wireless communication, communication circuits and systems design and radio sensing.
\end{IEEEbiography}
	
	\begin{IEEEbiography}[{\includegraphics[width=1in,height =1.25in,clip,keepaspectratio]{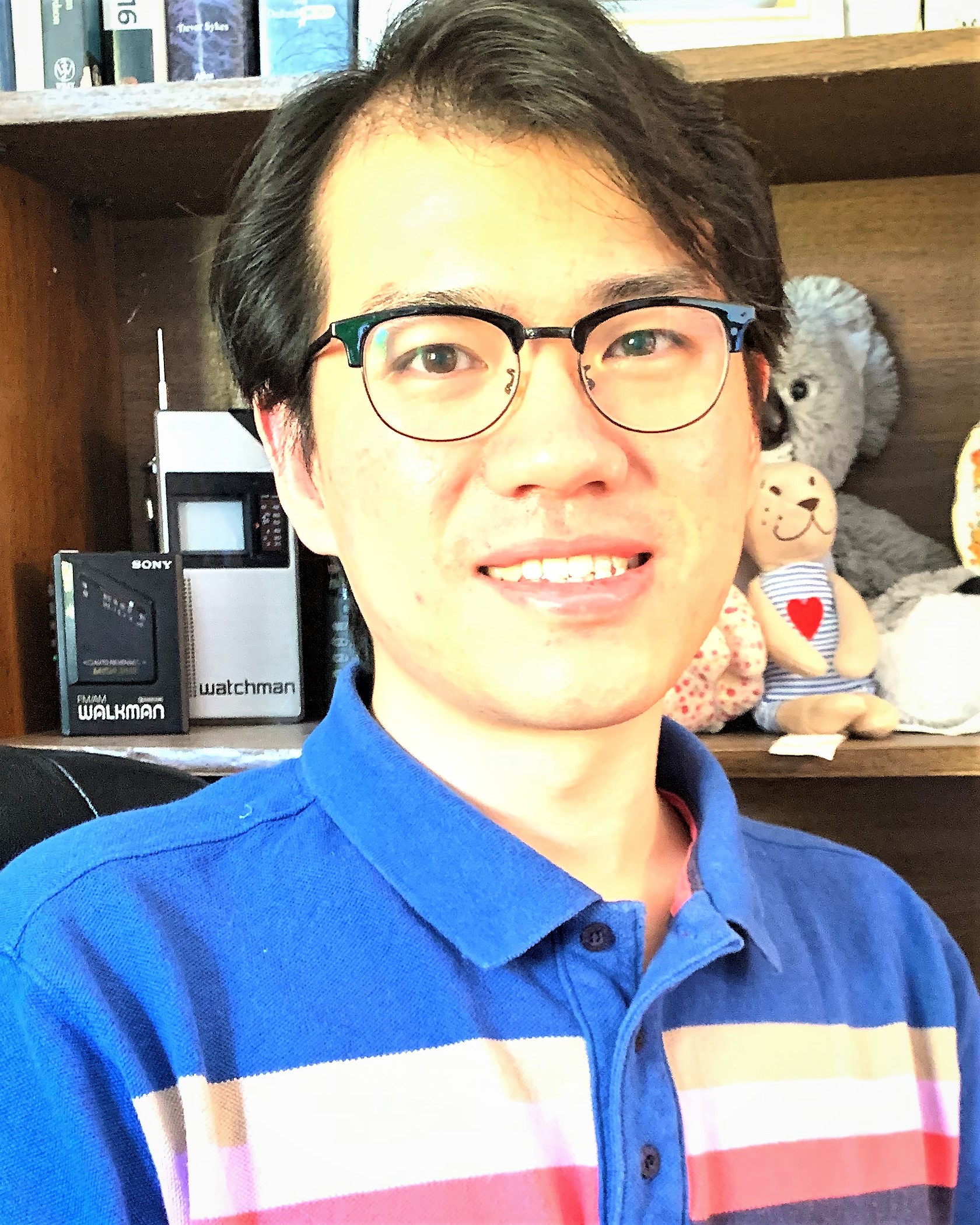}}]{Kai Wu} received the B.E. from Xidian University, Xi'an, China, in 2012, PhD from Xidian University in 2019, and PhD from the University of Technology Sydney (UTS), Sydney, Australia, in 2020. He is now a research fellow at the Global Big Data Technologies Centre, UTS. From Nov. 2017 to April 2018, he was a research assistant at the same centre. Before that, he was a visiting scholar at DATA61, CSIRO, Australia (Nov. 2016 - Nov. 2017). His research interests include array signal processing, and its applications in radar and communications.
	\end{IEEEbiography}

	\begin{IEEEbiography}
		[{\includegraphics[width=1in,height=1.25in,clip,keepaspectratio]{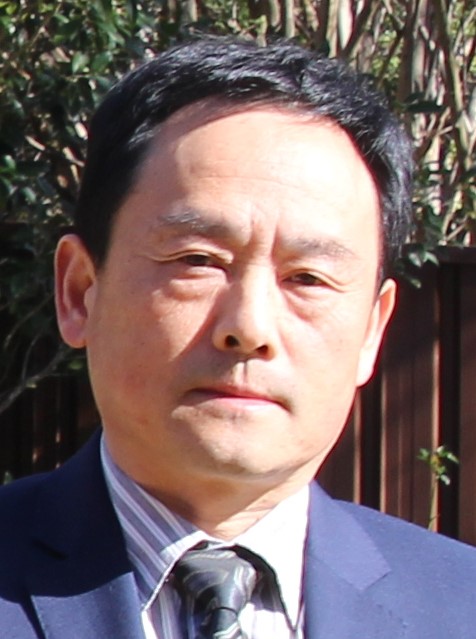}}]{Xiaojing Huang} (M'99-SM'11) received the B.Eng., M.Eng., and Ph.D. degrees in electronic engineering from Shanghai Jiao Tong University, Shanghai, China, in 1983, 1986, and 1989, respectively. He was a Principal Research Engineer with the Motorola Australian Research Center, Botany, NSW, Australia, from 1998 to 2003, and an Associate Professor with the University of Wollongong, Wollongong, NSW, Australia, from 2004 to 2008. He had been a Principal Research Scientist with the Commonwealth Scientific and Industrial Research Organisation (CSIRO), Sydney, NSW, Australia, and the Project Leader of the CSIRO Microwave and mm-Wave Backhaul projects since 2009.  He is currently a Professor of Information and Communications Technology with the School of Electrical and Data Engineering and the Program Leader for Mobile Sensing and Communications with the Global Big Data Technologies Center, University of Technology Sydney (UTS), Sydney, NSW, Australia. His research interests include high-speed wireless communications, digital and analog signal processing, and synthetic aperture radar imaging. With over 32 years of combined industrial, academic, and scientific research experience, he has authored over 330 book chapters, refereed journal and conference papers, major commercial research reports, and filed 31 patents. Prof. Huang was a recipient of the CSIRO Chairman's Medal and the Australian Engineering Innovation Award in 2012 for exceptional research achievements in multigigabit wireless communications.  
\end{IEEEbiography}

\begin{IEEEbiography}[{\includegraphics[width=1in,height =1.25in,clip,keepaspectratio]{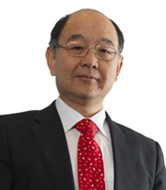}}]{Y. Jay Guo} (Fellow’2014) received a Bachelor Degree and a Master Degree from Xidian University in 1982 and 1984, respectively, and a PhD Degree from Xian Jiaotong University in 1987, all in China. His research interest includes antennas, mm-wave and THz communications and sensing systems as well as big data technologies. He has published four books and over 550 research papers including 300 journal papers, most of which are in IEEE Transactions, and he holds 26 patents. He is a Fellow of the Australian Academy of Engineering and Technology, a Fellow of IEEE and a Fellow of IET, and was a member of the College of Experts of Australian Research Council (ARC, 2016-2018). He has won a number of most prestigious Australian Engineering Excellence Awards (2007, 2012) and CSIRO Chairman’s Medal (2007, 2012). He was named one of the most influential engineers in Australia in 2014 and 2015, respectively, and one of the top researchers in Australia in 2020. 
	
Currently, he is a Distinguished Professor and the Director of Global Big Data Technologies Centre (GBDTC) at the University of Technology Sydney (UTS), Australia. Prior to this appointment in 2014, he served as a Director in CSIRO for over nine years. Before joining CSIRO, he held various senior technology leadership positions in Fujitsu, Siemens and NEC in the U.K. 
	
Prof Guo has chaired numerous international conferences and served as guest editors for a number of IEEE publications. He is the Chair of International Steering Committee, International Symposium on Antennas and Propagation (ISAP). He was the International Advisory Committee Chair of IEEE VTC2017, General Chair of ISAP2022, ISAP2015, iWAT2014 and WPMC'2014, and TPC Chair of 2010 IEEE WCNC, and 2012 and 2007 IEEE ISCIT. 
\end{IEEEbiography}
	
\begin{IEEEbiography}
[{\includegraphics[width=1in,height=1.25in,clip,keepaspectratio]{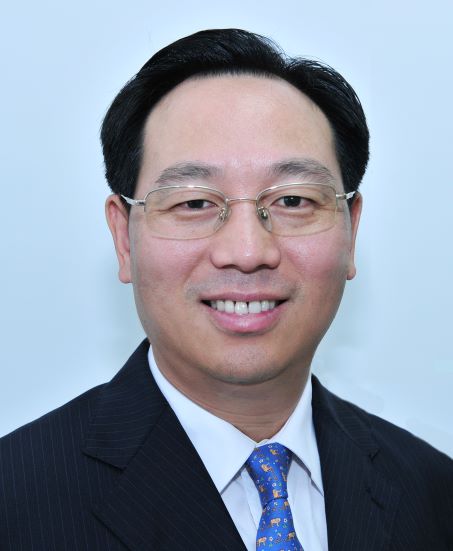}}]{Shanzhi Chen} (Fellow) received the Bachelor and Ph.D. degree from Xidian University and BUPT, China, in 1991 and 1997, respectively. He joined Datang Telecom Group and CATT in 1994, and has served
as EVP R\&D since 2008, and the director of State Key Laboratory of Wireless Mobile Communications. He has contributed to the design, standardization, and development of 4G TD-LTE, 5G and C-V2X system. His current research interests include B5G and 6G, C-V2X.	
\end{IEEEbiography}

\begin{IEEEbiography}			
		[{\includegraphics[width=1in,height=1.25in,clip,keepaspectratio]{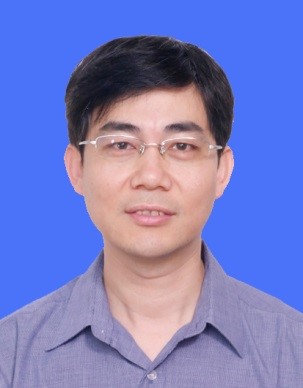}}]{Jinhong Yuan} (M'02--SM'11--F'16) received the B.E. and Ph.D. degrees in electronics engineering from the Beijing Institute of Technology, Beijing, China, in 1991 and 1997, respectively. From 1997 to 1999, he was a Research Fellow with the School of Electrical Engineering, University of Sydney, Sydney, Australia. In 2000, he joined the School of Electrical Engineering and Telecommunications, University of New South Wales, Sydney, Australia, where he is currently a Professor and Head of Telecommunication Group with the School. He has published two books, five book chapters, over 300 papers in telecommunications journals and conference proceedings, and 50 industrial reports. He is a co-inventor of one patent on MIMO systems and four patents on low-density-parity-check codes. He has co-authored four Best Paper Awards and one Best Poster Award, including the Best Paper Award from the IEEE International Conference on Communications, Kansas City, USA, in 2018, the Best Paper Award from IEEE Wireless Communications and Networking Conference, Cancun, Mexico, in 2011, and the Best Paper Award from the IEEE International Symposium on Wireless Communications Systems, Trondheim, Norway, in 2007. He is an IEEE Fellow and currently serving as an Associate Editor for the IEEE Transactions on Wireless Communications and IEEE Transactions on Communications. He served as the IEEE NSW Chapter Chair of Joint Communications/Signal Processions/Ocean Engineering Chapter during 2011-2014 and served as an Associate Editor for the IEEE Transactions on Communications during 2012-2017. His current research interests include error control coding and information theory, communication theory, and wireless communications.
\end{IEEEbiography}

 \end{document}